\documentclass[widetext,showpacs,preprintnumbers,superscriptaddress,nofootinbib]{revtex4}

\usepackage[dvipdfmx]{graphicx}
\usepackage{amssymb}
\usepackage{amsmath}
\usepackage{dcolumn}
\usepackage{bm}
\usepackage{natbib}
\usepackage{multirow}
\usepackage{subfigure}
\usepackage{url}

\makeatletter
\@addtoreset{equation}{section}
\def\theequation{\arabic{section}.\arabic{equation}}
\makeatother

\begin{document}

\preprint{ICRR-Report-696-2014-22,\ IPMU14-0348}

\title{Axion dark matter from topological defects}

\author{Masahiro Kawasaki}
\email{kawasaki@icrr.u-tokyo.ac.jp}
\affiliation{Institute for Cosmic Ray Research, The University of Tokyo, 
5-1-5 Kashiwa-no-ha, Kashiwa City, Chiba 277-8582, Japan} 
\affiliation{Kavli Institute for the Physics and Mathematics of the Universe (WPI), Todai Institutes for Advanced Study, The University of Tokyo,
5-1-5 Kashiwa-no-ha, Kashiwa City, Chiba 277-8582, Japan}

\author{Ken'ichi Saikawa}
\email{saikawa@th.phys.titech.ac.jp}
\affiliation{Department of Physics, Tokyo Institute of Technology,
2-12-1 Ookayama, Meguro-ku, Tokyo 152-8551, Japan}

\author{Toyokazu Sekiguchi}
\email{toyokazu.sekiguchi@helsinki.fi}
\affiliation{Helsinki Institute of Physics, University of Helsinki, PO Box 64, FIN-00014, Finland} 

\date{\today}

\begin{abstract}
The cosmological scenario where the Peccei-Quinn symmetry is broken after inflation is investigated.
In this scenario, topological defects such as strings and domain walls produce 
a large number of axions, which contribute to the cold dark matter of the universe.
The previous estimations of the cold dark matter abundance are updated and refined based on the field-theoretic simulations
with improved grid sizes. The possible uncertainties originated in the numerical calculations are also discussed. 
It is found that axions can be responsible for the cold dark matter in the mass range $m_a=(0.8\textendash 1.3)\times 10^{-4}\mathrm{eV}$
for the models with the domain wall number $N_{\rm DW}=1$, and $m_a\approx\mathcal{O}(10^{-4}\textendash 10^{-2})\mathrm{eV}$
with a mild tuning of parameters for the models with $N_{\rm DW}>1$.
Such higher mass ranges can be probed in future experimental studies.
\end{abstract}


\maketitle

\begin{widetext}
\tableofcontents\vspace{5mm}
\end{widetext}

\section{\label{sec1}Introduction}
Due to the developments of astronomical and cosmological observations in recent years,
it was revealed that our universe is filled by a nonbaryonic and highly nonrelativistic (cold) matter component called dark matter.
The nature of dark matter cannot be explained in the framework of the Standard Model (SM) of particle physics,
which strongly suggests that new physics exists beyond the SM.
So far, the weakly interacting massive particles (WIMPs) motivated by supersymmetry (SUSY) are
regarded as leading candidates of the dark matter.
However, this WIMP scenario becomes less plausible after the recent results of the Large Hadron Collider that
no evidence of SUSY has been observed yet [e.g., Ref.~\cite{Melzer-Pellmann:2014eta}].
In this context, the axion, which is an alternative possibility to explain dark matter, is now getting more attention.

The axion~\cite{Weinberg:1977ma,Wilczek:1977pj} appears as a (pseudo) Nambu-Goldstone boson associated with the spontaneous breaking of the Peccei-Quinn (PQ)
symmetry, which is introduced as a solution of the strong CP problem of quantum chromodynamics (QCD)~\cite{Peccei:1977hh,Peccei:1977ur}.
In order to compensate the CP-violating term in the QCD Lagrangian, the axion field $a$ has a following coupling with the gluon field,
\begin{equation}
\mathcal{L}_{agg} = -\frac{g^2}{32\pi^2}\frac{a}{F_a}G^{a\mu\nu}\tilde{G}^a_{\mu\nu}, \label{L_agg}
\end{equation}
where $g$ is the gauge coupling constant, $G^{a\mu\nu}$ is the gluon field strength, and $\tilde{G}^a_{\mu\nu}$ is its dual.
$F_a$ is so called the axion decay constant, whose value must be much higher than the electroweak scale
in order to avoid experimental constraints~\cite{Kim:1979if}.
It was pointed out that the axions are produced in the early universe due to the misalignment mechanism~\cite{Preskill:1982cy,Abbott:1982af,Dine:1982ah},
and that they behave like the nonrelativistic matter.

The physics of the axion is closely related with the early history of the universe.
In particular, the cosmological scenario becomes different according to whether the PQ symmetry is broken after inflation or not.
If the PQ symmetry is restored during inflation, the isocurvature perturbations induced by the quantum fluctuations of the axion field
during inflation affect the observational results of the cosmic microwave background, putting a stringent constraint on the axion models if the inflationary scale is sufficiently 
high~\cite{Beltran:2006sq,Kawasaki:2007mb,Hertzberg:2008wr,Visinelli:2009zm,Hamann:2009yf,Wantz:2009it,Hikage:2012be,Kawasaki:2013iha}.
On the other hand, if the PQ symmetry is broken after inflation,
topological defects such as strings and domain walls are formed, and we must take account of their evolution in the early universe.

The interesting consequence of the axion models is that the formation of hybrid networks of topological defects, where
strings are attached by domain walls, occurs at the epoch of the QCD phase transition
if the PQ symmetry is broken after inflation.
Let us call such configurations the string-wall systems.
The evolution of these string-wall systems becomes different depending on the value of an integer number $N_{\rm DW}$
called the ``domain wall number".
It is known that they are short-lived if $N_{\rm DW}=1$, and long-lived if $N_{\rm DW}>1$.

In the scenario described above, it is expected that an additional number of axions are produced because of the decay of strings and domain walls,
and that the estimation of the axion abundance is different from the usual one predicted by the misalignment mechanism~\cite{Davis:1986xc,Lyth:1991bb}.
This possibility was investigated extensively by several groups, but there were some controversies about the significance of the contribution from
strings and domain walls (see descriptions in the subsequent sections).
Recently, these controversies have been addressed by developing field-theoretic simulations of topological defects in the expanding universe.
In Ref.~\cite{Hiramatsu:2010yu}, the evolution of global strings was investigated with 3D simulations in a box of $512^3$ grids,
and the spectrum of axions produced from them was estimated.
Then, in Ref.~\cite{Hiramatsu:2010yn} the evolution and the decay of string-wall systems were studied with 2D simulations in a box of $4096^2$ grids.
Furthermore, in Ref.~\cite{Hiramatsu:2012gg} the formation and the decay of the string-wall systems for the models with $N_{\rm DW}=1$ were investigated with 3D simulations in a box of $512^3$ grids,
and the spectrum of axions radiated from them was estimated.
The similar analysis was performed for the models with $N_{\rm DW}>1$ in Ref.~\cite{Hiramatsu:2012sc}.
As a result of the series of studies, it becomes clear that the contributions from strings and domain walls can be larger than the misalignment component,
and that the constraints on the model parameters become more severe than before.

The computational methods established in the previous studies enable us to estimate the total abundance of the axion dark matter including the contributions from
topological defects. Then, it is important to ask for the accuracy of these theoretical estimations, since their results might have a relevance to various
experimental researches on dark matter axions.
In light of this fact, here we aim to update the previous results and to clarify the sources of uncertainty on the determination of the axion abundance
in the scenario where the PQ symmetry is broken after inflation.
The main improvement achieved in this paper is that 
the decay time of string-wall systems in the models with $N_{\rm DW}>1$ is estimated in detail
by the use of 2D simulations with grid sizes $8192^2$, $16384^2$, and $32768^2$.
Due to the large box sizes of the present simulations, it becomes possible to investigate the decay of string-wall systems in the cases with $N_{\rm DW}=5$ and $6$,
which were not studied in the previous work~\cite{Hiramatsu:2010yn} because of the limitation of dynamical ranges.
By performing additional 3D simulations in a box of $512^3$ grids, we also discuss the uncertainties contained in the estimated mean energy of radiated axions,
which were not addressed in Refs.~\cite{Hiramatsu:2010yu,Hiramatsu:2012gg,Hiramatsu:2012sc}.
Furthermore, for the sake of completeness, we reevaluate the contributions from the misalignment mechanism and strings in addition to
those form string-wall systems.

The outline of this paper is as follows.
In Sec.~\ref{sec2}, all production mechanisms in the scenario where the PQ symmetry is broken after inflation are described in detail.
Some new results of the numerical simulations are shown in Sec.~\ref{sec3}.
Then, we describe observational constraints 
and briefly comment on the prospects
in comparison with future experiments in Sec.~\ref{sec4}.
Finally, we make conclusion and discussion in Sec.~\ref{sec5}.

Throughout the paper, we work in the spatially flat Friedmann-Robertson-Walker (FRW) background
with a metric given by
\begin{equation}
ds^2 = -dt^2 + R(t)^2[dx^2+dy^2+dz^2], \nonumber
\end{equation}
where $t$ represents the cosmic time, and $R(t)$ is the scale factor of the universe.
A dot represents a derivative with respect to the cosmic time, i.e. $\dot{}=\partial/\partial t$.

\section{\label{sec2}Production mechanisms of axion cold dark matter}
In this section, we review the cosmological aspects of axions and describe production mechanisms of them.
If we assume that the PQ symmetry is broken after inflation, we must take account of three production mechanisms: misalignment mechanism,
decay of global strings, and decay of string-wall systems.
The energy density of relic axions can be estimated as a sum of these three contributions.
\subsection{\label{sec2-1}Misalignment mechanism}
In the invisible axion models, we introduce a complex scalar field $\Phi$, which is a singlet under the SM $SU(2)_L\times U(1)_Y$
gauge group and charged under the global $U(1)_{\rm PQ}$ symmetry. Let us call this scalar field PQ field.
The PQ symmetry is spontaneously broken when the PQ field acquires a vacuum expectation value $|\langle\Phi\rangle|=\eta$ at some high energy scale.
After that, the axion field $a(x)$ can be identified as a phase direction of the PQ field [i.e., $\Phi\propto \exp(ia/\eta)$].
This axion field acquires a periodic potential due to the nonperturbative effect of QCD when the temperature of the universe becomes less than 
$\mathcal{O}(0.1\textendash1)$GeV. At that time, the classical axion field is probably displaced from the minimum of its potential,
and this vacuum misalignment causes the coherent oscillation of the axion field~\cite{Preskill:1982cy,Abbott:1982af,Dine:1982ah}. 

The nonperturbative effect of QCD with finite temperature was discussed by several authors~\cite{Gross:1980br,Turner:1985si,Bae:2008ue,Wantz:2009it}
to model the temperature dependence of the axion mass $m_a(T)$.
Among them, in Ref.~\cite{Wantz:2009it} the axion mass was studied based on the interacting instanton liquid model (IILM)~\cite{Wantz:2009mi},
which gives a concrete framework to treat the QCD effect for all temperatures, in contrast to the earlier results~\cite{Turner:1985si,Bae:2008ue} 
based on the high temperature dilute gas approximation.
In this paper, we adopt the power law formula for $m_a(T)$ obtained in Ref.~\cite{Wantz:2009it} by fitting the result of the IILM calculation:
\begin{equation}
m_a(T)^2 = c_T\frac{\Lambda_{\rm QCD}^4}{F_a^2}\left(\frac{T}{\Lambda_{\rm QCD}}\right)^{-n}, \label{finite_temp_mass}
\end{equation}
with $c_T=1.68\times 10^{-7}$, $n=6.68$, and $\Lambda_{\rm QCD}=400\mathrm{MeV}$.
This power law behavior should be truncated when it exceeds the following zero temperature value
\begin{equation}
m_a(0)^2 = c_0\frac{\Lambda_{\rm QCD}^4}{F_a^2}, \label{zero_temp_mass}
\end{equation}
with $c_0=1.46\times 10^{-3}$.

We define the time $t_1$ corresponding to the beginning of the coherent oscillation from the following condition:
\begin{equation}
m_a(T_1) = 3H(t_1), \label{condition_t1}
\end{equation}
where $T_1$ is the temperature at the time $t_1$, and $H(t_1)$ is the Hubble parameter at that time. In the radiation dominated background, 
the Hubble parameter $H(t)$ can be related to the cosmic temperature $T$ via the Friedmann equation,
\begin{equation}
H^2 = \frac{8\pi^3G}{90}g_*(T)T^4, \label{Friedmann}
\end{equation}
where $G$ is the Newton's constant, and $g_*(T)$ is the relativistic degrees of freedom at the temperature $T$~\cite{Kolb:1990vq}.
From Eqs.~\eqref{finite_temp_mass},~\eqref{condition_t1}, and~\eqref{Friedmann}, we obtain
\begin{equation}
T_1 = 2.29\mathrm{GeV}\left(\frac{g_{*,1}}{80}\right)^{-1/(4+n)}\left(\frac{F_a}{10^{10}\mathrm{GeV}}\right)^{-2/(4+n)}\left(\frac{\Lambda_{\rm QCD}}{400\mathrm{MeV}}\right), \label{T1}
\end{equation}
where $g_{*,1}=g_*(T_1)$.\footnote{\label{fngs1}The temperature dependence of $g_*$ causes an ambiguity on the determination of $T_1$.
Within the range of the axion decay constant $10^8\mathrm{GeV}<F_a<10^{11}\mathrm{GeV}$, the value of $T_1$ given by Eq.~\eqref{T1}
varies from $1.5$ to $5.4\mathrm{GeV}$ for fixed values of $g_{*,1}=80$ and $\Lambda_{\rm QCD}=400\mathrm{MeV}$.
However, the value of $g_*(T)$ changes from $85$ to $80$ as the temperature decreases from $5.4$ to $1.5\mathrm{GeV}$~\cite{Wantz:2009it}.
This ambiguity leads to a minor correction to the result for the relic axion abundance, at most by a factor $(80/85)^{-(2+n)/2(4+n)}\simeq1.02$ [see, e.g., Eq.~\eqref{Omega_a_mis_2}].} Note that
Eq.~\eqref{T1} is valid only if the condition [Eq.~\eqref{condition_t1}] is satisfied before $m_a(T)$ reaches the zero temperature value [Eq.~\eqref{zero_temp_mass}].
This requirement is always satisfied for the range of the axion decay constant $10^8\mathrm{GeV}<F_a<10^{11}\mathrm{GeV}$ considered in this paper.

Let us estimate the energy density of axions produced by the misalignment mechanism.
Assuming that the potential for the axion field is dominated by the quadratic term of the form $\frac{1}{2}m_a^2a^2$,
we can write the energy density of these axions at the initial time $t_1$ as
\begin{equation}
\rho_{a,\mathrm{mis}}(t_1) = \frac{1}{2}m_a(T_1)^2\bar{\theta}_{\rm ini}^2F_a^2, \label{rho_a_mis_t1}
\end{equation}
where $\bar{\theta}_{\rm ini}$ is the initial misalignment angle.
Note that the quantity $R^3\rho_{a,\mathrm{mis}}/m_a$ is conserved over time once the adiabatic condition $H\ll m_a$ is satisfied~\cite{Kolb:1990vq}.
Therefore, the energy density at the present time $t_0$ can be estimated as
\begin{equation}
\rho_{a,\mathrm{mis}}(t_0) = \rho_{a,\mathrm{mis}}(t_1)\frac{m_a(0)}{m_a(T_1)}\left(\frac{R(t_1)}{R(t_0)}\right)^3. \label{rho_a_mis_t0}
\end{equation}
The dilution factor can be computed in terms of the entropy conservation: $(R(t_1)/R(t_0))^3=45s_0/2\pi^2 g_{*,1}T_1^3$, with $s_0$ being the entropy density at the present time.

So far we have assumed that the potential of the axion field is quadratic
and that the quantity $R^3\rho_{a,\mathrm{mis}}/m_a$ is exactly conserved just after the time $t_1$.
Strictly speaking, these two conditions are inaccurate, and the deviations from these approximations lead to some correction factors~\cite{Turner:1985si,Lyth:1991ub,Strobl:1994wk,Bae:2008ue,Visinelli:2009zm}.
First, the anharmonic effect becomes important for large values of $\bar{\theta}_{\rm ini}$. 
We can model this effect by replacing a factor $\bar{\theta}_{\rm ini}^2$ in Eq.~\eqref{rho_a_mis_t1}
with $f(\bar{\theta}_{\rm ini})\bar{\theta}_{\rm ini}^2$, where $f(\bar{\theta}_{\rm ini})$ is a function that approaches 1 for $|\bar{\theta}_{\rm ini}|\ll 1$
but takes a value greater than 1 for $|\bar{\theta}_{\rm ini}|\gtrsim 1$.
Second, the deviation from the adiabatic approximation at the initial stage of the coherent oscillation leads to lager energy density than the naive estimation performed in Eq.~\eqref{rho_a_mis_t0}.
This effect was calculated in Ref.~\cite{Bae:2008ue} and it was shown that the correction factor $f(\bar{\theta}_{\rm ini})\bar{\theta}_{\rm ini}^2$ should be multiplied by a factor of $1.85$.
Taking account of these correction factors, we estimate the fraction between the energy density of coherently oscillating axions and
the critical density of the universe today $\rho_{c,0}$ as
\begin{equation}
\Omega_{a,\mathrm{mis}}h^2 \equiv \frac{\rho_{a,\mathrm{mis}}(t_0)h^2}{\rho_{c,0}} 
= 7.03\times10^{-4}\times f(\bar{\theta}_{\rm ini})\bar{\theta}_{\rm ini}^2\left(\frac{g_{*,1}}{80}\right)^{-(2+n)/2(4+n)}\left(\frac{F_a}{10^{10}\mathrm{GeV}}\right)^{(6+n)/(4+n)}\left(\frac{\Lambda_{\rm QCD}}{400\mathrm{MeV}}\right),
\label{Omega_a_mis_1}
\end{equation}
where $h$ is the parameter for the Hubble constant ($H_0 = 100h~\mathrm{km}\cdot\mathrm{sec}^{-1}\mathrm{Mpc}^{-1}$).

If the PQ symmetry is broken after inflation, we expect that the value of $\bar{\theta}_{\rm ini}$ varies randomly over the horizon scale $\sim t_1$
at the QCD phase transition. Therefore, we can replace the factor $f(\bar{\theta}_{\rm ini})\bar{\theta}_{\rm ini}^2$ with its average
\begin{equation}
f(\bar{\theta}_{\rm ini})\bar{\theta}_{\rm ini}^2 \to \langle f(\bar{\theta}_{\rm ini})\bar{\theta}_{\rm ini}^2 \rangle_{\rm av} \equiv \frac{1}{2\pi}\int^{\pi}_{-\pi}f(\bar{\theta}_{\rm ini})\bar{\theta}_{\rm ini}^2d\bar{\theta}_{\rm ini}
= c_{\rm av}\frac{\pi^2}{3},
\end{equation}
where the coefficient $c_{\rm av} \equiv \langle f(\bar{\theta}_{\rm ini})\bar{\theta}_{\rm ini}^2 \rangle_{\rm av}/\langle \bar{\theta}_{\rm ini}^2 \rangle_{\rm av}$
represents the deviation from the value $\langle \bar{\theta}_{\rm ini}^2 \rangle_{\rm av}=\pi^2/3$
in the absence of the anharmonic effect [$f(\bar{\theta}_{\rm ini})=1$].
The numerical calculation in Ref.~\cite{Turner:1985si} showed that $c_{\rm av}=1.9\textendash 2.4$.
Similar results were obtained in Refs.~\cite{Lyth:1991ub,Strobl:1994wk,Visinelli:2009zm} by using some analytical modelings for the behavior of $f(\bar{\theta}_{\rm ini})$
around $\bar{\theta}_{\rm ini}\sim \pi$. Here we use $c_{\rm av}=2$ as a typical value for the anharmonic correction.
Using this value in Eq.~\eqref{Omega_a_mis_1}, we finally obtain
\begin{equation}
\Omega_{a,\mathrm{mis}}h^2 = 
4.63\times10^{-3}\times\left(\frac{c_{\rm av}}{2}\right)
\left(\frac{g_{*,1}}{80}\right)^{-(2+n)/2(4+n)}\left(\frac{F_a}{10^{10}\mathrm{GeV}}\right)^{(6+n)/(4+n)}\left(\frac{\Lambda_{\rm QCD}}{400\mathrm{MeV}}\right).
\label{Omega_a_mis_2}
\end{equation}

\subsection{\label{sec2-2}Global strings}
The global $U(1)_{\rm PQ}$ symmetry is broken when the temperature of the universe becomes $T\lesssim\eta$,
and the PQ field acquires a vacuum expectation value $|\langle\Phi\rangle|=\eta$. This process can be modeled by
the dynamics of the complex scalar field $\Phi$ with the following potential
\begin{equation}
V(\Phi) = \frac{\lambda}{4}(|\Phi|^2-\eta^2)^2. \label{V_phi_string}
\end{equation}
Spontaneous breaking of the global $U(1)_{\rm PQ}$ symmetry induced by this potential leads to the formation of line-like objects called global strings.
These strings continuously produce axions, which can contribute to the cold dark matter abundance~\cite{Davis:1986xc}.

The results of various numerical studies~\cite{Bennett:1989yp,Allen:1990tv,Yamaguchi:1998gx,Yamaguchi:1999yp,Yamaguchi:1999dy,Moore:2001px,Yamaguchi:2002zv,Yamaguchi:2002sh,Hiramatsu:2010yu}
indicate that the evolution of the strings can be described by the scaling solution, in which the energy density of strings is given by
\begin{equation}
\rho_{\rm string}(t) = \frac{\xi\mu_{\rm string}}{t^2}, \label{rho_string}
\end{equation}
where
\begin{equation}
\mu_{\rm string} = \pi\eta^2\ln\left(\frac{t}{\delta_s\sqrt{\xi}}\right) \label{mu_string}
\end{equation}
is the energy of a string per unit length,
$\delta_s\simeq(\sqrt{\lambda}\eta)^{-1}$ is the core width of the string,
and $\xi$ is a numerical coefficient which we call the length parameter.

The value for the length parameter $\xi$ can be determined from the result of the simulation of
global strings~\cite{Bennett:1989yp,Allen:1990tv,Yamaguchi:1998gx,Yamaguchi:1999yp,Moore:2001px,Yamaguchi:2002zv,Yamaguchi:2002sh,Hiramatsu:2010yu}.
However, the obtained value of $\xi$ for global strings contains a large systematic uncertainty, because of the poor understanding of the emission rate
of Nambu-Goldstone bosons [see Refs.~\cite{Yamaguchi:2002sh,Martins:2000cs,Moore:2001px,Hiramatsu:2012gg} for detailed discussions].
Following Ref.~\cite{Hiramatsu:2012gg}, here we adopt the rough estimate that $\xi$ has a central value 1 with 50\% uncertainty (i.e., $\xi=1.0\pm 0.5$).

The strings continue to radiate axions from the time of the PQ phase transition, which we denote as $t_c$, to the time of
the QCD phase transition $\sim t_1$.
Here we use the approximation that axions are treated as massless particles for $t<t_1$.\footnote{The production of axions
from strings for $t>t_1$, where the mass of the axion cannot be ignored, is discussed in Sec.~\ref{sec2-3}.}
The time evolution of the energy density of strings due to the radiation of axions can be modeled by the following equations:
\begin{align}
\frac{d\rho_{\rm string}}{dt} &= -2H\rho_{\rm string} - \left.\frac{d\rho_{\rm string}}{dt}\right|_{\rm emission}, \label{drho_string_dt}\\
\frac{d\rho_{a,\mathrm{string}}}{dt} &= -4H\rho_{a,\mathrm{string}} + \left.\frac{d\rho_{\rm string}}{dt}\right|_{\rm emission}, \label{drho_a_string_dt}
\end{align}
where $\rho_{a,\mathrm{string}}$ is the energy density of axions radiated from strings,
and $(d\rho_{\rm string}/dt)|_{\rm emission}$ is the energy loss rate of the strings due to the radiation of axions.
From Eqs.~\eqref{rho_string} and~\eqref{drho_string_dt}, we obtain
\begin{equation}
\left.\frac{d\rho_{\rm string}}{dt}\right|_{\rm emission} = \frac{\pi\eta^2\xi}{t^3}\left[\ln\left(\frac{t}{\delta_s\sqrt{\xi}}\right)-1\right]. \label{drho_string_dt_emission}
\end{equation}
On the other hand, Eq.~\eqref{drho_a_string_dt} can be reduced to
\begin{equation}
\frac{d E_{a,\mathrm{string}}}{dt} = R(t)^4\left.\frac{d\rho_{\rm string}}{dt}\right|_{\rm emission}, \label{dE_a_string_dt}
\end{equation}
where $E_{a,\mathrm{string}}(t)=R(t)^4\rho_{a,\mathrm{string}}$ is the comoving energy of radiated axions at the time $t$.
Combining Eqs.~\eqref{drho_string_dt_emission} and~\eqref{dE_a_string_dt}, we estimate the comoving number of radiated axions
at the time $t>t_1$ as
\begin{align}
N_{a,\mathrm{string}}(t>t_1) &= \int^{t_1}_{t_c}dt'\frac{1}{R(t')\bar{\omega}_a(t')}\frac{d E_{a,\mathrm{string}}}{dt} \nonumber\\
&= \int^{t_1}_{t_c}dt'\frac{R(t')^3}{\bar{\omega}_a(t')}\frac{\pi\eta^2\xi}{t'^3}\left[\ln\left(\frac{t'}{\delta_s\sqrt{\xi}}\right)-1\right], \label{N_a_string_t>t1}
\end{align}
where $\bar{\omega}_a(t)$ is a mean energy of axions radiated at the time $t$.

In the literature, there is a controversy on the determination of the mean energy of radiated axions $\bar{\omega}_a(t)$.
In Refs.~\cite{Davis:1986xc,Davis:1989nj,Dabholkar:1989ju,Battye:1993jv,Battye:1994au}, it was claimed that $\bar{\omega}_a(t)$ is comparable to the horizon scale at the time $t$.
However, authors of Refs.~\cite{Harari:1987ht,Hagmann:1990mj,Hagmann:2000ja} suggested that the spectrum of radiated axions becomes hard because of a turbulent decay process,
and that $\bar{\omega}_a(t)$ can become larger than the value of the order of the horizon scale.
Later, the evolution of global strings in the expanding universe was investigated based on the field theoretic simulations in Refs.~\cite{Yamaguchi:1998gx,Hiramatsu:2010yu},
and the spectrum of axions radiated from string networks was estimated.
The spectrum peaked at the scale corresponding to the horizon ($\sim 2\pi/t$), 
which supported the claim of Refs.~\cite{Davis:1986xc,Davis:1989nj,Dabholkar:1989ju,Battye:1993jv,Battye:1994au}.
Here, we follow this hypothesis and parametrize the mean energy of radiated axions as
\begin{equation}
\bar{\omega}_a(t) = \epsilon\frac{2\pi}{t}, \label{mean_omega_a_string}
\end{equation}
where $\epsilon$ is some numerical factor.

Using Eqs.~\eqref{N_a_string_t>t1} and~\eqref{mean_omega_a_string}, we can estimate the number density of radiated axions
at the present time $t_0$:
\begin{equation}
n_{a,\mathrm{string}}(t_0) = \frac{N_{a,\mathrm{string}}(t>t_1)}{R(t_0)^3} \simeq \left(\frac{R(t_1)}{R(t_0)}\right)^3\frac{\eta^2\xi}{t_1\epsilon}
\left[\ln\left(\frac{t_1}{\delta_s\sqrt{\xi}}\right)-3\right],
\end{equation}
where we ignored the contribution at $t=t_c$ in the last equality.
The ratio between the present energy density of axions radiated from strings $\rho_{a,\mathrm{string}}(t_0)=m_a(0)n_{a,\mathrm{string}}(t_0)$ and the critical density is given by
\begin{equation}
\Omega_{a,\mathrm{string}}h^2 = 2.94\times 10^{-2}\times\frac{\xi N_{\rm DW}^2}{\epsilon}\left(\frac{\beta'}{58}\right)\left(\frac{g_{*,1}}{80}\right)^{-(2+n)/2(4+n)}\left(\frac{F_a}{10^{10}\mathrm{GeV}}\right)^{(6+n)/(4+n)}\left(\frac{\Lambda_{\rm QCD}}{400\mathrm{MeV}}\right),
\label{Omega_a_string_1}
\end{equation}
where
\begin{align}
\beta' &\equiv \ln\left(\frac{t_1}{\delta_s\sqrt{\xi}}\right) -3 \equiv \beta_1 - 3, \\
\beta_1 &\simeq 60.8 -\frac{1}{2}\ln\left(\frac{\xi}{1.0}\right) + \ln N_{\rm DW} + \frac{1}{2}\ln\left(\frac{\lambda}{0.1}\right) -\frac{n}{2(4+n)}\ln\left(\frac{g_{*,1}}{80}\right)
+\frac{8+n}{4+n}\ln\left(\frac{F_a}{10^{10}\mathrm{GeV}}\right) - 2\ln \left(\frac{\Lambda_{\rm QCD}}{400\mathrm{MeV}}\right).
\end{align}
In Eq.~\eqref{Omega_a_string_1}, we used the relation between $\eta$ and $F_a$ [see Eq.~\eqref{N_DW-eta-Fa}].

\subsection{\label{sec2-3}String-wall systems}
When the temperature of the universe becomes $T\lesssim \mathcal{O}(0.1\textendash1)\mathrm{GeV}$,
axions acquire the mass because of the QCD effect, and the formation of domain walls occurs at that time~\cite{Sikivie:1982qv}.
The structure of domain walls is specified by the integer number $N_{\rm DW}$, whose value is
related to the degree of degeneracy of the low energy vacua.
Recall that the quantity $\bar{\theta}=a/F_a$ has a periodicity $2\pi$
because of the periodicity of the QCD $\theta$-vacuum.
On the other hand, generically the field $a$ might have a periodicity greater than $2\pi F_a$.
Hence the domain wall number corresponding to the degree of degeneracy of vacua can be counted as
\begin{equation}
N_{\rm DW} \equiv \frac{(\mathrm{periodicity\ of\ }a)}{2\pi F_a}.
\end{equation}
If the axion field corresponds to the phase of a single complex scalar field $\Phi\propto\exp(ia/\eta)$,
the field $a$ has a periodicity $2\pi\eta$, which leads to the following relation:
\begin{equation}
N_{\rm DW} = \frac{\eta}{F_a}. \label{N_DW-eta-Fa}
\end{equation}
For simplicity, in this paper we consider the models in which the relation~\eqref{N_DW-eta-Fa} holds.\footnote{In the case where
the axion field is represented as a combination of multiple scalar fields, the estimation of $N_{\rm DW}$
is not so straightforward as Eq.~\eqref{N_DW-eta-Fa}~\cite{Choi:1985iv}.}

The axion field shifts as $a\to a+c\eta$ with an arbitral constant $c$ under the PQ symmetry, which induces the shift $\propto c(\eta/F_a)G^{a\mu\nu}\tilde{G}^a_{\mu\nu}$
in the effective Lagrangian [see Eq.~\eqref{L_agg}].
This change in the Lagrangian must be compensated by the anomaly induced by quarks charged under $U(1)_{\rm PQ}$.
Therefore, Eq.~\eqref{N_DW-eta-Fa} implies that the domain wall number can be calculated in terms of
the $U(1)_{\rm PQ}$-$SU(3)_C$-$SU(3)_C$ anomaly coefficient~\cite{Sikivie:1982qv,Georgi:1982ph,Kim:1986ax}.
For instance, in the Kim-Shifman-Vainshtein-Zakharov (KSVZ) model~\cite{Kim:1979if,Shifman:1979if} we have $N_{\rm DW}=1$,
while in the Dine-Fischler-Srednicki-Zhitnitsky (DFSZ) model~\cite{Zhitnitsky:1980tq,Dine:1981rt} we have $N_{\rm DW}=2N_g$ with
$N_g$ being the number of generations.

Taking account of the degeneracy of vacua, we can describe the potential for the axion field $a$ as follows:
\begin{equation}
V(a) = \frac{m_a^2\eta^2}{N_{\rm DW}^2}\left\{1-\cos\left(N_{\rm DW}\frac{a}{\eta}\right)\right\}. \label{V_axion}
\end{equation}
This potential has $N_{\rm DW}$ degenerate minima given by $\Phi_k=\eta\exp(2\pi ik/N_{\rm DW})$ with $k = 0,1,\dots,N_{\rm DW}-1$,
and it explicitly breaks the global $U(1)_{\rm PQ}$ symmetry into the discrete subgroup $Z_{N_{\rm DW}}$.
This $Z_{N_{\rm DW}}$ symmetry is also spontaneously broken at $t\sim t_1$, and then $N_{\rm DW}$ domain walls
corresponding to the boundaries of $N_{\rm DW}$ degenerate vacua are attached to strings.
The width of domain walls is estimated as $\delta_w\simeq m_a^{-1}$ and the surface mass density of domain walls is given by
\begin{equation}
\sigma_{\rm wall} \simeq 9.23 m_a F_a^2, \label{sigma_wall}
\end{equation}
where the coefficient 9.23 includes the contribution from the structure of the neutral pion field~\cite{Huang:1985tt,Hiramatsu:2012sc}.

Just after the time of the QCD phase transition $\sim t_1$, the tension of strings dominates over that of domain walls.
However, they become comparable at the time $t_2$ defined by the following condition:
\begin{equation}
\sigma_{\rm wall}(t_2) = \frac{\mu_{\rm string}(t_2)}{t_2}. \label{condition_t2}
\end{equation}
Denoting $T_2$ as the temperature at the time $t_2$, from Eq.~\eqref{condition_t2} we find
\begin{equation}
T_2 = 1.41\mathrm{GeV}\left(\frac{\beta_2}{62}\right)^{-2/(4+n)}
\left(\frac{g_{*,2}}{75}\right)^{-1/(4+n)}\left(\frac{F_a}{10^{10}\mathrm{GeV}}\right)^{-2/(4+n)}\left(\frac{\Lambda_{\rm QCD}}{400\mathrm{MeV}}\right), \label{T2}
\end{equation}
where $g_{*,2}=g_*(T_2)$~\footnote{For a similar reason with the footnote~\ref{fngs1}, within the range $10^8\mathrm{GeV}<F_a<10^{11}\mathrm{GeV}$
the values of $T_2$ and $g_{*,2}$ vary as $0.9\mathrm{GeV}<T_2<3.4\mathrm{GeV}$ and $75<g_{*,2}<84$, respectively.} and
\begin{align}
\beta_2 &\equiv \ln\left(\frac{t_2}{\delta_s\sqrt{\xi}}\right) \nonumber\\
&\simeq 61.8 +\frac{4}{4+n}\ln\left(\frac{\beta'}{62}\right) -\frac{1}{2}\ln\left(\frac{\xi}{1.0}\right) + \ln N_{\rm DW} + \frac{1}{2}\ln\left(\frac{\lambda}{0.1}\right) \nonumber\\
&\quad -\frac{n}{2(4+n)}\ln\left(\frac{g_{*,2}}{75}\right) +\frac{8+n}{4+n}\ln\left(\frac{F_a}{10^{10}\mathrm{GeV}}\right) - 2\ln \left(\frac{\Lambda_{\rm QCD}}{400\mathrm{MeV}}\right).
\end{align}
After the time $t_2$, the dynamics of the system is dominated by the tension of domain walls.

From the sequence of the formation, evolution, and decay of the string-wall systems, additional axions are expected to be produced~\cite{Lyth:1991bb}.
The fate of these string-wall systems is different between the case with $N_{\rm DW}=1$ and that with $N_{\rm DW}>1$.
In what follows, we discuss these two cases separately.

\subsubsection{\label{sec2-3-1}Short-lived domain walls}
First, we consider the models with $N_{\rm DW}=1$.
For the case with $N_{\rm DW}=1$, one domain wall is attached to each string.
Such string-wall systems are unstable, and decay because of the tension of walls soon after the formation~\cite{Barr:1986hs}.

We expect that the curvature radius of domain walls at the formation time is comparable to the horizon scale at that time $\sim t_1$~\cite{2000csot.book.....V}.
In other words, the energy of domain walls per horizon volume is estimated as $\sim\sigma_{\rm wall}t_1^2$, and their energy density is given by
\begin{equation}
\rho_{\rm wall}(t_1) = \frac{\mathcal{A}\sigma_{\rm wall}}{t_1}, \label{rho_wall_t1}
\end{equation}
where $\mathcal{A}$ is a numerical coefficient.
The value of $\mathcal{A}$, which we call the area parameter, can be estimated in the numerical simulations~\cite{Hiramatsu:2012gg,Hiramatsu:2012sc}.
The string-wall systems collapse around the time $t_d\simeq t_2$ due to the tension of domain walls.
Assuming that their energy density is well approximated by extrapolating Eqs.~\eqref{rho_string} and~\eqref{rho_wall_t1} to $t=t_d\simeq t_2$,
we estimate that\footnote{It is not straightforward to estimate the exact behavior of $\rho_{\mathrm{string}\mathchar`-\mathrm{wall}}(t)$ for $t>t_1$.
In the previous paper~\cite{Hiramatsu:2012gg}, we assumed that $\rho_{\mathrm{string}\mathchar`-\mathrm{wall}}(t)$ is diluted as $\propto R(t)^{-3}$ from $t_1$ until $t_d$,
but it might be more reasonable to use Eq.~\eqref{rho_string-wall} since the results of numerical simulations in Ref.~\cite{Hiramatsu:2012gg}
indicate that $\rho_{\rm string}(t)$ and $\rho_{\rm wall}(t)$ do not deviate significantly from the expressions given by Eqs.~\eqref{rho_string} and~\eqref{rho_wall_t1}
even for $t_1\lesssim t < t_d$.}
\begin{equation}
\rho_{\mathrm{string}\mathchar`-\mathrm{wall}}(t_d) \simeq \frac{\mathcal{A}\sigma_{\rm wall}(t_2)}{t_2} + \frac{\xi\mu_{\rm string}(t_2)}{t_2^2}.
\label{rho_string-wall}
\end{equation}
Then, the number density of axions produced by the decay of string-wall systems is given by
\begin{equation}
n_{a,\mathrm{dec}}(t) = \frac{\rho_{\mathrm{string}\mathchar`-\mathrm{wall}}(t_d)}{\bar{\omega}_a}\left(\frac{R(t_d)}{R(t)}\right)^3, \label{n_a_dec_t}
\end{equation}
where $\bar{\omega}_a$ is the mean energy of axions produced by this decay process.

Similarly to the case of the axion production from strings,
the determination of the mean energy $\bar{\omega}_a$ becomes a controversial issue.
In Ref.~\cite{Nagasawa:1994qu}, it was claimed that $\bar{\omega}_a$ is comparable to the mass of the axion $m_a$.
On the other hand, in Ref.~\cite{Chang:1998tb} it was argued that $\bar{\omega}_a$ becomes larger than the naive estimation in Ref.~\cite{Nagasawa:1994qu},
and that the number density of axions estimated by Eq.~\eqref{n_a_dec_t} is suppressed by a factor of $\mathcal{O}(10)$.
Later, the spectrum of axions produced from string-wall systems was computed in Ref.~\cite{Hiramatsu:2012gg}, which revealed that
most axions are mildly relativistic and agreed with the claim of Ref.~\cite{Nagasawa:1994qu}.
Following the results of these previous studies, we parameterize the average of the energy of axions
$\bar{\omega}_a(t_d)$ at the time of the decay of string-wall systems $t_d$ as\footnote{\label{on_tilde_epsilon_w}In Ref.~\cite{Hiramatsu:2012gg},
the average of the momentum of axions $\bar{k}(t_d)$ at the time of 
the decay of string-wall systems $t_d$ was computed, and the following parameterization was introduced:
\begin{equation}
\epsilon_w = \frac{\bar{k}(t_d)/R(t_d)}{m_a(T_d)}. \label{epsilon_w_def}
\end{equation}
By using the value of $\epsilon_w$ obtained above, $\bar{\omega}_a$ was estimated as
\begin{equation}
\bar{\omega}_a = \sqrt{1+\epsilon_w^2}m_a(T_d). \nonumber
\end{equation}
Strictly speaking, the above equation does not correctly represent the mean energy, since it does not correspond to the value of the energy
averaged over the momentum distribution of radiated axions.
Instead of using such a parameterization, in this paper we compute $\bar{\omega}_a$ directly from the power spectrum of radiated axions [see Eq.~\eqref{mean_omega}] and estimate
the present energy density of radiated axions by using Eq.~\eqref{tilde_epsilon_w_def}.}
\begin{equation}
\tilde{\epsilon}_w = \frac{\bar{\omega}_a(t_d)}{m_a(T_d)}, \label{tilde_epsilon_w_def}
\end{equation}
where $T_d$ is the temperature at the time $t_d$.

Using Eqs.~\eqref{rho_string-wall},~\eqref{n_a_dec_t}, and~\eqref{tilde_epsilon_w_def} with
the approximations that $R(t_d)\simeq R(t_2)$ and $m_a(T_d)\simeq m_a(T_2)$,\footnote{\label{fnma1ma2}In Ref.~\cite{Hiramatsu:2012gg},
it was assumed that $m_a(T_d)\simeq m_a(T_1)$,
but it is appropriate to use $m_a(T_d)\simeq m_a(T_2)$ since the change of $m_a(T)$ during the decay process is remarkable.
Indeed, this change leads to the correction by a factor of $m_a(T_1)/m_a(T_2)=(T_2/T_1)^{n/2}\simeq 0.2$.
We thank Asimina Arvanitaki, Sergei Dubovsky, and Giovanni Villadoro for pointing out this correction.} we can estimate
the present energy density of axions radiated after $t_1$ as
\begin{align}
\rho_{a,\mathrm{dec}}(t_0) &= m_a(0)n_{a,\mathrm{dec}}(t_0) \nonumber\\
&= \frac{m_a(0)}{\tilde{\epsilon}_w m_a(T_2)}
\left[\frac{\mathcal{A}\sigma_{\rm wall}(t_2)}{t_2} + \frac{\xi\mu_{\rm string}(t_2)}{t_2^2}\right]\left(\frac{R(t_2)}{R(t_0)}\right)^3. 
\label{rho_a_dec_t0_short}
\end{align}
Then, its ratio to the critical density today is given by
\begin{align}
\Omega_{a,\mathrm{dec}}h^2 = 
7.88\times 10^{-3}\times \frac{\mathcal{A}+\xi}{\tilde{\epsilon}_w}\left(\frac{\beta_2}{62}\right)^{2/(4+n)}\left(\frac{g_{*,2}}{75}\right)^{-(2+n)/2(4+n)}\left(\frac{F_a}{10^{10}\mathrm{GeV}}\right)^{(6+n)/(4+n)}\left(\frac{\Lambda_{\rm QCD}}{400\mathrm{MeV}}\right).
\label{Omega_a_dec_1}
\end{align}
We use this estimation together with Eqs.~\eqref{Omega_a_mis_2} and~\eqref{Omega_a_string_1} to obtain the observational constraint on the model parameter in Sec.~\ref{sec4-1}.

\subsubsection{\label{sec2-3-2}Long-lived domain walls}
Next, let us turn our attention to the models with $N_{\rm DW}>1$.
If $N_{\rm DW}>1$, more than two domain walls are attached to a single string.
Such string-wall networks are stable and long-lived, since the strings are sustained by tension of the walls from multiple directions.

After the time $t_2$, we can ignore the effect of the strings on the dynamics of the string-wall systems.
In particular, the energy density of the string-wall systems can be estimated in terms of that of domain walls:
\begin{equation}
\rho_{\mathrm{string}\mathchar`-\mathrm{wall}}(t) \simeq \rho_{\rm wall}(t) \quad \mathrm{for}\quad t>t_2.
\end{equation}
Similarly to the case of the cosmic strings, results of various numerical
studies [Refs.~\cite{Press:1989yh,Garagounis:2002kt,Hiramatsu:2010yz,Kawasaki:2011vv,Leite:2011sc,Leite:2012vn,Hiramatsu:2013qaa}
for domain wall networks in the $Z_2$ symmetric model and Refs.~\cite{Ryden:1989vj,Hiramatsu:2010yn,Hiramatsu:2012sc} for string-wall networks in the axionic model]
indicate that the evolution of domain walls is described by the scaling solution, in which the energy density of the walls is given by
\begin{equation}
\rho_{\rm wall}(t) = \frac{\mathcal{A}\sigma_{\rm wall}}{t}. \label{rho_wall_t}
\end{equation}
Again, the area parameter $\mathcal{A}$ can be estimated in the numerical simulations~\cite{Hiramatsu:2012sc}.

From Eq.~\eqref{rho_wall_t}, we see that the energy density of domain walls decreases as $\propto 1/t \propto R^{-2}$ in the radiation-dominated universe.
Since this decrement is slower than those of dusts ($\propto R^{-3}$) and radiations ($\propto R^{-4}$),
domain walls eventually overclose the universe, which leads to a problem in the standard cosmology~\cite{Zeldovich:1974uw}.
However, such a problem can be avoided if there exists a ``bias" term in the potential, which slightly breaks the discrete symmetry~\cite{Vilenkin:1981zs,Gelmini:1988sf,Sikivie:1982qv}.
Here we model this effect by introducing the following term in the potential of the PQ field $\Phi$~\cite{Sikivie:1982qv,Hiramatsu:2010yn,Hiramatsu:2012sc}:
\begin{equation}
V_{\rm bias}(\Phi) = -\Xi\eta^3\left(\Phi e^{-i\delta}+{\rm h.c.}\right), \label{V_phi_bias}
\end{equation}
where $\Xi$ and $\delta$ are dimensionless parameters.
If $\Xi$ takes a small but nonvanishing value, $N_{\rm DW}$ degenerate vacua are lifted by a quantity proportional to $\Xi$.
Eventually a domain having the lowest energy dominates over others, which causes the annihilation of the walls.

Let us estimate the time scale for the annihilation of domain walls.
The difference in the potential energy between the minimum with the lowest energy $\Phi_0=\eta\exp(i\delta)$ and its neighbor $\Phi_1=\eta\exp[i\delta+(2\pi i/N_{\rm DW})]$ is given by
\begin{equation}
\Delta V = V_{\rm bias}(\Phi_1) - V_{\rm bias}(\Phi_0) = 2\Xi\eta^4\left[1-\cos\left(\frac{2\pi}{N_{\rm DW}}\right)\right]. \label{Delta_V}
\end{equation}
This energy difference can be regarded as a volume pressure $p_V\sim \Delta V$ acting on domain walls.
The evolution of domain walls is also affected by a tension force $p_T$, which makes them to stretch up to the horizon scale.
The decay of domain walls occurs when the volume pressure $p_V$ dominates over the tension $p_T$.
In order to estimate $p_T$, we define the effective curvature radius $R_{\rm wall}$ of domain walls such that $\rho_{\rm wall}\sim \sigma_{\rm wall}/R_{\rm wall}$.
Equation~\eqref{rho_wall_t} indicates that $R_{\rm wall}\sim t/\mathcal{A}$.
Then, the tension of domain walls is estimated as $p_T \sim \sigma_{\rm wall}/R_{\rm wall}\sim\mathcal{A}\sigma_{\rm wall}/t$.
The decay time $t_{\rm dec}$ of the domain walls is obtained from the condition $p_T\sim p_V$:
\begin{equation}
t_{\rm dec} = C_d \frac{\mathcal{A}\sigma_{\rm wall}}{\Xi\eta^4 (1-\cos(2\pi/N_{\rm DW}))}, \label{t_dec_exact_scaling}
\end{equation}
where $C_d$ is a numerical coefficient.\footnote{Equation~\eqref{t_dec_exact_scaling} is slightly different from a naive estimation,
\begin{equation}
t_{\rm dec} = \alpha\frac{m_a}{N_{\rm DW}\Xi\eta^2}\quad \mathrm{with}\quad \alpha\simeq 18, \label{t_dec_previous}
\end{equation}
used in the previous papers~\cite{Hiramatsu:2010yn,Hiramatsu:2012sc}. In particular, Eq.~\eqref{t_dec_previous} does not take account of the additional factor $\mathcal{A}$
appearing in Eq.~\eqref{t_dec_exact_scaling}. Furthermore, the dependence on $N_{\rm DW}$ in Eq.~\eqref{t_dec_previous} is different from that in Eq.~\eqref{t_dec_exact_scaling}, because of
a rough (and perhaps inappropriate) estimation $p_V\sim\Delta V\sim 4\pi\Xi\eta^4/N_{\rm DW}$ in Ref.~\cite{Hiramatsu:2010yn}.
} We will determine the value of $C_d$ from the results of numerical simulations in Sec.~\ref{sec3-2}.

As we will see in Sec.~\ref{sec3-2}, some results of numerical simulations show slight deviations from the scaling behavior [i.e. Eq.~\eqref{rho_wall_t} with $\mathcal{A}$ being constant].
If we take account of these deviations from the scaling solution, we model the evolution of the domain wall networks as~\cite{Jeong:2013oza}
\begin{equation}
\rho_{\rm wall}(t) = \frac{\mathcal{A}(t)\sigma_{\rm wall}}{t}\quad \mathrm{with}\quad \mathcal{A}(t) = \mathcal{A}_{\rm form}\left(\frac{t}{t_{\rm form}}\right)^{1-p}, \label{rho_wall_t_dev}
\end{equation}
instead of Eq.~\eqref{rho_wall_t}.
Here, $t_{\rm form}$ is the time of the formation of domain walls, which will be specified later, and $\mathcal{A}_{\rm form}$ is the area parameter at $t=t_{\rm form}$.
The parameter $p$ will be fixed by the results of numerical simulations. The exact scaling solution corresponds to the case with $p=1$.
If we assume the deviation from the scaling solution [Eq.~\eqref{rho_wall_t_dev}], the estimation for the decay time of domain walls becomes
slightly different from Eq.~\eqref{t_dec_exact_scaling}:
\begin{equation}
t_{\rm dec} = C_d\left[\frac{\mathcal{A}_{\rm form}\sigma_{\rm wall}}{t_{\rm form}\Xi\eta^4(1-\cos(2\pi/N_{\rm DW}))}\right]^{1/p}t_{\rm form}.
\label{t_dec_dev_scaling}
\end{equation}
Again, $C_d$ is a numerical coefficient whose value is fixed from the results of numerical simulations.

Until the time $t_{\rm dec}$, the domain wall networks continuously radiate axion particles, which give
an additional contribution to the cold dark matter abundance.
In the previous study~\cite{Hiramatsu:2012sc}, we confirmed that the physical mean momentum $\bar{k}/R(t)$ of radiated axions
hardly varies with time, and it was conjectured that the value of the mean momentum is given by the mass of the axion.
Therefore, it is convenient to define the ratio between the mean momentum and the axion mass:
\begin{equation}
\epsilon_a = \frac{\bar{k}(t)/R(t)}{m_a},
\label{epsilon_a_def}
\end{equation}
or that between the mean energy $\bar{\omega}_a(t)$ and the axion mass:
\begin{equation}
\tilde{\epsilon}_a = \frac{\bar{\omega}_a(t)}{m_a}.
\label{tilde_epsilon_a_def}
\end{equation}
The calculation of the cold dark matter abundance $\Omega_{a,\mathrm{dec}}h^2$ from these long-lived domain wall networks is performed in Appendix~\ref{secA}.
The result for $\Omega_{a,\mathrm{dec}}h^2$ becomes different
depending on whether we assume the exact scaling solution [Eq.~\eqref{rho_wall_t}] or not [Eq.~\eqref{rho_wall_t_dev}].
Here and hereafter we consider both cases. The result derived from the assumption of Eq.~\eqref{rho_wall_t} should be regarded as a more conservative constraint
in comparison with that derived from the assumption of Eq.~\eqref{rho_wall_t_dev}.

\section{\label{sec3} Numerical simulations}
In order to estimate the energy density of relic axions correctly, we need to know the values of some numerical coefficients such as the area parameter $\mathcal{A}$
of domain walls, the coefficient $C_d$ related to the decay time of domain walls,
and the parameters ($\epsilon$,\ $\tilde{\epsilon}_w$,\ $\tilde{\epsilon}_a$) related to the mean energy of radiated axions.
These parameters can be estimated by performing numerical simulations on the cosmological evolutions of topological defects.
The purpose of this section is to refine the results of the previous numerical studies~\cite{Hiramatsu:2010yu,Hiramatsu:2010yn,Hiramatsu:2012gg,Hiramatsu:2012sc}.
In particular, we focus on the following issues:

\begin{enumerate}
\item The evolution and decay of the long-lived string-wall systems were studied in Ref.~\cite{Hiramatsu:2010yn}
based on the 2D lattice simulations, and the decay time of the networks was estimated.
However, the results suffered from large systematic uncertainties, which were perhaps caused by the limitation of dynamical ranges in the numerical simulations.
Here we improve dynamical ranges of the simulations and update the profile of the decay time of string-wall networks.
\item The spectra of axions produced from global strings, short-lived string-wall systems, and long-lived string-wall systems were calculated
in Refs.~\cite{Hiramatsu:2010yu},~\cite{Hiramatsu:2012gg}, and~\cite{Hiramatsu:2012sc}, respectively.
Later, in Ref.~\cite{Saikawa:2013thesis}, it was found that the outcome of the calculation of the mean momentum of radiated axions
depends on the choice of the number of bins $n_{\rm bin}$ in the power spectrum, which causes a large systematic uncertainty in the case
of the short-lived string-wall systems. It is probable that a similar uncertainty exists in the case of the strings and the long-lived string-wall systems.
With an aim to resolve this subtlety, here we reevaluate the mean momentum or the mean energy of radiated axions by checking the dependence on $n_{\rm bin}$
more carefully.
\item In Ref.~\cite{Hiramatsu:2012gg}, numerical simulations of the short-lived string-wall systems were performed
with varying the ratio $\kappa\equiv \Lambda_{\rm QCD}/F_a$ from 0.3 to 0.4, and a slight dependence of $\epsilon_w$
on $\kappa$ was observed. This fact makes the estimation of the parameter $\epsilon_w$ obscure, since
we cannot perform the simulation with a realistic value of $\kappa\sim\mathcal{O}(10^{-11})$.
Here we reevaluate the significance of this $\kappa$ dependence after resolving the uncertainty due to the choice of $n_{\rm bin}$
described above.
\end{enumerate}

In order to investigate the issue 1, we perform numerical simulations on 2D lattice, since in 2D simulations we can improve the dynamical
range, which enables us to measure the decay time for various choices of the model parameters.
On the other hand, we perform 3D simulations to investigate the issues 2 and 3.
Although we use 2D simulations to calculate the time evolution of the area parameter $\mathcal{A}$ of domain walls and to determine their decay time,
we expect that the results are not affected by the choice of the dimensionality of the simulations,
since we checked that at least in the short dynamical range, which 3D simulations can investigate, the behavior of $\mathcal{A}$ is almost unchanged between 2D and 3D simulations.

The outline of this section is as follows. First, in Sec.~\ref{sec3-1} we describe the setup of 2D and 3D simulations, respectively.
Next, we describe the results of 2D simulations and determine the decay time of long-lived string-wall systems in Sec.~\ref{sec3-2}.
Finally, in Sec.~\ref{sec3-3} we describe the results of 3D simulations and estimate the mean energy of axions radiated from
strings, short-lived string-wall systems, and long-lived string-wall systems, respectively.
\subsection{\label{sec3-1} Setup of the simulations}
\subsubsection{\label{sec3-1-1} 2D}
In 2D simulations, we solve the classical field equation of the scalar field $\Phi$ in the FRW background with the potential given by
Eqs.~\eqref{V_phi_string},~\eqref{V_axion}, and~\eqref{V_phi_bias}:
\begin{equation}
V(\Phi) = \frac{\lambda}{4}(|\Phi|^2-\eta^2)^2 +\frac{m_a^2\eta^2}{N_{\rm DW}^2}\left\{1-\frac{|\Phi|}{\eta}\cos \left(N_{\rm DW}(\mathrm{Arg}(\Phi))\right)\right\}-\Xi\eta^3(\Phi e^{-i\delta}+\Phi^*e^{i\delta}).
\label{V_phi_simulation}
\end{equation}
In Eq.~\eqref{V_phi_simulation}, the coefficient of the term proportional to $\cos \left(N_{\rm DW}(\mathrm{Arg}(\Phi))\right)$ is modified as
$1\to |\Phi|/\eta$, since otherwise we suffer from instabilities occurring around $|\Phi|\simeq 0$~\cite{Hiramatsu:2010yn,Hiramatsu:2012sc}.
The spatial configuration of the scalar field $\Phi({\bf x})$ is obtained in terms of discrete coordinates [i.e. the continuous label ${\bf x}$ is replaced with some integers $(i,j,k)$].
The simulations are performed in comoving coordinates, and periodic boundary conditions are imposed in the simulation box.
The time evolution is solved by using the fourth-order symplectic integration scheme~\cite{Yoshida:1990zz}, while the spatial derivative is
computed by using the fourth-order finite difference method.
Initial conditions for $\Phi({\bf x},\tau_i)$ and $\dot{\Phi}({\bf x},\tau_i)$ are generated as random Gaussian fluctuations in the momentum space.
See Ref.~\cite{Hiramatsu:2012sc} for further details on the simulation techniques.

In the numerical studies of the model given by Eq.~\eqref{V_phi_simulation}, we normalize all dimensionful quantities in the unit of $\eta=1$.
Values of some parameters are fixed as $\lambda=0.1$, $\delta=0$, and $m_a=0.1$, but those for $\Xi$ and $N_{\rm DW}$ are varied.
The time integration is performed in terms of the conformal time $\tau$ defined by $d\tau = dt/R(t)$, and the initial time of the simulation is fixed as $\tau_i=2$.
We assume the radiation-dominated universe, where the scale factor evolves as $R(\tau) \propto \tau \propto t^{1/2}$.
This scale factor is normalized such that $R(\tau_i)=1$.

Let us denote the length of a side of the simulation box as $L$ and the number of grid points along one coordinate axis as $N$.
The lattice spacing in the comoving coordinate is estimated by $\Delta x = L/N$, and
the scale of Hubble radius $H^{-1}$ and the size of the core of strings $\delta_s$ divided by the physical lattice spacing $\Delta x_{\rm phys}=R(\tau)\Delta x$ are given by
\begin{equation}
\frac{H^{-1}}{\Delta x_{\mathrm{phys}}} = \frac{N}{L}\tau \qquad \mathrm{and} \qquad \frac{\delta_s}{\Delta x_{\mathrm{phys}}} = \frac{N}{L\sqrt{\lambda}}\left(\frac{\tau_i}{\tau}\right).
\label{conditions_resolution}
\end{equation}
In order to resolve the width of topological defects, we must require that $\delta_s/\Delta x_{\rm phys}\gtrsim1$ at the end of the simulation.
Furthermore, $H^{-1}/\Delta x_{\rm phys}$ must be smaller than $N$, since otherwise the Hubble radius exceeds the simulation box and
we cannot follow the evolution of the defect networks correctly.

We perform simulations for three different simulation boxes, $(N,L,\tau_f)=(8192,320,160)$, $(16384,460,230)$, and $(32768,640,320)$,
where $\tau_f$ is the final time of the simulations.
Dynamical ranges of these simulations are $\tau_f/\tau_i=80$, $115$, and $160$, respectively, and they are much larger than that of the previous study ($\tau_f/\tau_i=55$)~\cite{Hiramatsu:2010yn}.
The interval $\Delta \tau$ for each time integration step is fixed as $0.01$ for the cases with $N=8192$ and $N=16384$, and as $0.005$ for the case with $N=32768$.
Note that the final time is given by $\tau_f=\tau_i+(\mathrm{time\ steps})\times \Delta\tau$.
Given a setup for the simulation box, we perform simulations with a range of values of $N_{\rm DW}$ and $\Xi$.
Then, for each choice of the parameters we execute 10 realizations with $N=8192$ and $N=16384$ and 1 realization with $N=32768$.
The parameters used in 2D simulations are summarized as cases (a)-(c) in Table~\ref{tab1}.
In Table~\ref{tab1}, the values for ratios $H^{-1}/\Delta x_{\rm phys}|_{\rm \tau_f}$ and $\delta_s/\Delta x_{\rm phys}|_{\rm \tau_f}$ at the final time of the simulations
are also shown. We see that the conditions $H^{-1}/\Delta x_{\rm phys}<N$ and $\delta_s/\Delta x_{\rm phys}\gtrsim 1$ described below Eq.~\eqref{conditions_resolution}
are marginally satisfied in these simulations.
Table~\ref{tab2} shows the values for $(N_{\rm DW},\Xi)$ used in the 2D simulations. In total, we perform simulations in 40 different sets of parameters as shown in
case (a-1) to (c-5) of Table~\ref{tab2}.

{\tabcolsep = 0.9mm
\begin{table}[h]
\begin{center} \scriptsize
\caption{Sets of parameters used in 2D numerical simulations. For other parameters, we used the common values $\lambda=0.1$, $\delta=0$,
$m_a=0.1$, and $\tau_i=2$.}
\vspace{3mm}
\begin{tabular}{c c c c c c c c c c}
\hline\hline
Case & Grid size ($N^2$) & Box size ($L$) & Time interval ($\Delta\tau$) & Time steps & Final time ($\tau_f$) & $N_{\rm DW}$ & $\Xi$ & $\left.H^{-1}/\Delta x_{\rm phys}\right|_{\tau_f}$ & $\left.\delta_s/\Delta x_{\rm phys}\right|_{\tau_f}$ \\
\hline 
(a) & $8192^2$ & 320 & 0.01 & 15800 & 160 & varying & varying & 4096 & 1.01 \\
(b) & $16384^2$ & 460 & 0.01 & 22800 & 230 & varying & varying & 8192 & 0.98 \\
(c) & $32768^2$ & 640 & 0.005 & 63600 & 320 & varying & 0 & 16384 & 1.01 \\
\hline\hline
\label{tab1}
\end{tabular}
\end{center}
\end{table}
}

{\tabcolsep = 2mm
\begin{table}[h]
\begin{center} \scriptsize
\caption{The values of the parameters $N_{\rm DW}$ and $\Xi$ used in 2D simulations, and the number of realizations executed for each choice of them.}
\vspace{3mm}
\begin{tabular}{ c c c c | c c c c | c c c c }
\hline\hline
\multicolumn{4}{c|}{$N=8192$} &
\multicolumn{4}{c|}{$N=16384$} &
\multicolumn{4}{c}{$N=32768$} \\
\hline
Case & $N_{\rm DW}$ & $\Xi$ & Realization & Case & $N_{\rm DW}$ & $\Xi$ & Realization & Case & $N_{\rm DW}$ & $\Xi$ & Realization \\
\hline 
(a-1) & 2 & 0 & 10 & (b-1) & 2 & 0 & 10 & (c-1) & 2 & 0 & 1 \\
(a-2) & 2 & 0.0002 & 10 & (b-2) & 2 & 0.0001 & 10 & (c-2) & 3 & 0 & 1 \\
(a-3) & 2 & 0.0003 & 10 & (b-3) & 2 & 0.0002 & 10 & (c-3) & 4 & 0 & 1 \\
(a-4) & 2 & 0.0004 & 10 & (b-4) & 2 & 0.0003 & 10 & (c-4) & 5 & 0 & 1 \\
(a-5) & 2 & 0.0005 & 10 & (b-5) & 2 & 0.0004 & 10 & (c-5) & 6 & 0 & 1 \\
(a-6) & 2 & 0.0006 & 10 & (b-6) & 3 & 0 & 10 &  &  &  &  \\
(a-7) & 3 & 0 & 10 & (b-7) & 3 & 0.00006 & 10 &  &  &  &  \\
(a-8) & 3 & 0.00015 & 10 & (b-8) & 3 & 0.0001 & 10 &  &  &  &  \\
(a-9) & 3 & 0.0002 & 10 & (b-9) & 3 & 0.00014 & 10 &  &  &  &  \\
(a-10) & 3 & 0.00025 & 10 & (b-10) & 4 & 0 & 10 &  &  &  &  \\
(a-11) & 4 & 0 & 10 & (b-11) & 4 & 0.00005 & 10 &  &  &  &  \\
(a-12) & 4 & 0.0001 & 10 & (b-12) & 4 & 0.00007 & 10 &  &  &  &  \\
(a-13) & 5 & 0 & 10 & (b-13) & 4 & 0.00009 & 10 &  &  &  &  \\
(a-14) & 5 & 0.00009 & 10 & (b-14) & 5 & 0 & 10 &  &  &  &  \\
(a-15) & 6 & 0 & 10 & (b-15) & 5 & 0.00004 & 10 &  &  &  &  \\
(a-16) & 6 & 0.00006 & 10 & (b-16) & 5 & 0.00006 & 10 &  &  &  &  \\
&  &  &  & (b-17) & 6 & 0 & 10 &  &  &  &  \\
&  &  &  & (b-18) & 6 & 0.00003 & 10 &  &  &  &  \\
&  &  &  & (b-19) & 6 & 0.00004 & 10 &  &  &  &  \\
\hline\hline
\label{tab2}
\end{tabular}
\end{center}
\end{table}
}

We note that the value of the bias parameter $\Xi$ cannot be arbitrarily large.
This is because domain walls can decay owing to the nonuniform initial probability distribution of percolated vacua~\cite{Coulson:1995nv,Larsson:1996sp}
rather than the volume pressure $p_V$ induced by the energy difference $\Delta V$, if $\Xi$ is sufficiently large.
Let us assume that there are two degenerate vacua (i.e. the case with $N_{\rm DW}=2$) lifted by the bias term.
The energy density of one vacuum (true vacuum) is smaller than that of another vacuum (false vacuum), and their energy difference is given by $\Delta V_{\rm bias}\simeq 2\Xi\eta^4$.
Then, if $\Xi\ne 0$, we expect that the probability to find a true vacuum $p_t$ is different from that to find a false vacuum $p_f$ at the time of the formation of domain walls.
The ratio between these two probabilities can be estimated as~\cite{Hiramatsu:2010yz,Hiramatsu:2010yn}
\begin{equation}
\frac{p_f}{p_t} = \exp\left(-\frac{\Delta V_{\rm bias}}{\Delta V_{\rm pot}}\right) \simeq \exp\left(-\frac{\Xi\eta^2 N_{\rm DW}^2}{m_a^2}\right),
\end{equation}
where $\Delta V_{\rm pot}\simeq 2m_a^2\eta^2/N_{\rm DW}^2$ is the height of the potential barrier of domain walls.
This initial biased distribution of percolated vacua leads to the collapse of domain walls in the (conformal) time scale $\tau_{\rm dec,prob}$,
which is given by~\cite{Larsson:1996sp}
\begin{equation}
\frac{\tau_{\rm prob,dec}}{\tau_{\rm form}} \simeq \varepsilon^{-D/2}.
\end{equation}
Here, $\tau_{\rm form}$ is the conformal time corresponding to the formation time $t_{\rm form}$, $D$ is the spatial dimension, and $\varepsilon$ is given by
$p_t = 0.5+\varepsilon$ and $p_f = 0.5-\varepsilon$, which leads to
\begin{equation}
\varepsilon = \frac{1}{2}\frac{1-p_f/p_t}{1+p_f/p_t}.
\end{equation}
Requiring that $\tau_{\rm prob,dec}>\tau_f$ with $D=2$, we obtain the condition
\begin{equation}
\frac{\Xi\eta^2N_{\rm DW}^2}{m_a^2} < 2\tanh^{-1}\left(\frac{2\tau_{\rm form}}{\tau_f}\right). \label{condition_bias}
\end{equation}
It can be checked that all parameters shown in Table~\ref{tab2} satisfy the above condition with $\tau_{\rm form}\simeq \mathcal{O}(10)$.\footnote{
The condition [Eq.~\eqref{condition_bias}] is derived with the assumption of $N_{\rm DW}=2$, while it is not straightforward to apply this result to the
case with $N_{\rm DW}>2$. However, we expect that the constraint for $\Xi$ would be weaker than Eq.~\eqref{condition_bias} for the case with $N_{\rm DW}>2$,
since the magnitude of $\Delta V_{\rm bias}$
becomes smaller than $2\Xi\eta^4$ and the initial probability distribution is less sensitive to $\Xi$.
Hence, here we just consider the bound given by Eq.~\eqref{condition_bias} for any choice of $N_{\rm DW}$.}

\subsubsection{\label{sec3-1-2} 3D}
We also perform 3D simulations and use their results to estimate the spectrum of axions radiated by topological defects.
The power spectrum of axions can be defined by
\begin{equation}
\frac{1}{2}\langle\dot{a}({\bf k},\tau)^*\dot{a}({\bf k'},\tau)\rangle_{\rm en} = \frac{(2\pi)^3}{k^2}\delta^{(3)}({\bf k}-{\bf k'})P(k,\tau), \label{def_power_spectrum}
\end{equation}
where $\langle\dots\rangle_{\rm en}$ is an ensemble average.
The left-hand side of the above equation can be computed from the simulated data of $\Phi({\bf x},\tau)$ and $\dot{\Phi}({\bf x},\tau)$:
\begin{equation}
\dot{a}({\bf k},\tau) = \int d^3{\bf x} e^{i{\bf k\cdot x}}\dot{a}({\bf x},\tau)\qquad \mathrm{with} \qquad \dot{a}({\bf x},\tau) = \eta\mathrm{Im}\left[\frac{\dot{\Phi}}{\Phi}({\bf x},\tau)\right].
\end{equation}
To obtain a reliable estimation for $P(k,\tau)$, we excise the core of strings and that of domain walls from the map of $\dot{a}({\bf x},\tau)$
[see Refs.~\cite{Hiramatsu:2010yu,Hiramatsu:2012gg,Hiramatsu:2012sc} for details].
Furthermore, we subtract the contaminations caused by initial field fluctuations 
by computing the difference of power spectra evaluated at two different time steps $\tau_A$ and $\tau_B>\tau_A$:
\begin{equation}
\Delta P(k,\tau_B) = P(k,\tau_B) - \mathcal{R}(k,\tau_A,\tau_B) P(k,\tau_A),
\label{diff_spectrum}
\end{equation}
where $\mathcal{R}(k,\tau_A,\tau_B)$ is a reduction factor taking account of the effect of cosmic expansion.
We will specify $\tau_A$, $\tau_B$, and $\mathcal{R}(k,\tau_A,\tau_B)$ shortly.

The numerical calculation of the power spectrum is performed in terms of the plural bins, which are given by
\begin{align}
F_i &= \left\{{\bf k}\left| k_i^{(\mathrm{min})}\le |{\bf k}| \le k_i^{(\mathrm{max})} \right.\right\} \qquad (i=1,2,\dots, n_{\rm bin}), \nonumber\\
k_i^{(\mathrm{min})} &= \frac{i-1}{n_{\rm bin}}\frac{\pi N}{L}, \qquad \mathrm{and} \qquad k_i^{(\mathrm{max})} = \frac{i}{n_{\rm bin}}\frac{\pi N}{L},
\end{align}
with $n_{\rm bin}$ being some integer. Then, the power spectrum $P(k,\tau)$ appearing in the right-hand side of Eq.~\eqref{def_power_spectrum}
is replaced with the average over $i$-th bin $P(k_i,\tau)$, where $k_i\equiv (\sum_{{\bf k}\in F_i}|{\bf k}|)/(\sum_{{\bf k}\in F_i}1)$.
Using the difference of discretized power spectrum, we can compute the mean energy of radiated axions:\footnote{In Refs.~\cite{Hiramatsu:2012gg,Hiramatsu:2012sc},
the mean comoving momentum of radiated axions was incorrectly computed as
\begin{equation}
\bar{k}(\tau_B) = \frac{\sum_{i=1}^{n_{\rm bin}}\Delta P(k_i,\tau_B)}{\sum_{i=1}^{n_{\rm bin}}\frac{1}{k_i}\Delta P(k_i,\tau_B)}, \nonumber
\end{equation}
which should be replaced with
\begin{equation}
\bar{k}(\tau_B) = \frac{\sum_{i=1}^{n_{\rm bin}}\frac{k_i}{\omega_a(k_i,\tau_B)}\Delta P(k_i,\tau_B)}{\sum_{i=1}^{n_{\rm bin}}\frac{1}{\omega_a(k_i,\tau_B)}\Delta P(k_i,\tau_B)}. \nonumber
\end{equation}
The contributions of $k_i\lesssim m_a$ are suppressed in the latter case, and hence the results of Refs.~\cite{Hiramatsu:2012gg,Hiramatsu:2012sc} underestimate $\bar{k}(\tau_B)$,
which we aim to correct here.
Note that we compute $\bar{\omega}_a(\tau_B)$ rather than $\bar{k}(\tau_B)$ from the reason described in the footnote~\ref{on_tilde_epsilon_w}.}
\begin{equation}
\bar{\omega}_a(\tau_B) = \frac{\displaystyle{\sum_{i=1}^{n_{\rm bin}}\Delta P(k_i,\tau_B)}}{\displaystyle{\sum_{i=1}^{n_{\rm bin}}\frac{1}{\omega_a(k_i,\tau_B)}\Delta P(k_i,\tau_B)}},
\label{mean_omega}
\end{equation}
where $\omega_a(k_i,\tau_B)$ will be specified shortly.

In 3D simulations, we estimate the mean energy of radiated axions [Eq.~\eqref{mean_omega}] for three cases:
(i) global strings, (ii) short-lived string-wall systems, and (iii) long-lived string-wall systems.
Here we choose the grid size as $512^3$,
and use the same setups as the previous studies~\cite{Hiramatsu:2010yu,Hiramatsu:2012gg,Hiramatsu:2012sc}.
In what follows, we briefly describe the setups in the three different regimes:
\begin{enumerate}
\renewcommand{\labelenumi}{(\roman{enumi})}
\item {\it Global strings}~\cite{Hiramatsu:2010yu}. The evolution of the scalar field $\Phi$ is solved in the FRW background with the potential
given by Eq.~\eqref{V_phi_string} and also with the finite temperature correction of the form $\propto \lambda T^2|\Phi|^2$.
Global strings are formed around the conformal time $\tau=\tau_c$, which can be defined by the condition $T(\tau=\tau_c)=\sqrt{3}\eta$.
The reduction factor in Eq.~\eqref{diff_spectrum} can be specified as $\mathcal{R}(k,\tau_A,\tau_B)=(R(\tau_A)/R(\tau_B))^4$,
since axions are massless in this regime and the power spectrum scales as $\propto R(\tau)^{-4}$.
Following Ref.~\cite{Hiramatsu:2010yu}, we choose $\tau_A=3.5\tau_c$ and $\tau_B=5\tau_c$.
After calculating $\bar{\omega}_a(\tau_B)$ in Eq.~\eqref{mean_omega} with $\omega_a(k_i,\tau_B)=k_i/R(\tau_B)$,
we estimate the value of the parameter $\epsilon$ from Eq.~\eqref{mean_omega_a_string}.
\item {\it Short-lived string-wall systems}~\cite{Hiramatsu:2012gg}. The evolution of the scalar field $\Phi$ is solved in the FRW background with the potential
given by Eqs.~\eqref{V_phi_string} and~\eqref{V_axion} with $N_{\rm DW}=1$. 
We also include the finite temperature correction of the form $\propto \lambda T^2|\Phi|^2$.
Furthermore, we use the finite temperature axion mass [Eq.~\eqref{finite_temp_mass}], which can be rewritten as
\begin{equation}
\frac{m_a(T)^2}{F_a^2} = c_T\kappa^{n+4}\left(\frac{T}{F_a}\right)^{-n}\quad \mathrm{with} \quad \kappa = \frac{\Lambda_{\rm QCD}}{F_a}.
\label{finite_temp_mass_simulation}
\end{equation}
In the numerical simulations, we vary the value of $\kappa$.
The reduction factor in Eq.~\eqref{diff_spectrum} can be specified as $\mathcal{R}(k,\tau_A,\tau_B)=(\omega_a(k,\tau_B)/\omega_a(k,\tau_A))(R(\tau_A)/R(\tau_B))^3$
with $\omega_a(k,\tau)=\sqrt{m_a(T)^2+k^2/R(\tau)^2}$, which takes account of the finiteness of the axion mass.
Following Ref.~\cite{Hiramatsu:2012gg}, we choose $\tau_A$ and $\tau_B$ such that
$\tau_A$ corresponds to the cosmic time $t_1$ defined by Eq.~\eqref{condition_t1} and $\tau_B$ corresponds to the time at which
the area parameter of domain walls $\mathcal{A}$ becomes less than 0.01.
The later condition can be regarded as a definition of the decay time $t_d$ of the short-lived string-wall systems. 
After calculating $\bar{\omega}_a(\tau_B)$ in Eq.~\eqref{mean_omega} with $\omega_a(k_i,\tau_B)=\sqrt{m_a(\tau_B)^2+k_i^2/R(\tau_B)^2}$ and $m_a(\tau_B)$
being the value of the axion mass at $\tau=\tau_B$,
we estimate the value of the parameter $\tilde{\epsilon}_w$ from Eq.~\eqref{tilde_epsilon_w_def}.
\item {\it Long-lived string-wall systems}~\cite{Hiramatsu:2012sc}.
The evolution of the scalar field $\Phi$ is solved in the FRW background with the potential
given by Eqs.~\eqref{V_phi_string} and~\eqref{V_axion} with $N_{\rm DW}>1$.
Here we fix the axion mass as a constant value ($m_a/\eta = 0.1$) instead of using the temperature dependent mass [Eq.~\eqref{finite_temp_mass_simulation}],
since we are interested in the time scale much later than the epoch of the QCD phase transition.
Namely, the model is the same as that in 2D simulations [Eq.~\eqref{V_phi_simulation}] with $\Xi=0$.
The reduction factor in Eq.~\eqref{diff_spectrum} can be specified as $\mathcal{R}(k,\tau_A,\tau_B)=(\omega_a(k,\tau_B)/\omega_a(k,\tau_A))(R(\tau_A)/R(\tau_B))^3$
with $\omega_a(k,\tau)=\sqrt{m_a^2+k^2/R(\tau)^2}$. Following Ref.~\cite{Hiramatsu:2012sc}, we choose $\tau_A=14\eta^{-1}$ and $\tau_B=40\eta^{-1}$.
In the numerical simulations, we vary the value of $N_{\rm DW}$.
After calculating $\bar{\omega}_a(\tau_B)$ in Eq.~\eqref{mean_omega} with $\omega_a(k_i,\tau_B)=\sqrt{m_a^2+k_i^2/R(\tau_B)^2}$,
we estimate the value of the parameter $\tilde{\epsilon}_a$ from Eq.~\eqref{tilde_epsilon_a_def} for each value of $N_{\rm DW}$.
\end{enumerate}

We summarize the setup of 3D simulations in Table~\ref{tab3}.
For each choice of the parameters we execute 10 realizations.
In total, we perform simulations in 12 different sets of parameters as shown in cases (d)$\textendash$(f-5) of Table~\ref{tab3}.
In each case, we compute $\bar{\omega}_a(\tau_B)$ and its error by averaging the results of 10 realizations.
However, as we will see in Sec.~\ref{sec3-3}, the method used in Refs.~\cite{Hiramatsu:2010yu,Hiramatsu:2012gg,Hiramatsu:2012sc}
leads to some inappropriate results for averaged values.
To avoid this situation, in this work we introduce a new estimator described in Appendix~\ref{secB}.
In Sec.~\ref{sec3-3}, we will also check the difference between the old and new averaging methods.

{\tabcolsep = 2mm
\begin{table}[h]
\begin{center} 
\caption{Models and parameters used in 3D numerical simulations, and the number of realizations executed for each choice of them.
For other parameters, we used the same values as previous studies indicated in the ``Reference" column.}
\vspace{3mm}
\begin{tabular}{c c c c l c}
\hline\hline
Case & Model & Grid size ($N^3$) & Reference & Parameter & Realization \\
\hline 
(d) & Global strings & $512^3$ & \cite{Hiramatsu:2010yu} & & 10\\
(e-1) & Short-lived string-wall systems ($N_{\rm DW}=1$) & $512^3$ & \cite{Hiramatsu:2012gg} & $\kappa=0.275$ & 10\\
(e-2) & Short-lived string-wall systems ($N_{\rm DW}=1$) & $512^3$ & \cite{Hiramatsu:2012gg} & $\kappa=0.3$ & 10\\
(e-3) & Short-lived string-wall systems ($N_{\rm DW}=1$) & $512^3$ & \cite{Hiramatsu:2012gg} & $\kappa=0.325$ & 10\\
(e-4) & Short-lived string-wall systems ($N_{\rm DW}=1$) & $512^3$ & \cite{Hiramatsu:2012gg} & $\kappa=0.35$ & 10\\
(e-5) & Short-lived string-wall systems ($N_{\rm DW}=1$) & $512^3$ & \cite{Hiramatsu:2012gg} & $\kappa=0.375$ & 10\\
(e-6) & Short-lived string-wall systems ($N_{\rm DW}=1$) & $512^3$ & \cite{Hiramatsu:2012gg} & $\kappa=0.4$ & 10\\
(f-1) & Long-lived string-wall systems ($N_{\rm DW}>1$) & $512^3$ & \cite{Hiramatsu:2012sc} & $N_{\rm DW}=2$ & 10\\
(f-2) & Long-lived string-wall systems ($N_{\rm DW}>1$) & $512^3$ & \cite{Hiramatsu:2012sc} & $N_{\rm DW}=3$ & 10\\
(f-3) & Long-lived string-wall systems ($N_{\rm DW}>1$) & $512^3$ & \cite{Hiramatsu:2012sc} & $N_{\rm DW}=4$ & 10\\
(f-4) & Long-lived string-wall systems ($N_{\rm DW}>1$) & $512^3$ & \cite{Hiramatsu:2012sc} & $N_{\rm DW}=5$ & 10\\
(f-5) & Long-lived string-wall systems ($N_{\rm DW}>1$) & $512^3$ & \cite{Hiramatsu:2012sc} & $N_{\rm DW}=6$ & 10\\
\hline\hline
\label{tab3}
\end{tabular}
\end{center}
\end{table}
}
\subsection{\label{sec3-2} Estimation of the decay time of domain walls}
\begin{figure}[htbp]
\centering
$\begin{array}{cc}
\subfigure[$\Xi=0,\ \tau=42$]{
\includegraphics[width=0.36\textwidth]{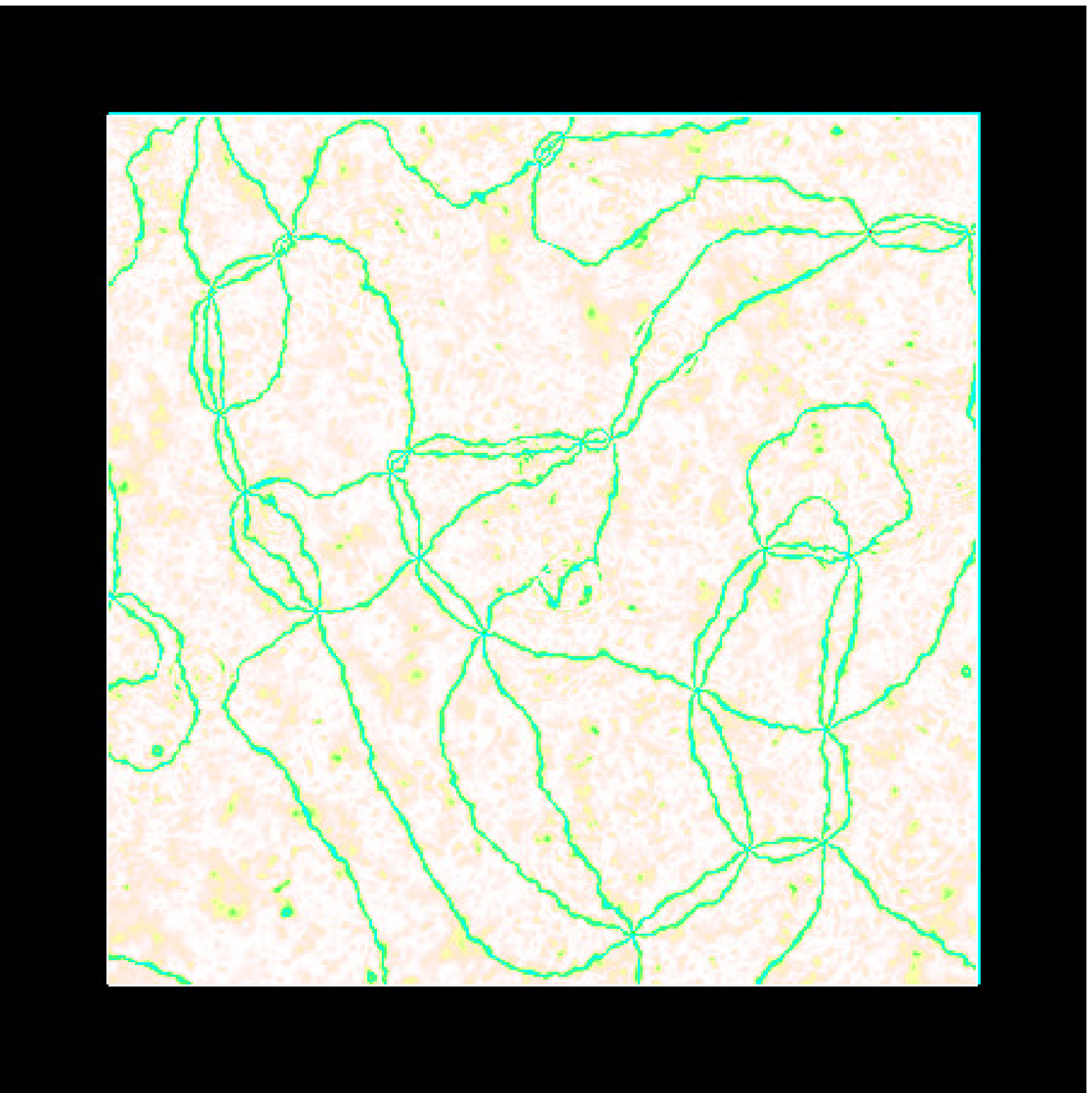}}
\hspace{20pt}
\subfigure[$\Xi=0.00006,\ \tau=42$]{
\includegraphics[width=0.36\textwidth]{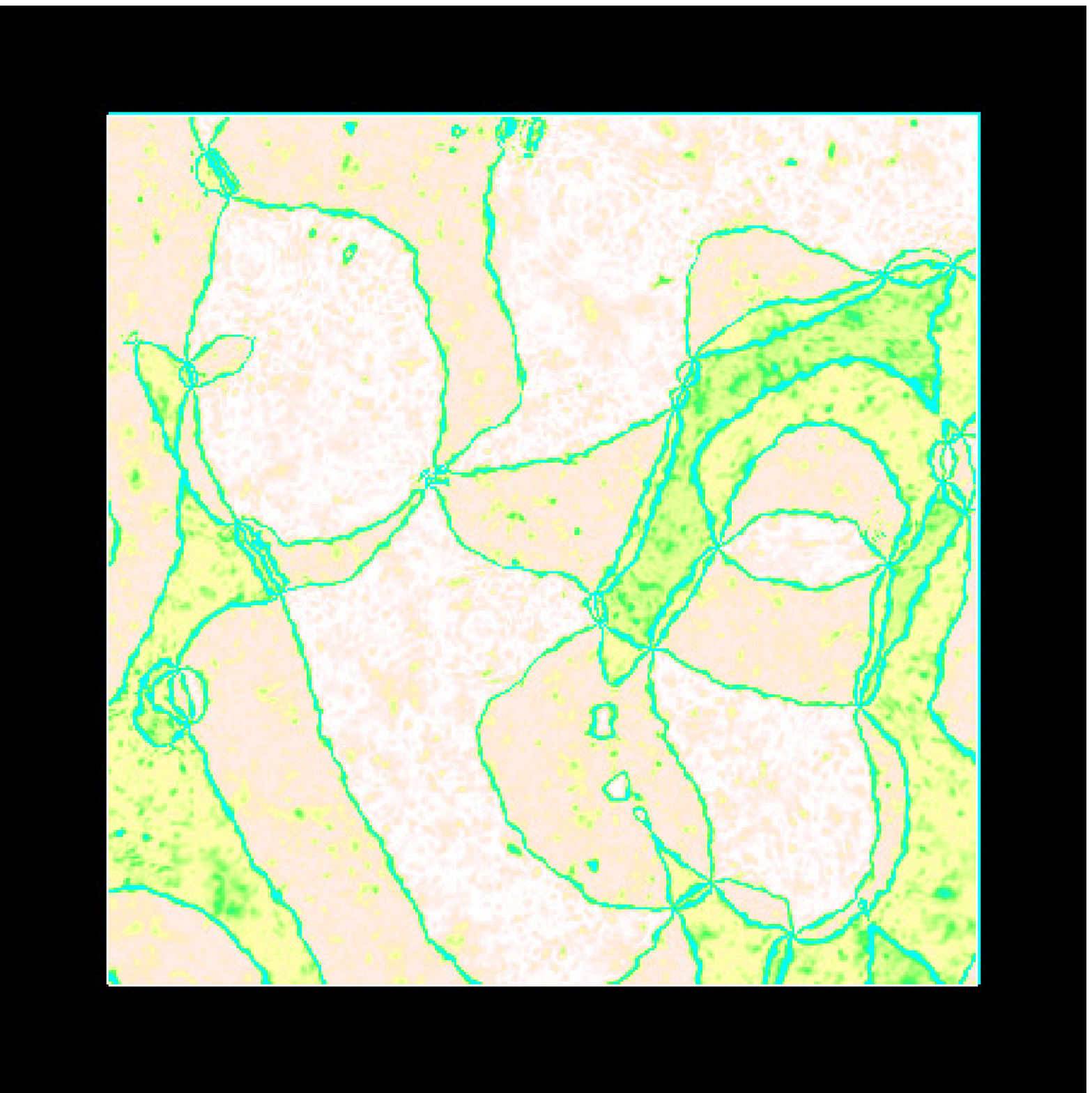}}
\\
\subfigure[$\Xi=0,\ \tau=62$]{
\includegraphics[width=0.36\textwidth]{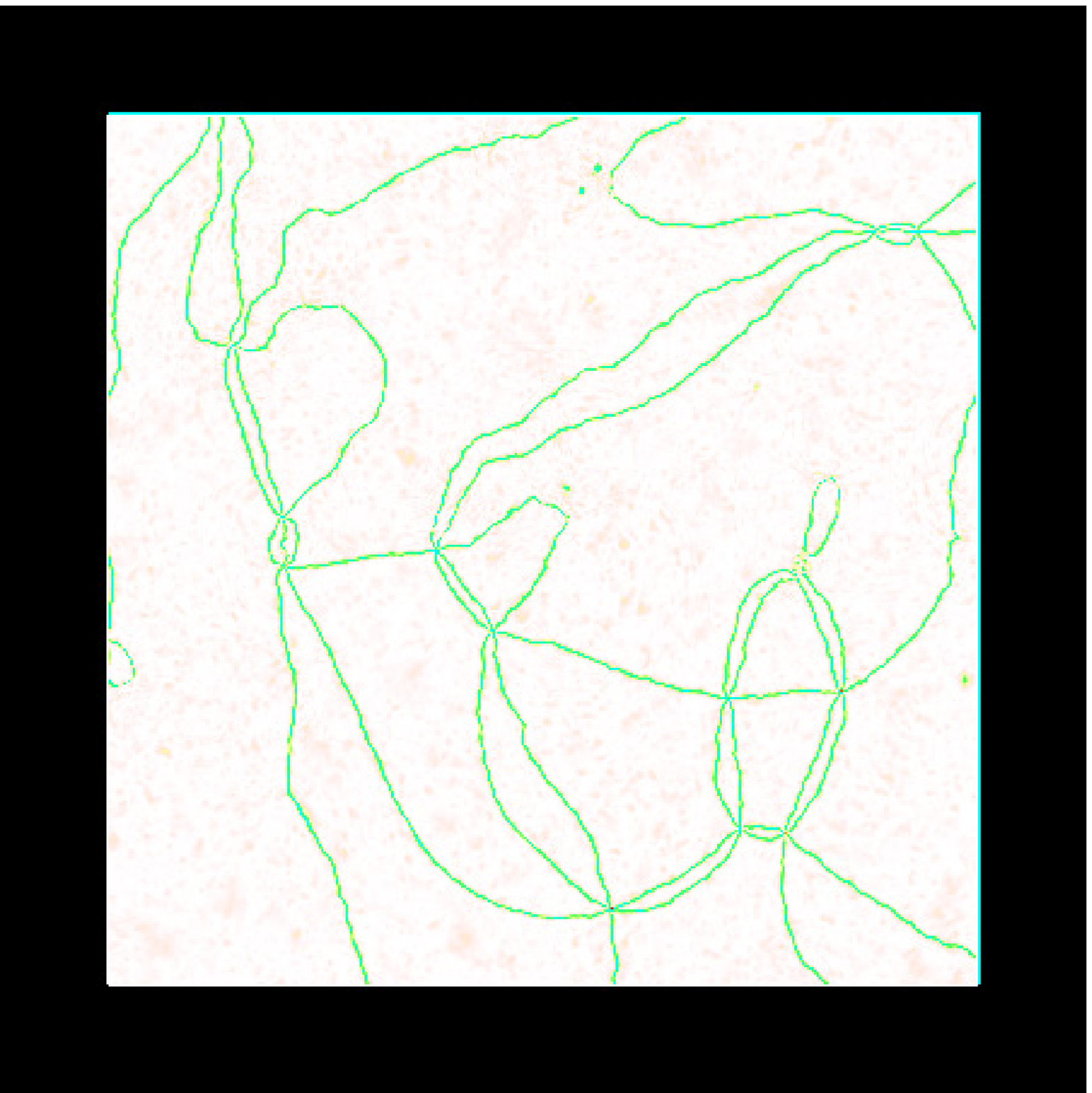}}
\hspace{20pt}
\subfigure[$\Xi=0.00006,\ \tau=62$]{
\includegraphics[width=0.36\textwidth]{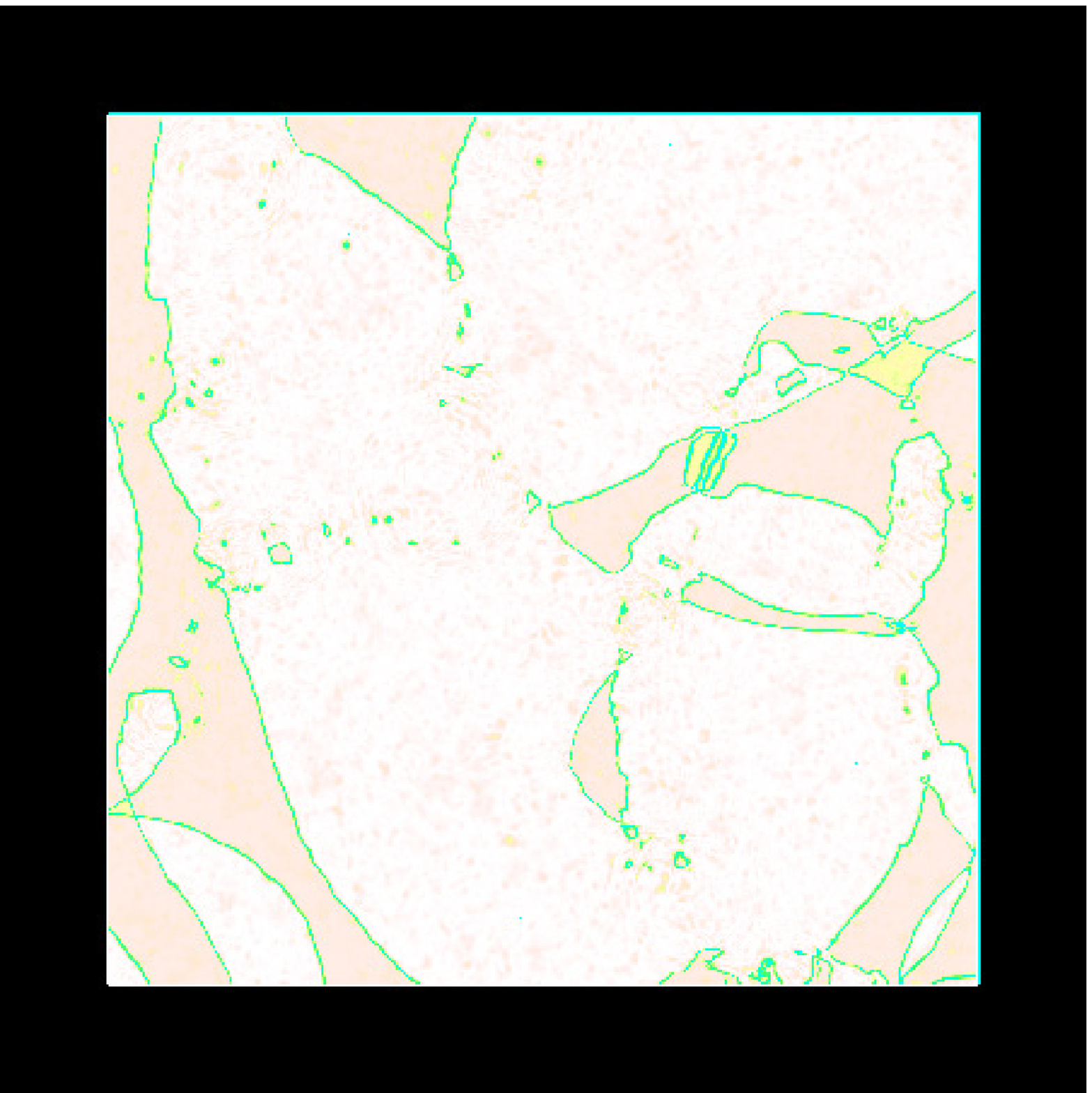}}
\\
\subfigure[$\Xi=0,\ \tau=82$]{
\includegraphics[width=0.36\textwidth]{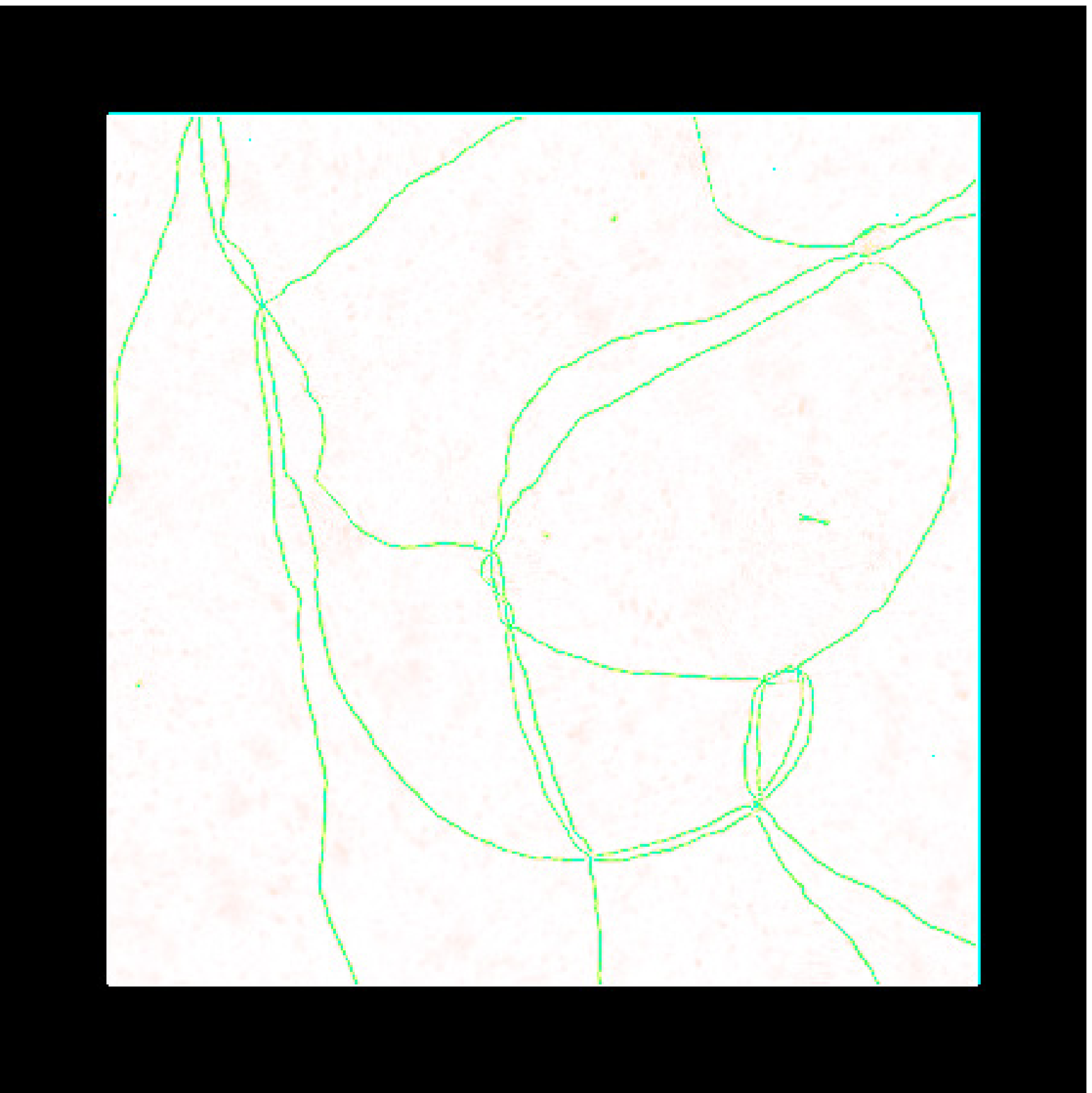}}
\hspace{20pt}
\subfigure[$\Xi=0.00006,\ \tau=82$]{
\includegraphics[width=0.36\textwidth]{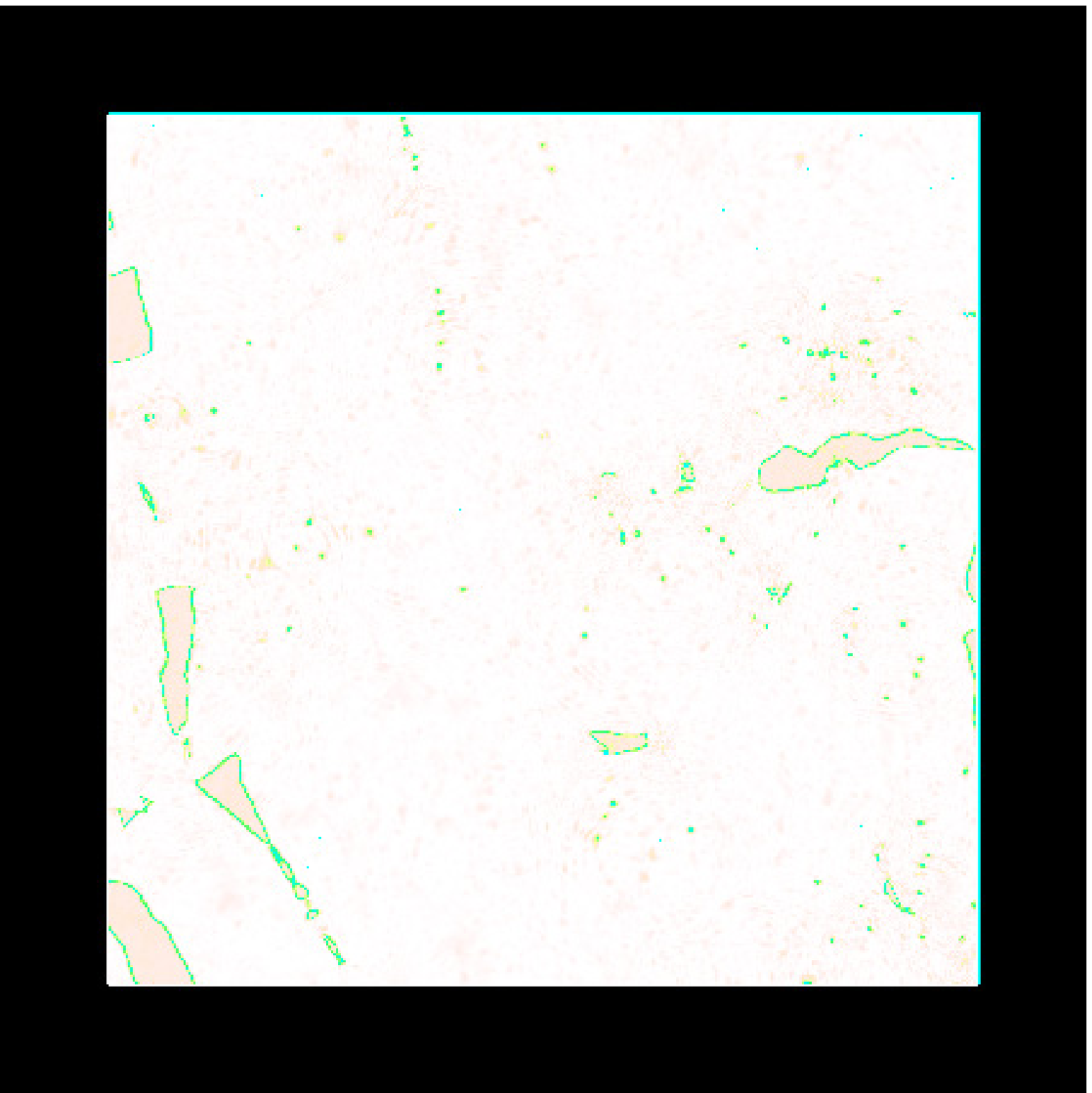}}
\end{array}$
\caption{The distribution of the potential energy of the scalar field on the simulation box.
These figures are generated from the results of 2D simulations with $N=8192$ and $N_{\rm DW}=6$ [cases (a-15) and (a-16) in Table~\ref{tab2}].
Each subfigure corresponds to a different choice of the bias parameter and the time step:
(a) $\Xi=0,\ \tau=42$, (b) $\Xi=0.00006,\ \tau=42$, (c) $\Xi=0,\ \tau=62$,
(d) $\Xi=0.00006,\ \tau=62$, (e) $\Xi=0,\ \tau=82$, and (f) $\Xi=0.00006,\ \tau=82$.
The size of these figures is set to be a quarter ($4096^2$) of the size of the simulation box ($8192^2$).
The green region corresponds to the core of domain walls $V(\Phi)= 2m_a^2\eta^2/N_{\rm DW}^2$, and the white region corresponds to the vacuum $V(\Phi)= 0$.
}
\label{fig1}
\end{figure}

\begin{figure}[htbp]
\centering
$\begin{array}{cc}
\subfigure[]{
\includegraphics[width=0.44\textwidth]{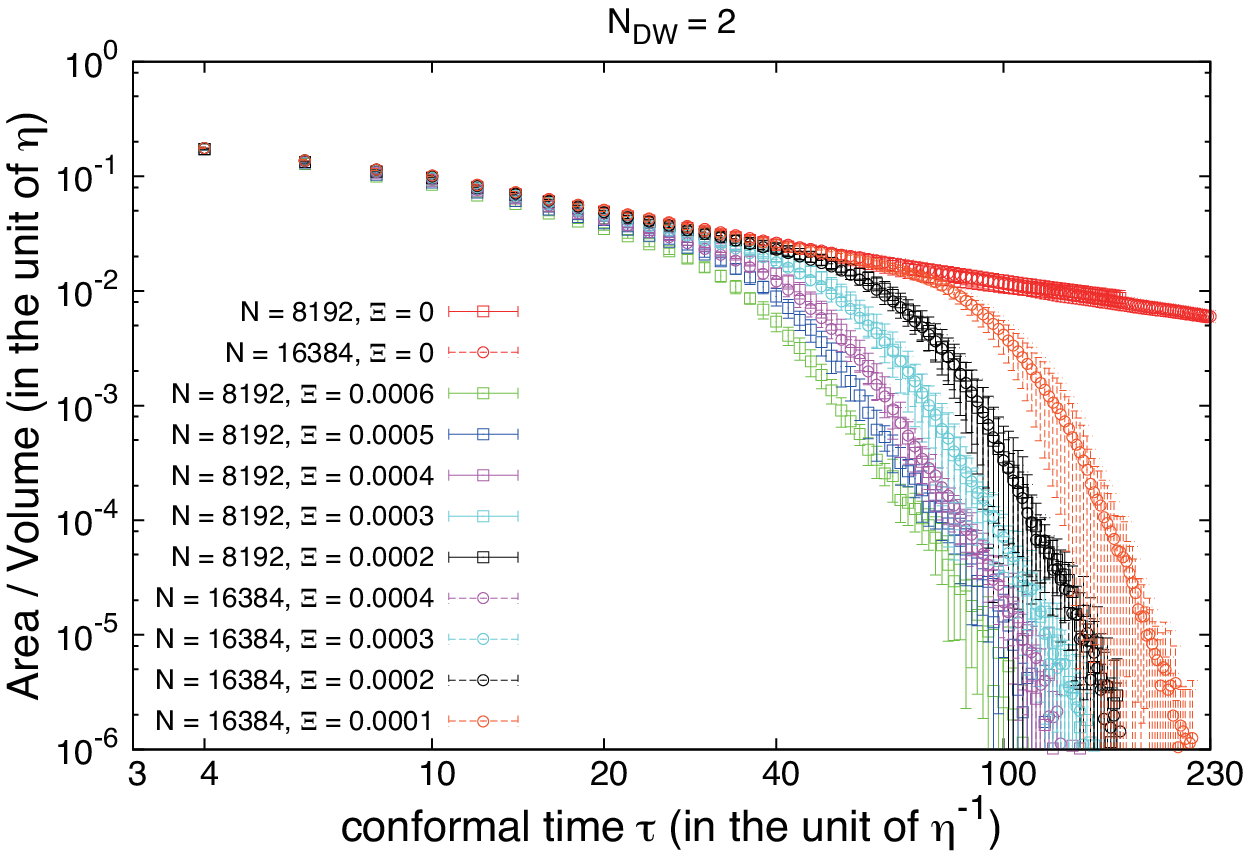}}
\hspace{20pt}
\subfigure[]{
\includegraphics[width=0.44\textwidth]{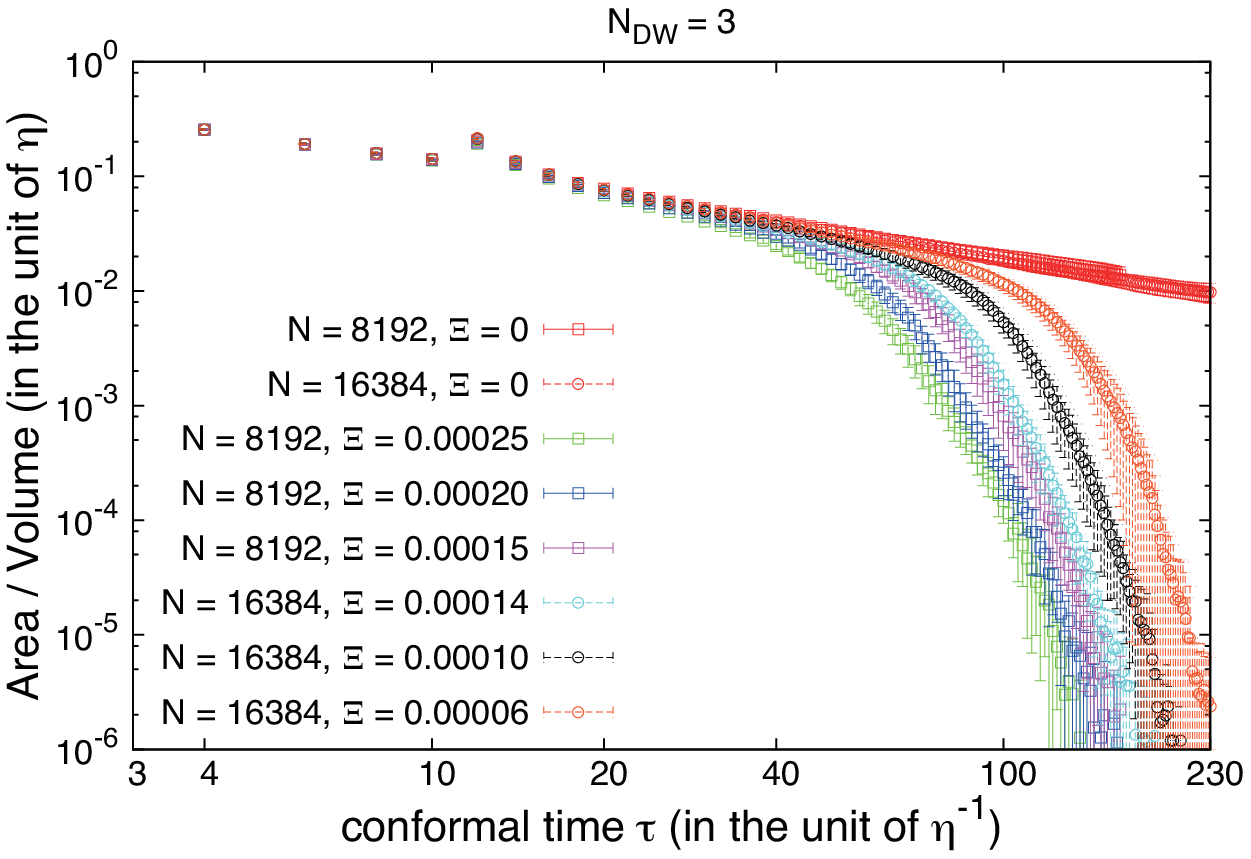}}
\\
\subfigure[]{
\includegraphics[width=0.44\textwidth]{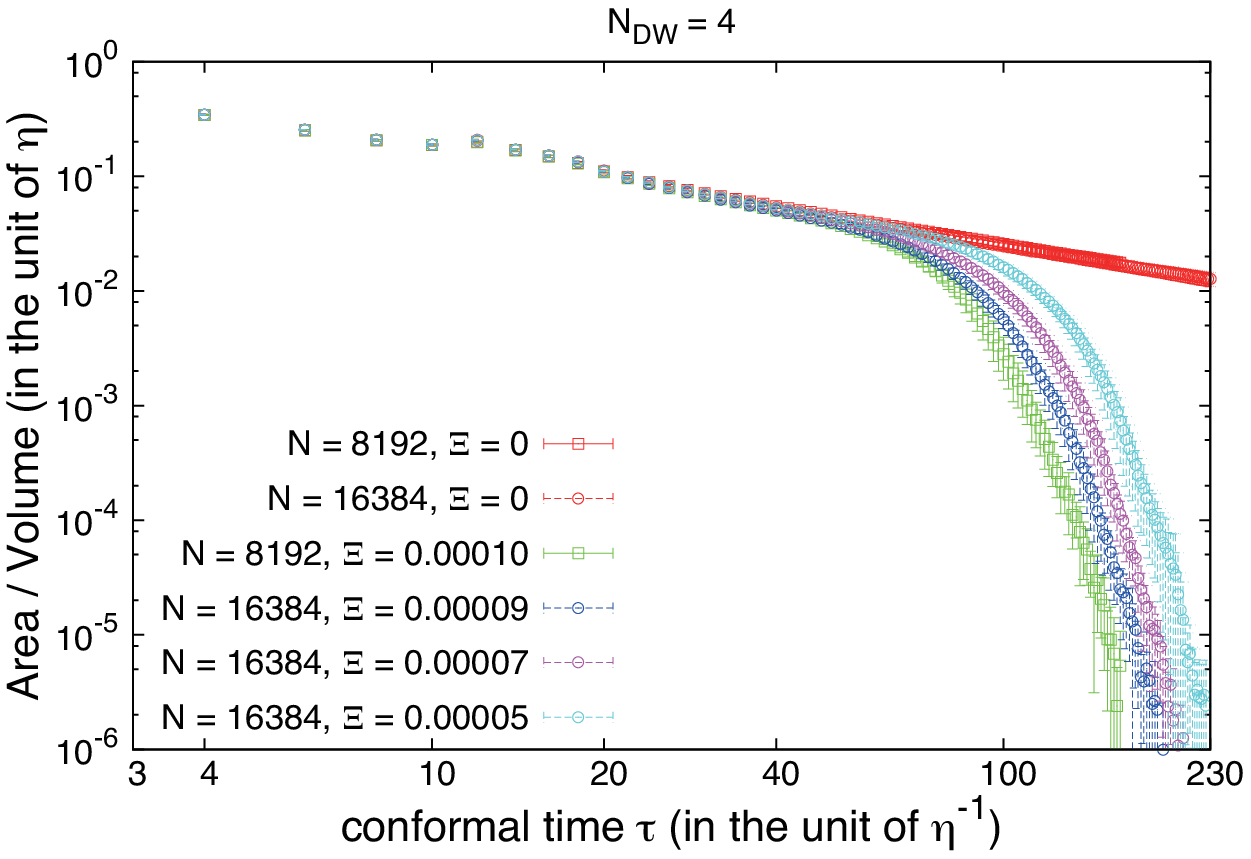}}
\hspace{20pt}
\subfigure[]{
\includegraphics[width=0.44\textwidth]{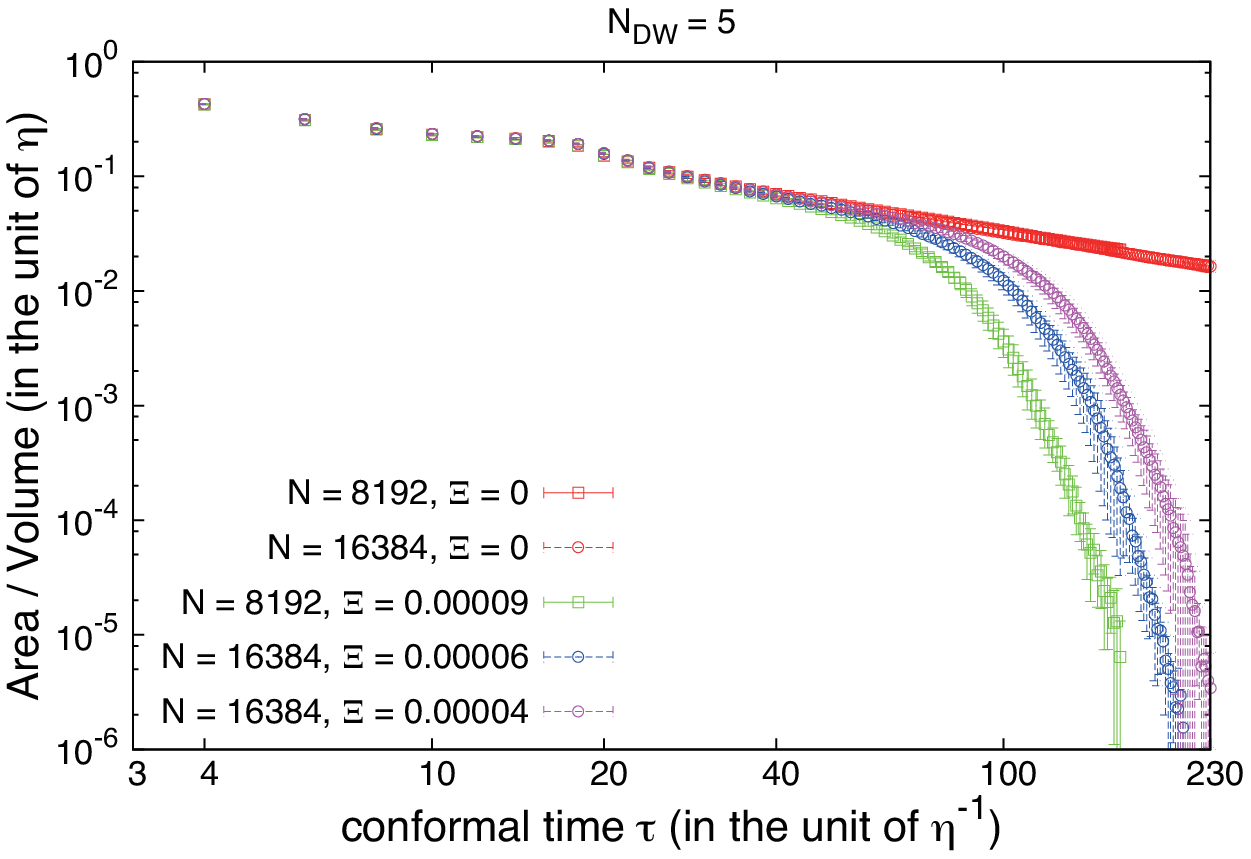}}
\end{array}$
\subfigure[]{
\includegraphics[width=0.44\textwidth]{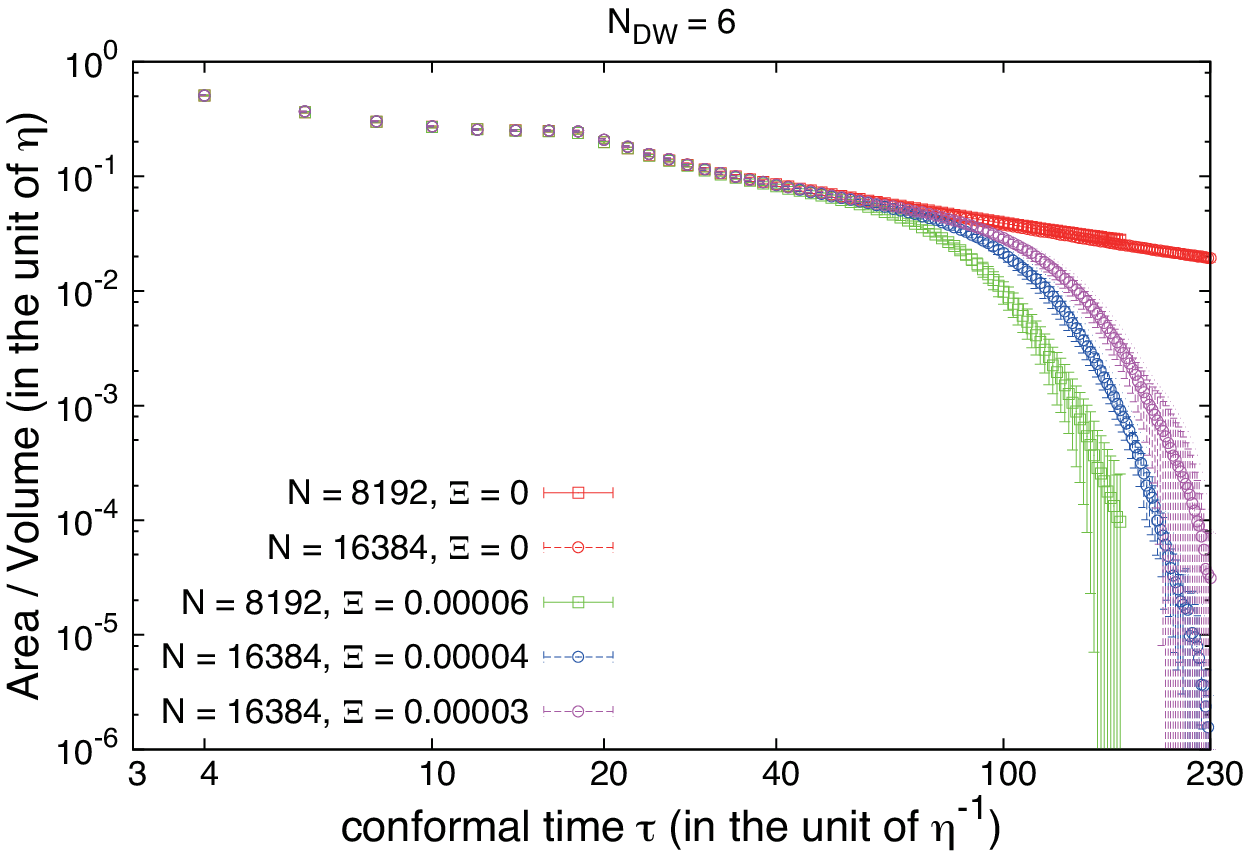}}
\caption{Time evolution of the comoving area density $A/V$ of domain walls for various values of $\Xi$.
Each panel shows the results for (a) $N_{\rm DW}=2$, (b) $N_{\rm DW}=3$, (c) $N_{\rm DW}=4$, (d) $N_{\rm DW}=5$, and (e) $N_{\rm DW}=6$.
In these figures, we plot the mean and standard deviation among 10 realizations of numerical simulations.}
\label{fig2}
\end{figure}

In Fig.~\ref{fig1}, we show the map of the potential energy of the field $\Phi$ as a visualization of the 2D simulations.
As shown in Figs.~\ref{fig1} (a), \ref{fig1} (c), and \ref{fig1} (e), green lines corresponding to the core of domain walls continue to exist for the case with $\Xi=0$.
On the other hand, Figs.~\ref{fig1} (b), \ref{fig1} (d), and \ref{fig1} (f) show that these domain walls tend to collapse for the case with $\Xi\ne 0$.
In this case, the domains with higher energies (colored regions) gradually disappear, and
the domain with the lowest energy (white region) dominates over the simulation box at late times.
In Fig.~\ref{fig2}, we also plot the time evolution of the quantity $A/V$, where $A$ is the comoving area occupied by domain walls,
and $V=L^3$ is the comoving volume of the simulation box.
The results of the simulations with $\Xi=0$ indicate the behavior $A/V\propto \tau^{-1}$, which corresponds to the scaling solution,
while the values of $A/V$ rapidly fall off at late times for the cases with $\Xi \ne 0$.


\begin{figure}[htbp]
\begin{center}
\includegraphics[scale=1.0]{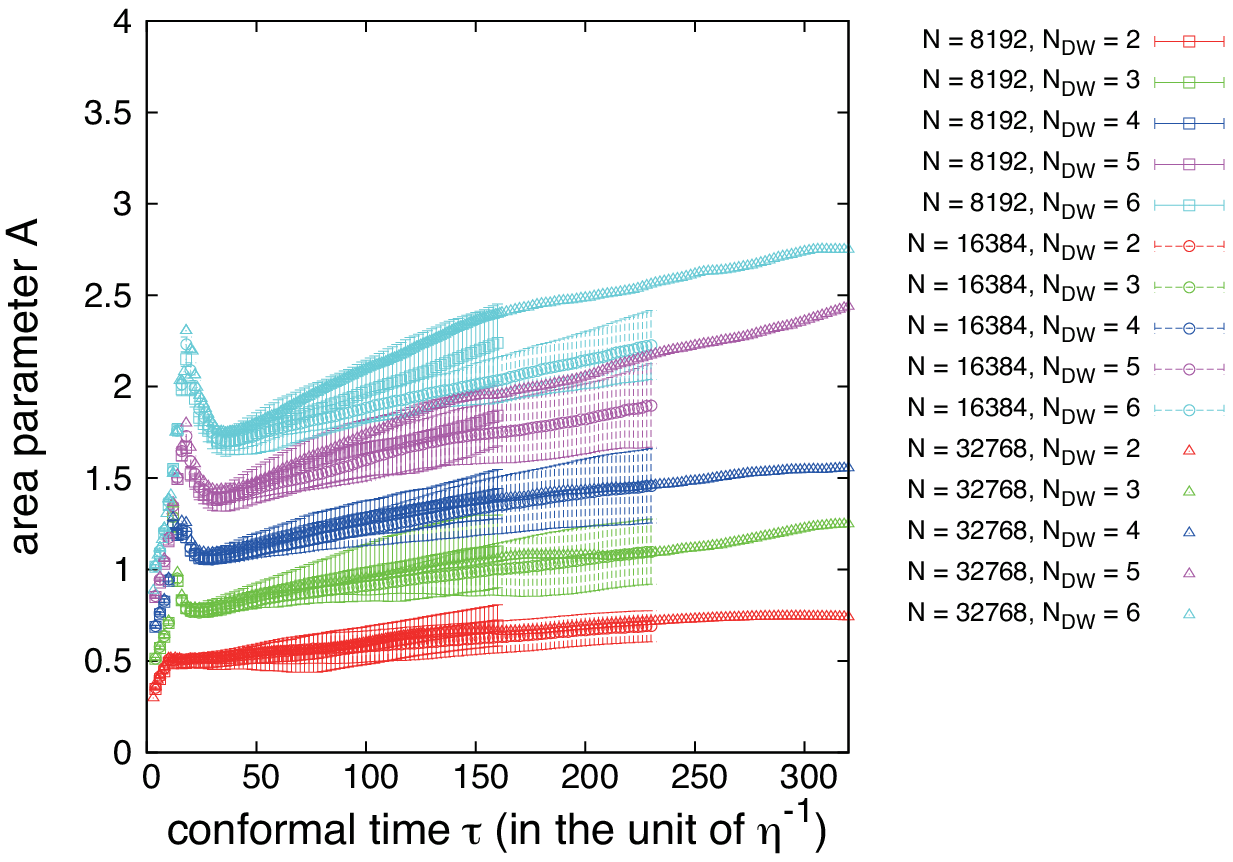}
\end{center}
\caption{Time evolution of the area parameter $\mathcal{A}$ for various values of $N_{\rm DW}$ without the bias ($\Xi=0$).
For the simulations with $N=8192$ and $N=16384$, we plot the mean and standard deviation among 10 realizations.
For those with $N=32768$, we just plot the result of 1 realization.
}
\label{fig3}
\end{figure}

In order to estimate the energy density of axions produced from long-lived domain walls,
we must estimate some numerical quantities, which cannot be predicted in the analytical calculations.
First, we determine the value of the area parameter $\mathcal{A}$ defined by Eq.~\eqref{rho_wall_t}.
In a similar way to Ref.~\cite{Hiramatsu:2012sc},
this quantity can be computed in terms of the comoving area density $A/V$ obtained from the results of the numerical simulations:
\begin{equation}
\mathcal{A} = \frac{At}{R(t)V}.
\end{equation}
In Fig.~\ref{fig3}, we show the results for the simulations with $\Xi=0$ [cases (a-1), (a-7), (a-11), (a-13), (a-15), (b-1), (b-6), (b-10), (b-14), (b-17), (c-1), (c-2), (c-3), (c-4), and (c-5) in Table~\ref{tab2}].
Without the bias term, the area parameter $\mathcal{A}$ takes almost constant values of $\mathcal{O}(1)$.
The values of $\mathcal{A}$ at the final time of the simulations are shown in Table~\ref{tab4}.
We see that the value of $\mathcal{A}$ increases for large $N_{\rm DW}$, which agrees with the results of the previous study~\cite{Hiramatsu:2012sc}.


{\tabcolsep = 2mm
\begin{table}[h]
\begin{center} 
\caption{The value of $\mathcal{A}$ at the final time $\tau_f$ of numerical simulations for various values of $N_{\rm DW}$.}
\vspace{3mm}
\begin{tabular}{ c c c }
\hline\hline
$N_{\rm DW}$ & $\mathcal{A}(\tau_f)$ ($N=8192$, $\tau_f=160$) & $\mathcal{A}(\tau_f)$ ($N=16384$, $\tau_f=230$) \\
\hline 
2 & $0.694 \pm 0.113$ & $0.690 \pm 0.085$ \\
3 & $1.10 \pm 0.20$ & $1.10 \pm 0.18$ \\
4 & $1.41 \pm 0.13$ & $1.46 \pm 0.20$ \\
5 & $1.84 \pm 0.17$ & $1.90 \pm 0.23$ \\
6 & $2.24 \pm 0.21$ & $2.23 \pm 0.19$ \\
\hline\hline
\label{tab4}
\end{tabular}
\end{center}
\end{table}
}

We note that there is a subtlety in the results shown in Fig.~\ref{fig3}.
The values of $\mathcal{A}$ plotted in this figure slightly increase with time, deviating from the exact scaling behavior ($\mathcal{A}=\mathrm{constant}$).
At this stage it is not clear whether this slight increase of $\mathcal{A}$ continues in later times or not, because of the limitation of the dynamical range of the numerical simulations. 
Here, we just take account of the possibility of the deviation from the scaling solution, and fit the result of $\mathcal{A}(\tau)$ obtained from the simulations with $N=16384$
into the model function given by Eq.~\eqref{rho_wall_t_dev}. In terms of the conformal time, this model function can be rewritten as
\begin{equation}
\mathcal{A}(\tau) = \mathcal{A}_{\rm form}\left(\frac{\tau}{\tau_{\rm form}}\right)^{2(1-p)}. \label{area_t_model}
\end{equation}
We fix the value of $\tau_{\rm form}$ and seek for a value of $p$ which fits the data obtained from the simulations.
The value of $\tau_{\rm form}$ should be taken as the time when initial fluctuations of $\mathcal{A}$ is sufficiently moderated.
Here we choose $\tau_{\rm form}=50$ as a reference value.
For this choice of $\tau_{\rm form}$, the best fit values of $p$ is shown in Table~\ref{tab5}.
The preferred values of $p$ become slightly smaller than $1$, as was expected. 

{\tabcolsep = 2mm
\begin{table}[h]
\begin{center} 
\caption{The values of $\mathcal{A}_{\rm form}$ (with $\tau_{\rm form}=50$) and the best fit values of the exponent $p$
in the model function [Eq.~\eqref{area_t_model}] for various values of $N_{\rm DW}$.
These results are obtained from the numerical simulations with the grid size $N=16384$.}
\vspace{3mm}
\begin{tabular}{ c c c }
\hline\hline
$N_{\rm DW}$ & $\mathcal{A}_{\rm form}$ & $p$ \\
\hline 
2 & $0.540 \pm 0.052$ & 0.929 \\
3 & $0.828 \pm 0.032$ & 0.926 \\
4 & $1.10 \pm 0.04$ & 0.917 \\
5 & $1.44 \pm 0.05$ & 0.918 \\
6 & $1.73 \pm 0.06$ & 0.932 \\
\hline\hline
\label{tab5}
\end{tabular}
\end{center}
\end{table}
}

Next, let us estimate the decay time of domain walls.
Our purpose here is to obtain the value of the coefficient $C_d$ appearing in Eq.~\eqref{t_dec_exact_scaling} or Eq.~\eqref{t_dec_dev_scaling}.
By using the results of numerical simulations with $\Xi\ne 0$, we determine the value of $t_{\rm dec}$ as the time at which the value of $A/V$
becomes $10\%$ or $1\%$ of that with $\Xi=0$ (hereafter
we call these criteria ``$10\%$ criterion" and ``$1\%$ criterion," respectively).\footnote{In the previous study~\cite{Hiramatsu:2010yn}, we measured $t_{\rm dec}$ based on the $1\%$ criterion only.
However, it might be more appropriate to use a higher percentage as the criterion, since most axions are produced at the time when $A/V$ starts to fall off.
In this paper, we consider two cases ($10\%$ criterion and $1\%$ criterion) in order to see how the different choice of the criterion affects the final result.}
Furthermore, we adopt two different assumptions: One is to assume that the area parameter $\mathcal{A}$ takes a constant value at late times (we call this case ``exact scaling"),
and another is to assume that $\mathcal{A}$ increases with time according to Eq.~\eqref{area_t_model} (we call this case ``deviation from scaling").
For the assumption of exact scaling, we use the values of $\mathcal{A}(\tau_f)$ shown in Table~\ref{tab4} to estimate $C_d$.
On the other hand, for the assumption of deviation from scaling, we use the values of $\mathcal{A}_{\rm form}$ and $p$ shown in Table~\ref{tab5} with $\tau_{\rm form}=50$.

\begin{figure}[htbp]
\centering
$\begin{array}{cc}
\subfigure[]{
\includegraphics[width=0.47\textwidth]{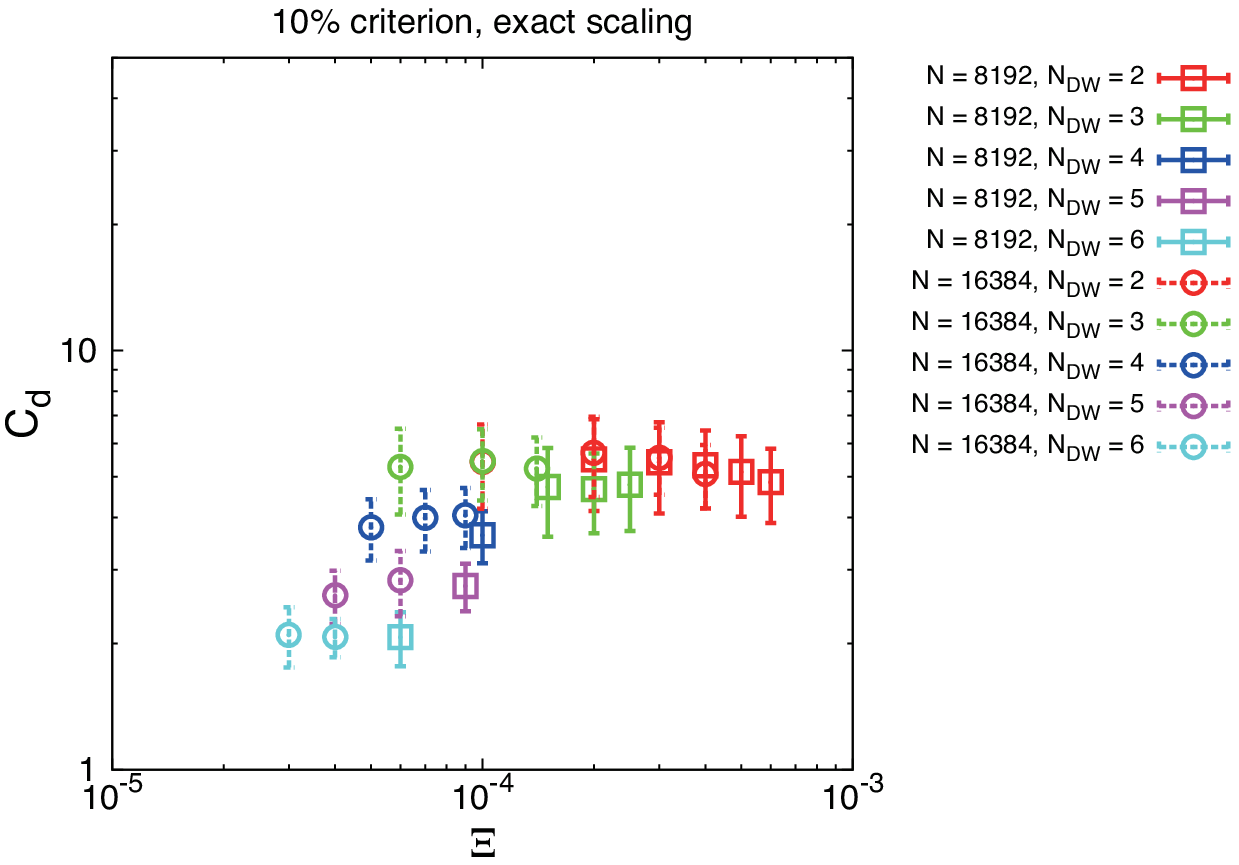}}
\hspace{15pt}
\subfigure[]{
\includegraphics[width=0.47\textwidth]{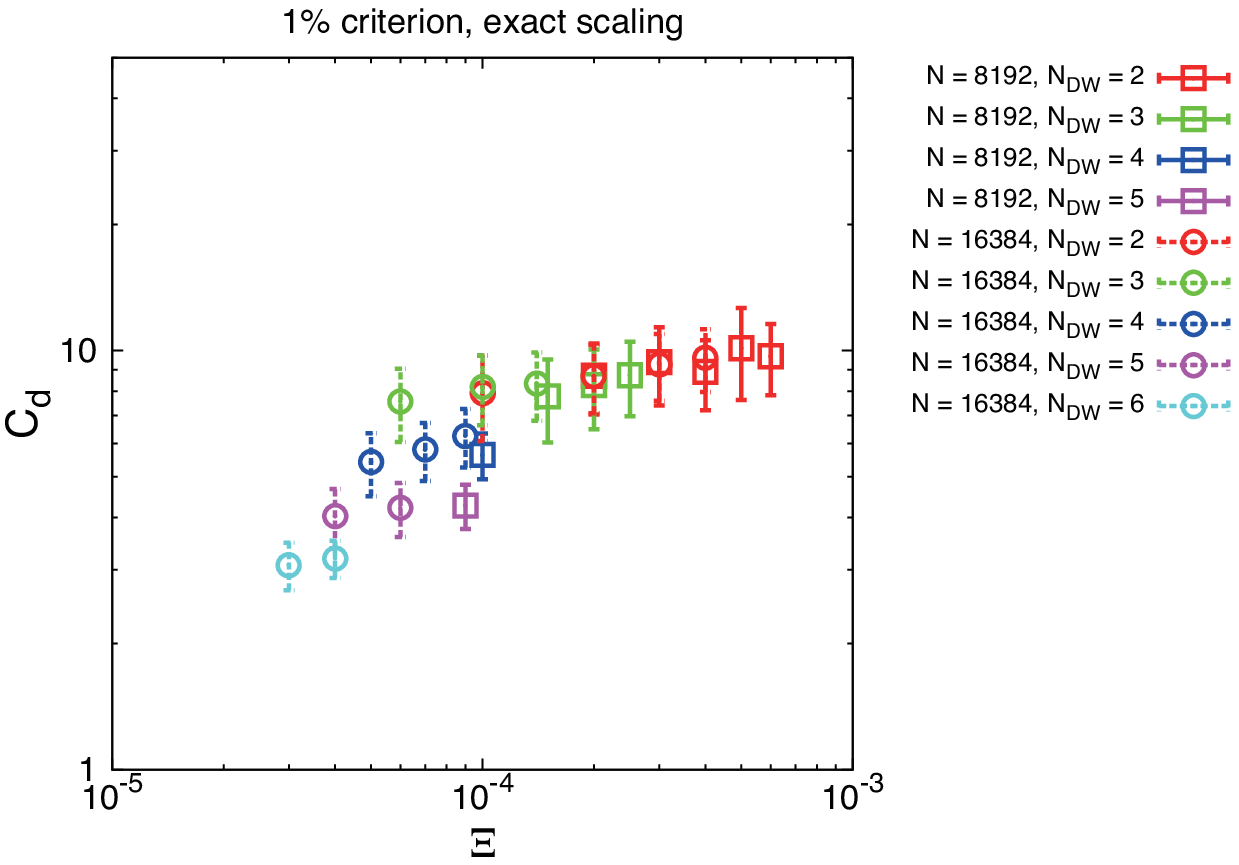}}
\\
\subfigure[]{
\includegraphics[width=0.47\textwidth]{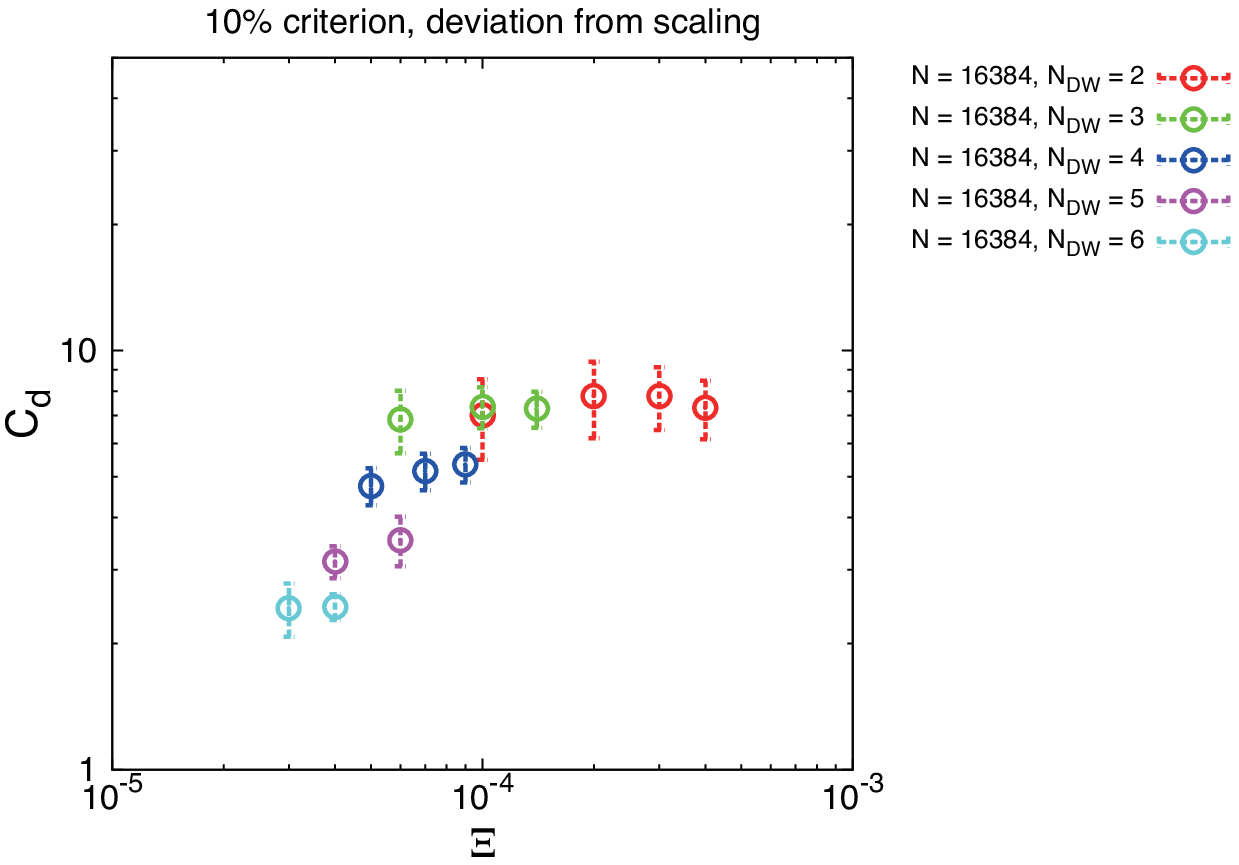}}
\hspace{15pt}
\subfigure[]{
\includegraphics[width=0.47\textwidth]{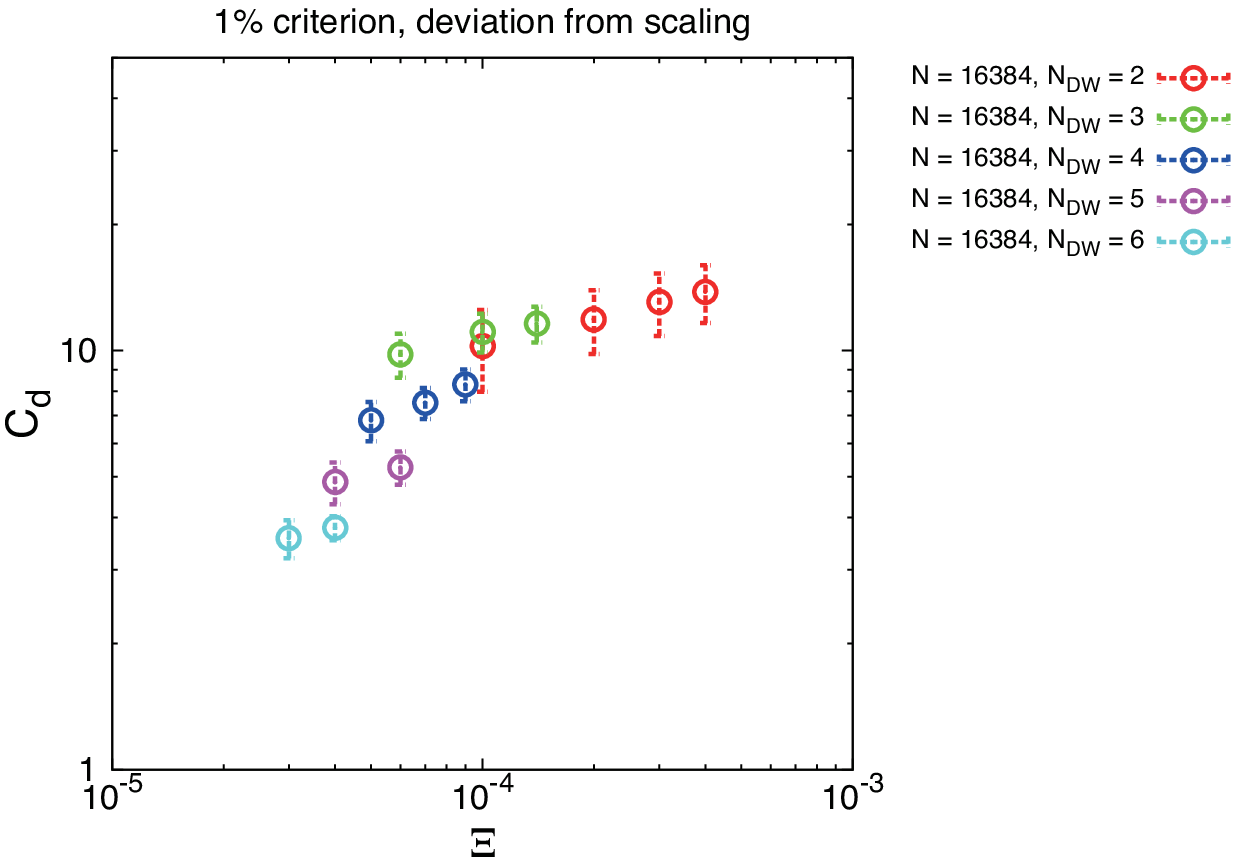}}
\end{array}$
\caption{Estimated values of $C_d$ for (a) the assumption of exact scaling with $10\%$ criterion,
(b) the assumption of exact scaling with $1\%$ criterion, (c) the assumption of deviation from scaling with $10\%$ criterion,
and (d) the assumption of deviation from scaling with $1\%$ criterion.
In the panels (c) and (d), we only show the results with $N=16384$, since the value of $p$ in Table~\ref{tab5} is obtained from the results of
the simulations with $N=16384$.
In the panel (b), the result of the simulations with $N=8192$ and $N_{\rm DW}=6$ is not shown, since in this case
the value of $A/V$ does not reach $1\%$ of that with $\Xi=0$ before the final time of the simulations.}
\label{fig4}
\end{figure}

Figure~\ref{fig4} shows the results of $C_d$ for every assumptions and criteria.
We see that there is no significant difference between the assumption of exact scaling and that of deviation from scaling.
In the results with $1\%$ criterion, the values of $C_d$ become about $\sim 2$ times larger than those with $10\%$ criterion,
since it takes longer times to satisfy the criterion.
The results hardly depend on the value of $\Xi$, but show a slight dependence on the value of $N_{\rm DW}$.
This can be interpreted as follows: If $N_{\rm DW}$ is small, the number of walls per a horizon volume becomes small, and
large planar walls are likely to be formed. Such large configurations take a longer time to collapse, leading to a larger value of $C_d$.

Since the results shown in Fig.~\ref{fig4} does not significantly depend on the choice of $\Xi$,
we estimate the value of $C_d$ by averaging over the plotted values for each choice of $N_{\rm DW}$.
The estimated values are shown in Table.~\ref{tab6}.


{\tabcolsep = 2mm
\begin{table}[h]
\begin{center} 
\caption{The values of $C_d$ for various assumptions.}
\vspace{3mm}
\begin{tabular}{l l l}
\hline\hline
$N_{\rm DW}$ & $C_d$ (exact scaling) & $C_d$ (deviation from scaling) \\
\hline
2 & $5.33\pm 0.38$ (for $10\%$ criterion) & $7.48\pm 0.71$ (for $10\%$ criterion) \\
 & $9.14\pm 0.62$ (for $1\%$ criterion) & $12.2\pm 1.1$ (for $1\%$ criterion) \\
3 & $5.02\pm 0.44$ (for $10\%$ criterion) & $7.16\pm 0.53$ (for $10\%$ criterion) \\
 & $8.15\pm 0.67$ (for $1\%$ criterion) & $10.8\pm 0.7$ (for $1\%$ criterion) \\ 
4 & $3.86\pm 0.31$ (for $10\%$ criterion) & $5.09\pm 0.29$ (for $10\%$ criterion) \\
 & $5.78\pm 0.45$ (for $1\%$ criterion) & $7.54\pm 0.40$ (for $1\%$ criterion) \\ 
5 & $2.72\pm 0.24$ (for $10\%$ criterion) & $3.34\pm 0.28$ (for $10\%$ criterion) \\
 & $4.17\pm 0.35$ (for $1\%$ criterion) & $5.06\pm 0.37$ (for $1\%$ criterion) \\ 
6 & $2.08\pm 0.17$ (for $10\%$ criterion) & $2.44\pm 0.20$ (for $10\%$ criterion) \\
 & $3.14\pm 0.26$ (for $1\%$ criterion) & $3.67\pm 0.22$ (for $1\%$ criterion) \\ 
\hline\hline
\label{tab6}
\end{tabular}
\end{center}
\end{table}
}

\subsection{\label{sec3-3} Estimation of the mean energy of radiated axions}
\begin{figure}[htbp]
\centering
\subfigure[]{
\includegraphics[width=0.44\textwidth]{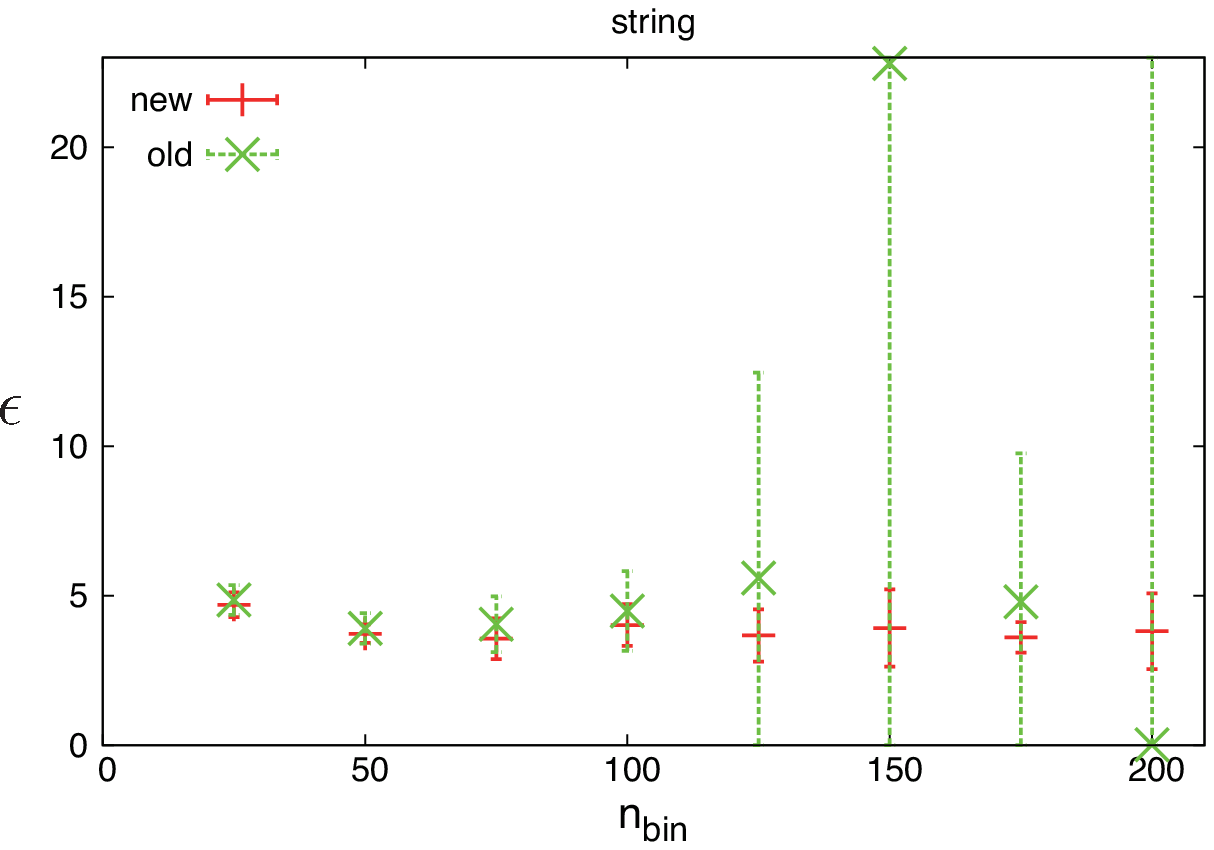}}
$\begin{array}{cc}
\subfigure[]{
\includegraphics[width=0.44\textwidth]{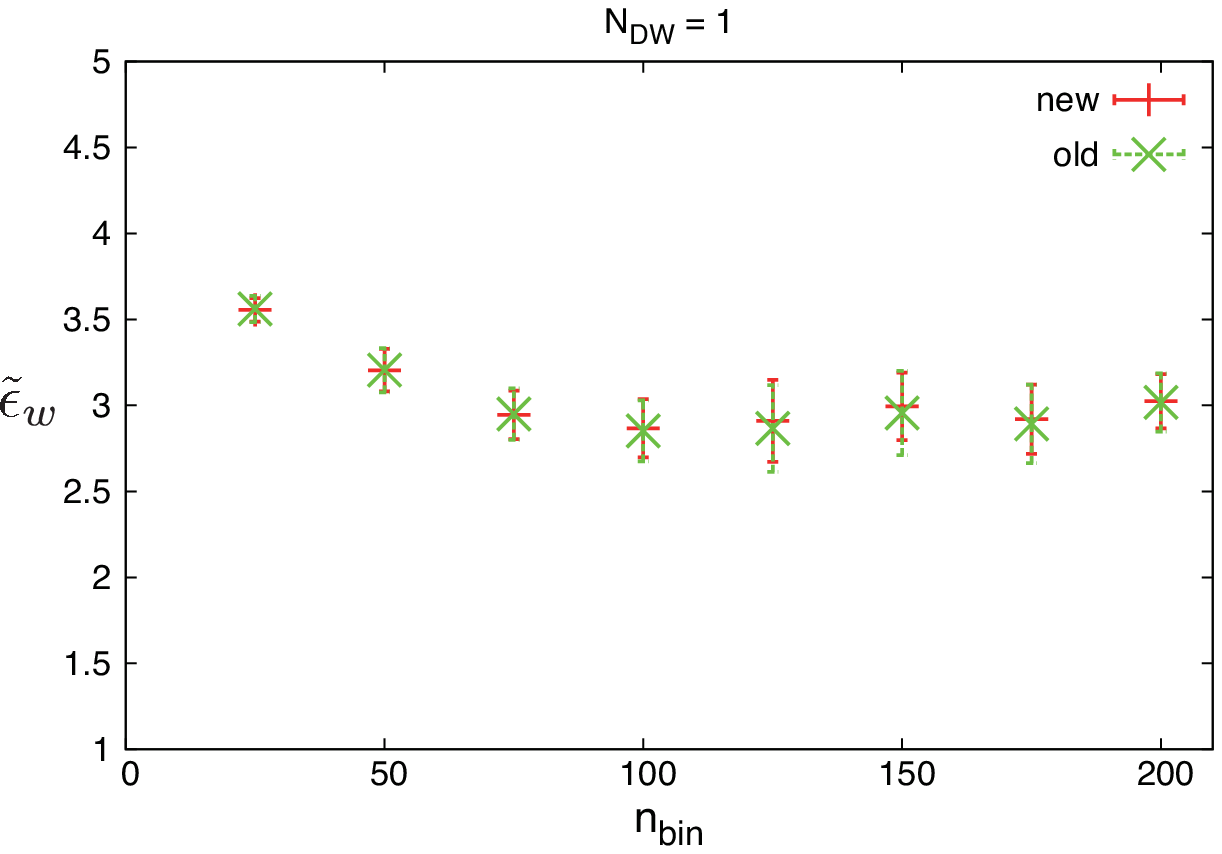}}
\hspace{20pt}
\subfigure[]{
\includegraphics[width=0.44\textwidth]{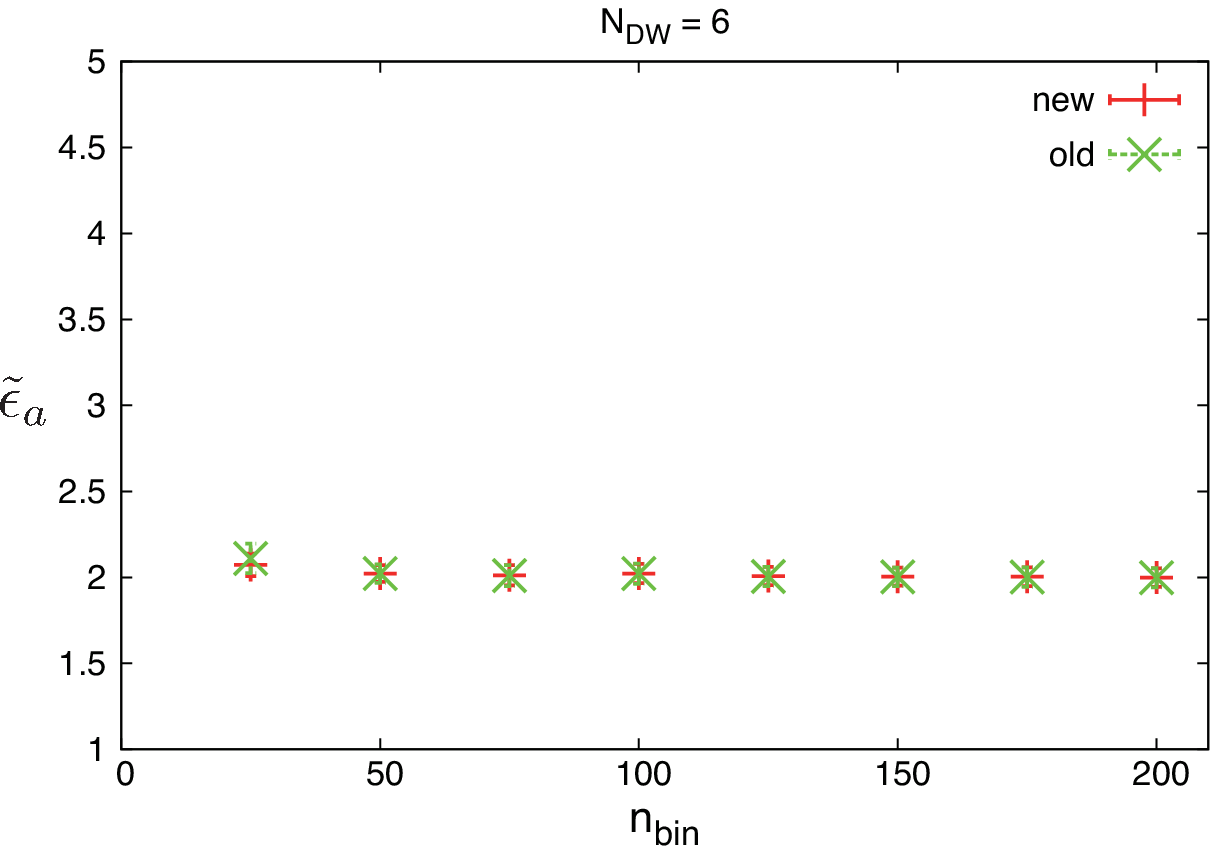}}
\end{array}$
\caption{Estimated values of the parameters related to the mean energy of radiated axions for various values of $n_{\rm bin}$.
Each panel shows the results for (a) $\epsilon$ (global strings), (b) $\tilde{\epsilon}_w$ (short-lived string-wall systems with $N_{\rm DW}=1$),
and (c) $\tilde{\epsilon}_a$ (long-lived string-wall systems with $N_{\rm DW}=6$).
The points indicated as ``old" correspond to the results obtained by the method used in the previous studies~\cite{Hiramatsu:2010yu,Hiramatsu:2012gg,Hiramatsu:2012sc},
while those indicated as ``new" correspond to the results obtained by the new estimators shown in Eq.~\eqref{new_average}.
The results shown in the panel (b) are obtained from the simulations with $\kappa=0.3$.
}
\label{fig5}
\end{figure}

Figure~\ref{fig5} shows the values of $\epsilon$, $\tilde{\epsilon}_w$, and $\tilde{\epsilon}_a$
obtained from the results of 3D simulations 
[cases (d), (e-2) and (f-5) in Table~\ref{tab3}]
for various choices of $n_{\rm bin}$.
Here we also compare the results between the old averaging method used in Refs.~\cite{Hiramatsu:2010yu,Hiramatsu:2012gg,Hiramatsu:2012sc}
and the new method introduced in Appendix~\ref{secB}.
We see that the error bars become large for the case of strings if we use the old averaging method.
These results can be understood as follows. 
When we compute the energy spectrum of radiated axions, we mask the grid points corresponding to the core of topological defects in the simulation box,
which leads to a systematic error in the final form for the energy spectrum [see Appendix~C of Ref.~\cite{Hiramatsu:2012sc} for details].
Since the configuration of the defects differs from one realization to another,
the magnitude of this systematic uncertainty varies accordingly.
Therefore, the error in the final result is determined by one particular realization, which gives the largest systematic uncertainty.
On the other hand, the new estimator optimizes the different systematic uncertainties obtained from 10 realizations of the simulations,
reducing the error bars for $\epsilon$ as shown in Fig.~\ref{fig5} (a).

We note that there is little difference between the old and new methods for the case of short-lived string-wall systems [Fig.~\ref{fig5} (b)]
and that of long-lived string-wall systems [Fig.~\ref{fig5} (c)].
These results can be understood as follows.
For the case of short-lived string-wall systems, the defects have mostly disappeared by the time $\tau_B$, at which we compute the spectrum of radiated axions,
and the masking process does not lead to significant systematic uncertainties.
On the other hand, for the case of long-lived string-wall systems, systematic uncertainties on the energy spectrum can be large
since the defects still exist at the time $\tau_B$, but such uncertainties only appear on the scales comparable to the size of the simulation box.
These uncertainties in small momenta ($k/R(\tau_B)\ll m_a$) do not significantly affect the computation of the mean energy,
since at these scales the mean energy is simply determined by the axion mass, $\omega_a(k,\tau_B)\simeq m_a$.
Therefore, we do not see the effect of the systematic uncertainties in the case of massive axions [Fig.~\ref{fig5} (c)],
while it appears in the case of massless axions [Fig.~\ref{fig5} (a)].


From Fig.~\ref{fig5} (b), we see that the value of $\tilde{\epsilon}_w$ varies with $n_{\rm bin}$ for $n_{\rm bin}<100$,
and that it tends to converge for $n_{\rm bin}\gtrsim 100$.
A similar trend is observed in Figs.~\ref{fig5} (a) and \ref{fig5} (c) for smaller values of $n_{\rm bin}$.
This fact implies that the choice of $n_{\rm bin}=25$ used in the previous study~\cite{Hiramatsu:2012gg}
is not good enough to resolve the peak of the power spectrum.
We must use the number of bins as large as $n_{\rm bin}\gtrsim 100$ to estimate the mean momentum of radiated axions correctly.

\begin{figure}[htbp]
\begin{center}
\includegraphics[scale=1.0]{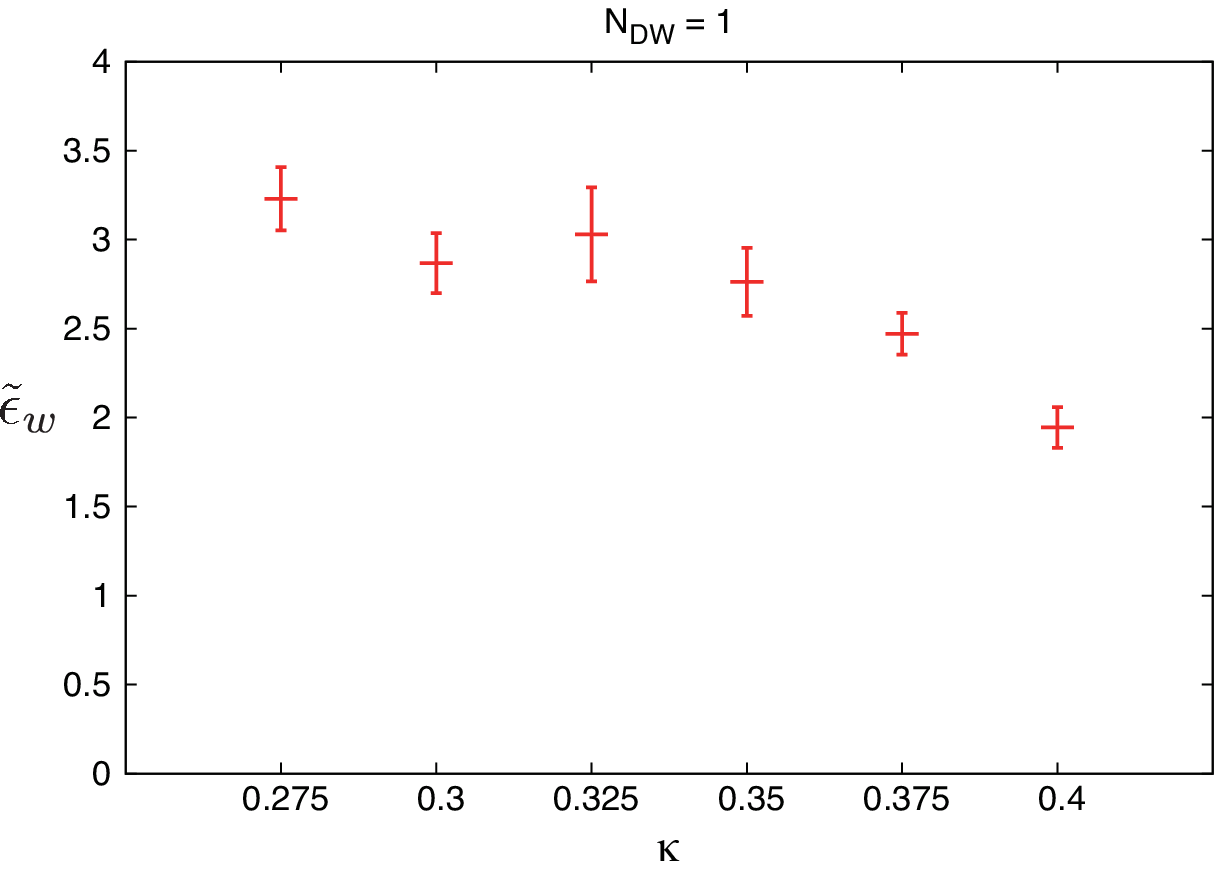}
\end{center}
\caption{The values of $\tilde{\epsilon}_w$ for various choices of $\kappa$ obtained
from the calculation with $n_{\rm bin} = 100$, where the average and error are estimated by using Eq.~\eqref{new_average}.
}
\label{fig6}
\end{figure}

Figure~\ref{fig6} shows the estimated values of $\tilde{\epsilon}_w$ for various choices of $\kappa$ [cases (e-1)$\textendash$(e-6) in Table~\ref{tab3}].
We see that the value of $\tilde{\epsilon}_w$ tends to become smaller as $\kappa$ takes a larger value,
which is an undesirable situation where two scales corresponding to the width of domain walls and that of strings become comparable.
On the other hand, it seems to converge for smaller values of $\kappa$.
This trend supports the extrapolation of the estimated value of $\tilde{\epsilon}_w$ to the limit $\kappa \ll 1$, which
should be further confirmed in the future numerical simulations with improved dynamical ranges.
For now we use the value estimated at $\kappa=0.275$ with $n_{\rm bin}=100$,
\begin{equation}
\tilde{\epsilon}_w =  3.23 \pm 0.18,
\label{epsilon_wall}
\end{equation}
to estimate the energy density of axions produced from short-lived string-wall systems.


The situation is more clear in the cases of global strings and long-lived string-wall systems.
From the result of the calculation with $n_{\rm bin} = 100$, where the averaged value and its error are estimated by using Eq.~\eqref{new_average}, we obtain
\begin{equation}
\epsilon = 4.02 \pm 0.70
\label{epsilon_string}
\end{equation}
for the axions radiated from global strings.
We also show the values of $\tilde{\epsilon}_a$ for those radiated from long-lived string-wall systems in Table~\ref{tab7}.

{\tabcolsep = 2mm
\begin{table}[h]
\begin{center} 
\caption{The values of $\tilde{\epsilon}_a$ for various choices of $N_{\rm DW}$ obtained
from the calculation with $n_{\rm bin} = 100$, where the average and the error are estimated by using Eq.~\eqref{new_average}.
}
\vspace{3mm}
\begin{tabular}{ c c }
\hline\hline
$N_{\rm DW}$ & $\tilde{\epsilon}_a$ \\
\hline 
2 & $ 1.96 \pm 0.13 $ \\
3 & $ 1.85 \pm 0.06 $ \\
4 & $ 1.87 \pm 0.07 $ \\
5 & $ 1.93 \pm 0.06 $ \\
6 & $ 2.02 \pm 0.06 $ \\
\hline\hline
\label{tab7}
\end{tabular}
\end{center}
\end{table}
}

\section{\label{sec4} Observational constraints}
\subsection{\label{sec4-1} Models with $N_{\rm DW}=1$}
We now discuss the constraints on the model parameters based on the results obtained in the previous sections.
First of all, let us consider the case with short-lived domain walls ($N_{\rm DW}=1$).
Here we must require that the total axion abundance at the present time does not exceed the observed cold dark matter abundance $\Omega_{\rm CDM}$:
\begin{align}
\Omega_{a,\mathrm{tot}}h^2 &= \Omega_{a,\mathrm{mis}}h^2 + \Omega_{a,\mathrm{string}}h^2 + \Omega_{a,\mathrm{dec}}h^2 \le \Omega_{\rm CDM}h^2, \label{constraint_abundance}
\end{align}
where $\Omega_{\rm CDM}h^2\simeq 0.12$~\cite{Ade:2013zuv}, and
$\Omega_{a,\mathrm{mis}}h^2$, $\Omega_{a,\mathrm{string}}h^2$, and $\Omega_{a,\mathrm{dec}}h^2$ are given by Eqs.~\eqref{Omega_a_mis_2},~\eqref{Omega_a_string_1}, and \eqref{Omega_a_dec_1}, respectively.

The contribution from strings $\Omega_{a,\mathrm{string}}h^2$ depends on two numerical coefficients $\xi$ and $\epsilon$.
As was mentioned in Sec.~\ref{sec2-2}, $\xi$ contains a large systematic uncertainty, and here we adopt a conservative estimation $\xi\simeq1.0\pm0.5$.
The value of $\epsilon$ is estimated in Eq.~\eqref{epsilon_string}.
Substituting these values into Eq.~\eqref{Omega_a_string_1}, we obtain
\begin{equation}
\Omega_{a,\mathrm{string}}h^2 = (7.3\pm3.9)\times 10^{-3}\times N_{\rm DW}^2\left(\frac{\beta'}{58}\right)\left(\frac{g_{*,1}}{80}\right)^{-(2+n)/2(4+n)}\left(\frac{F_a}{10^{10}\mathrm{GeV}}\right)^{(6+n)/(4+n)}\left(\frac{\Lambda_{\rm QCD}}{400\mathrm{MeV}}\right).
\label{Omega_a_string_2}
\end{equation}

The contribution from the decay of string-wall systems $\Omega_{a,\mathrm{dec}}h^2$ depends on three numerical coefficients $\xi$, $\mathcal{A}$, and $\tilde{\epsilon}_w$.
For the area parameter $\mathcal{A}$, we use the value $\mathcal{A}\simeq 0.50 \pm 0.25$ obtained in Ref.~\cite{Hiramatsu:2012gg}.
The value of $\tilde{\epsilon}_w$ is given by Eq.~\eqref{epsilon_wall}. Using these values and $\xi\simeq 1.0\pm0.5$, we find 
\begin{align}
\Omega_{a,\mathrm{dec}}h^2 = (3.7\pm1.4)\times 10^{-3}\times\left(\frac{\beta_2}{62}\right)^{2/(4+n)}\left(\frac{g_{*,2}}{75}\right)^{-(2+n)/2(4+n)}\left(\frac{F_a}{10^{10}\mathrm{GeV}}\right)^{(6+n)/(4+n)}\left(\frac{\Lambda_{\rm QCD}}{400\mathrm{MeV}}\right).
\label{Omega_a_dec_2}
\end{align}

Here and hereafter we ignore the weak dependence on the parameters $g_{*,1}$, $g_{*,2}$, $\beta_1$, and $\beta_2$,
and fix their values as $g_{*,1}=80$, $g_{*,2}=75$, $\beta_1=61$, and $\beta_2=62$.
From the sum of Eqs.~\eqref{Omega_a_mis_2},~\eqref{Omega_a_string_2}, and~\eqref{Omega_a_dec_2},
we estimate the total relic abundance of cold dark matter axions for the models with $N_{\rm DW}=1$ as
\begin{align}
\Omega_{a,\mathrm{tot}}h^2 &= \Omega_{a,\mathrm{mis}}h^2 + \Omega_{a,\mathrm{string}}h^2 + \Omega_{a,\mathrm{dec}}h^2 \nonumber\\
&= (1.6\pm0.4)\times 10^{-2}\times\left(\frac{F_a}{10^{10}\mathrm{GeV}}\right)^{(6+n)/(4+n)}\left(\frac{\Lambda_{\rm QCD}}{400\mathrm{MeV}}\right),
\label{Omega_a_tot_NDW=1}
\end{align}
where we used $c_{\rm av}=2$ in Eq.~\eqref{Omega_a_mis_2} and $N_{\rm DW}=1$ in Eq.~\eqref{Omega_a_string_2}.
From the requirement given by Eq.~\eqref{constraint_abundance}, we find
\begin{equation}
F_a \lesssim (4.6 \textendash 7.2)\times 10^{10}\mathrm{GeV}, \label{constraint_Fa}
\end{equation}
for the QCD scale $\Lambda_{\rm QCD}=400\mathrm{MeV}$.
This constraint corresponds to the lower bound on the axion mass:
\begin{equation}
m_a \gtrsim (0.8 \textendash 1.3) \times 10^{-4}\mathrm{eV}. \label{constraint_ma}
\end{equation}
The bounds shown in Eqs.~\eqref{constraint_Fa} and~\eqref{constraint_ma} are slightly weaker than
those obtained in the previous paper~\cite{Hiramatsu:2012gg}, since we include the correction factor $m_a(T_1)/m_a(T_2)$ (see footnote~\ref{fnma1ma2})
and the approximation given by Eq.~\eqref{rho_string-wall}, which were not considered in Ref.~\cite{Hiramatsu:2012gg}.

\subsection{\label{sec4-2} Models with $N_{\rm DW}>1$}
Next, we consider the case with long-lived domain walls ($N_{\rm DW}>1$).
Again we apply the condition for the dark matter abundance given by Eq.~\eqref{constraint_abundance},
but in this case $\Omega_{a,\mathrm{dec}}h^2$ is replaced with the contribution from long-lived domain walls estimated in Appendix~\ref{secA}.
For the assumption of exact scaling, we have [Eq.~\eqref{Omega_a_dec_long_exact_scaling_A}],
\begin{equation}
\Omega_{a,\mathrm{dec}}h^2 = 0.756\times\frac{C_d^{1/2}}{\tilde{\epsilon}_a}\left[\frac{\mathcal{A}^3}{N_{\rm DW}^4(1-\cos(2\pi/N_{\rm DW}))}\right]^{1/2}
\left(\frac{\Xi}{10^{-52}}\right)^{-1/2}\left(\frac{F_a}{10^{10}\mathrm{GeV}}\right)^{-1/2}\left(\frac{\Lambda_{\rm QCD}}{400\mathrm{MeV}}\right)^3,
\label{Omega_a_dec_long_exact_scaling}
\end{equation}
and for the assumption of deviation from scaling, we have [Eq.~\eqref{Omega_a_dec_long_dev_scaling_A}],
\begin{align}
\Omega_{a,\mathrm{dec}}h^2
&= 1.23\times 10^{-6}\times[7.22\times 10^3]^{3/2p}\times\frac{1}{\tilde{\epsilon}_a}\frac{2p-1}{3-2p}C_d^{3/2-p}
\mathcal{A}_{\rm form}^{3/2p}\left[N_{\rm DW}^4\left(1-\cos\left(\frac{2\pi}{N_{\rm DW}}\right)\right)\right]^{1-3/2p} \nonumber\\
&\quad\times \left(\frac{g_{*,1}}{80}\right)^{3(1/p-1)n/4(4+n)}\left(\frac{\Xi}{10^{-52}}\right)^{1-3/2p}\left(\frac{F_a}{10^{10}\mathrm{GeV}}\right)^{4+3(4p-16-3n)/2p(4+n)}
\left(\frac{\Lambda_{\rm QCD}}{400\mathrm{MeV}}\right)^{-3+6/p}. \label{Omega_a_dec_long_dev_scaling}
\end{align}

The formulas for the abundance of axions produced from long-lived domain walls [Eqs.~\eqref{Omega_a_dec_long_exact_scaling} and~\eqref{Omega_a_dec_long_dev_scaling}]
contain four parameters, $\tilde{\epsilon}_a$, $\mathcal{A}$ (or $\mathcal{A}_{\rm form}$), $p$, and $C_d$, whose values are determined by the results of numerical simulations.
Here, we use the values shown in Table~\ref{tab7} for $\tilde{\epsilon}_a$, and those shown in Table~\ref{tab6} for $C_d$. 
We take $\mathcal{A}$ as the value at the final time of the simulations shown in Table~\ref{tab4} for Eq.~\eqref{Omega_a_dec_long_exact_scaling},
while we take $\mathcal{A}_{\rm form}$ and $p$ as the fit results shown in Table~\ref{tab5} for Eq.~\eqref{Omega_a_dec_long_dev_scaling}.
We also note that Eqs.~\eqref{Omega_a_dec_long_exact_scaling} and~\eqref{Omega_a_dec_long_dev_scaling} depend not only on $F_a$ but also on $\Xi$.
Therefore, we have the constraint on the two-dimensional parameter space $(F_a,\Xi)$.

In addition to the constraint of the axion dark matter abundance, there is another constraint coming from
a CP-violating effect induced by the bias term [Eq.~\eqref{V_phi_bias}] in the potential of the PQ field.
If the bias term exists in the theory, it shifts the minimum of the potential of the axion field from the CP conserving point $\bar{\theta}=0$.
This shift leads to a large amount of CP violation, which is tightly constrained from the observation of the neutron electric dipole moment~\cite{Baker:2006ts}.
Following Ref.~\cite{Hiramatsu:2012sc}, we describe this constraint as
\begin{equation}
\bar{\theta} = \frac{2\Xi N_{\rm DW}^3F_a^2\sin\delta}{m_a^2+2\Xi N_{\rm DW}^2 F_a^2\cos\delta} < 7\times 10^{-12}. \label{constraint_NEDM}
\end{equation}

Various astrophysical phenomena put stringent constraints on the couplings between axions and other species of particles 
such as photons, electrons, and nucleons~\cite{Raffelt:1990yz,Raffelt:2006cw}.
In particular, the observed burst duration of the supernova (SN) 1987A constrains the cooling rate due to the axions produced in the core of the SN,
which leads to the strongest bound~\cite{Raffelt:2006cw}:
\begin{equation}
F_a > 4\times 10^8\mathrm{GeV}. \label{constraint_astro}
\end{equation}
In this paper, we use the above value as the lower bound on the axion decay constant.

\begin{figure}[htbp]
\centering
$\begin{array}{c}
\subfigure[]{
\includegraphics[scale=0.90]{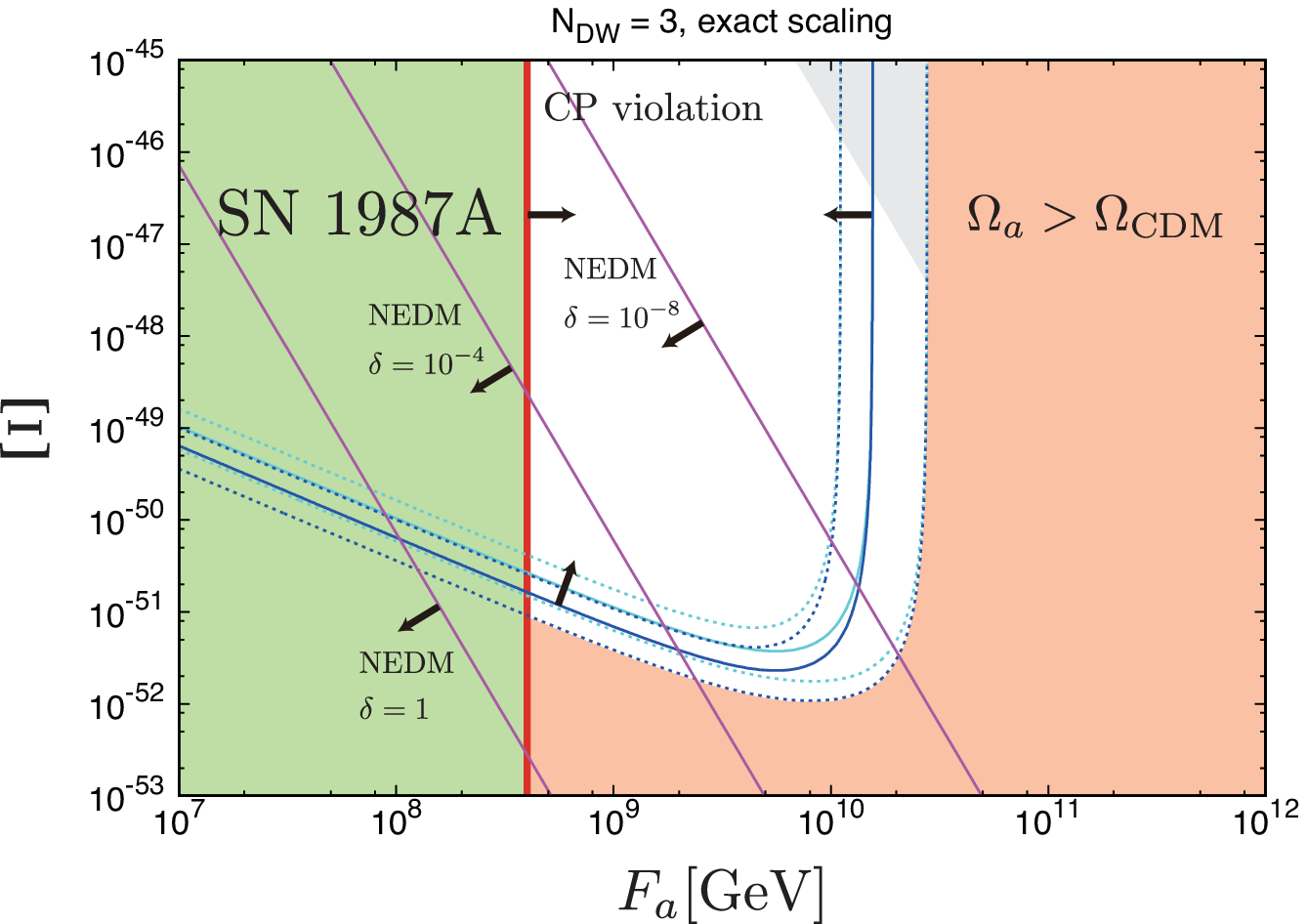}}
\vspace{5mm}
\\
\subfigure[]{
\includegraphics[scale=0.90]{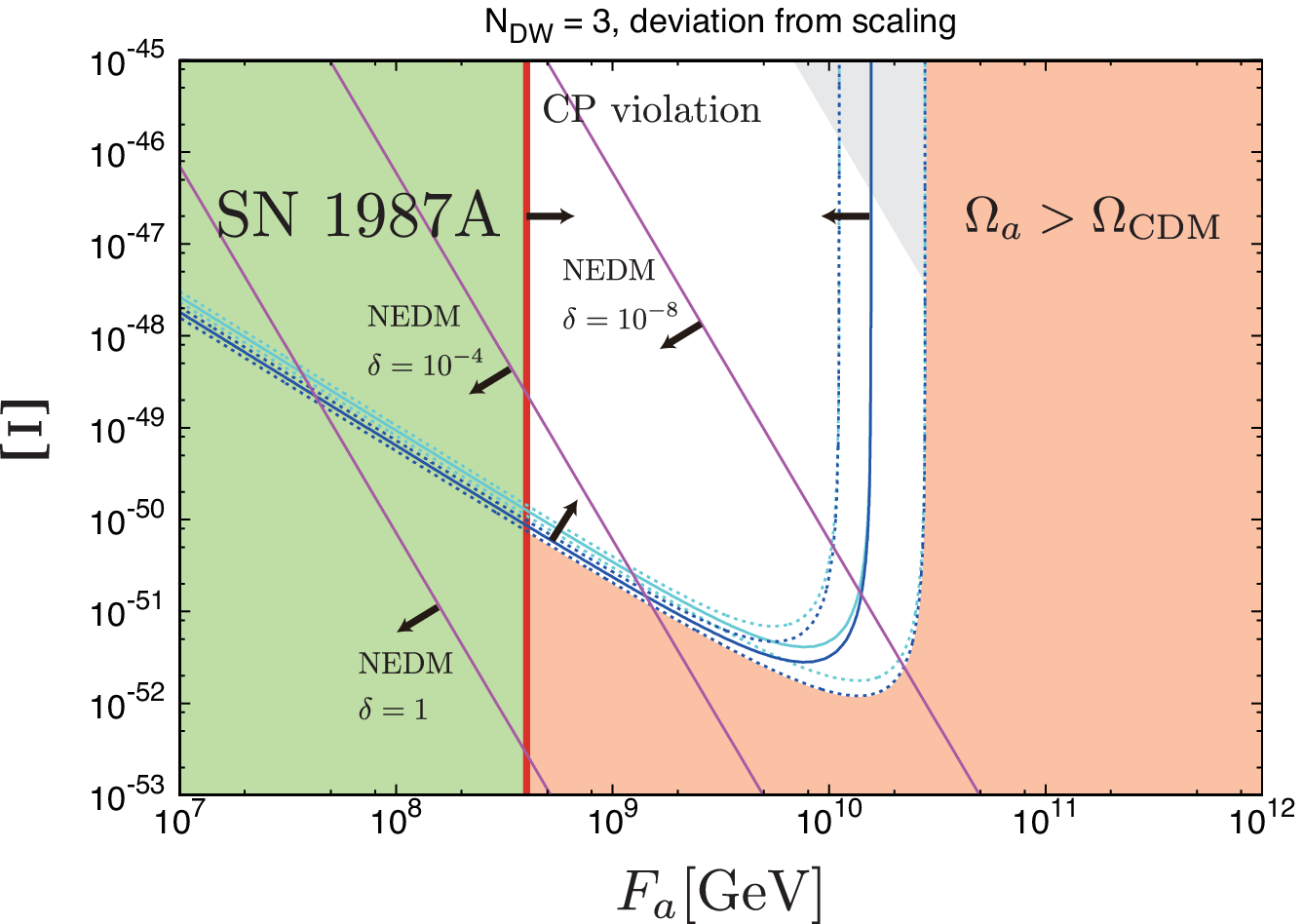}}
\end{array}$
\caption{
Observational constraints on the axion decay constant $F_a$ and the bias parameter $\Xi$ in the model with $N_{\rm DW}=3$ based on
(a) the assumption of exact scaling and (b) that of deviation from scaling.
The red solid line corresponds to the bound obtained from the burst duration of SN 1987A [Eq.~\eqref{constraint_astro}], and
the green area to the left side of this line is excluded.
The blue (cyan) solid line corresponds to the constraint of the overclosure of dark matter axions [Eq.~\eqref{constraint_abundance}]
with the coefficient $C_d$ estimated based on $10\%$ ($1\%$) criterion.
The dotted lines represent uncertainties of $\Omega_{a,\mathrm{tot}}h^2$ induced by the numerical parameters
$\epsilon$, $\xi$, $\tilde{\epsilon}_a$, $\mathcal{A}$ (or $\mathcal{A}_{\rm form}$), and $C_d$.
Except for these uncertainties, the red region below the blue (or cyan) line is excluded.
The purple solid lines correspond to the NEDM bounds [Eq.~\eqref{constraint_NEDM}] for $\delta=1$, $10^{-4}$, and $10^{-8}$.
The region above these lines is also excluded.
The shaded region corresponds to the parameters satisfying Eq.~\eqref{condition_Xi_dominate}, and in this region
the axion mass is dominated by the bias term.
The exclusion lines shown in these figures are obtained for $g_{*,1}=80$ and $\Lambda_{\rm QCD}=400\mathrm{MeV}$.
Furthermore, we use $\beta'=58$, $\xi=1.0\pm0.5$, and $\epsilon=4.02\pm 0.70$ to compute $\Omega_{a,\mathrm{string}}h^2$.
For parameters required to estimate $\Omega_{a,\mathrm{dec}}h^2$,
we take $\tilde{\epsilon}_a=1.85\pm 0.06$ (the result for $N_{\rm DW}=3$ in Table~\ref{tab7}), 
$\mathcal{A}=1.10\pm 0.18$ (the result for $N_{\rm DW}=3$ and $N=16384$ in Table~\ref{tab4}),
$\mathcal{A}_{\rm form}=0.828\pm 0.032$, and $p=0.926$ (the result for $N_{\rm DW}=3$ in Table~\ref{tab5}).
The value for $C_d$ is taken from Table~\ref{tab6}, such that $C_d=5.02\pm 0.44$ ($8.15\pm0.67$) for $10\%$ ($1\%$) criterion with the assumption of exact scaling [panel (a)]
and $C_d=7.16\pm 0.53$ ($10.8\pm0.7$) for $10\%$ ($1\%$) criterion with the assumption of deviation from scaling [panel (b)].}
\label{fig7}
\end{figure}

\begin{figure}[htbp]
\centering
$\begin{array}{c}
\subfigure[]{
\includegraphics[scale=0.90]{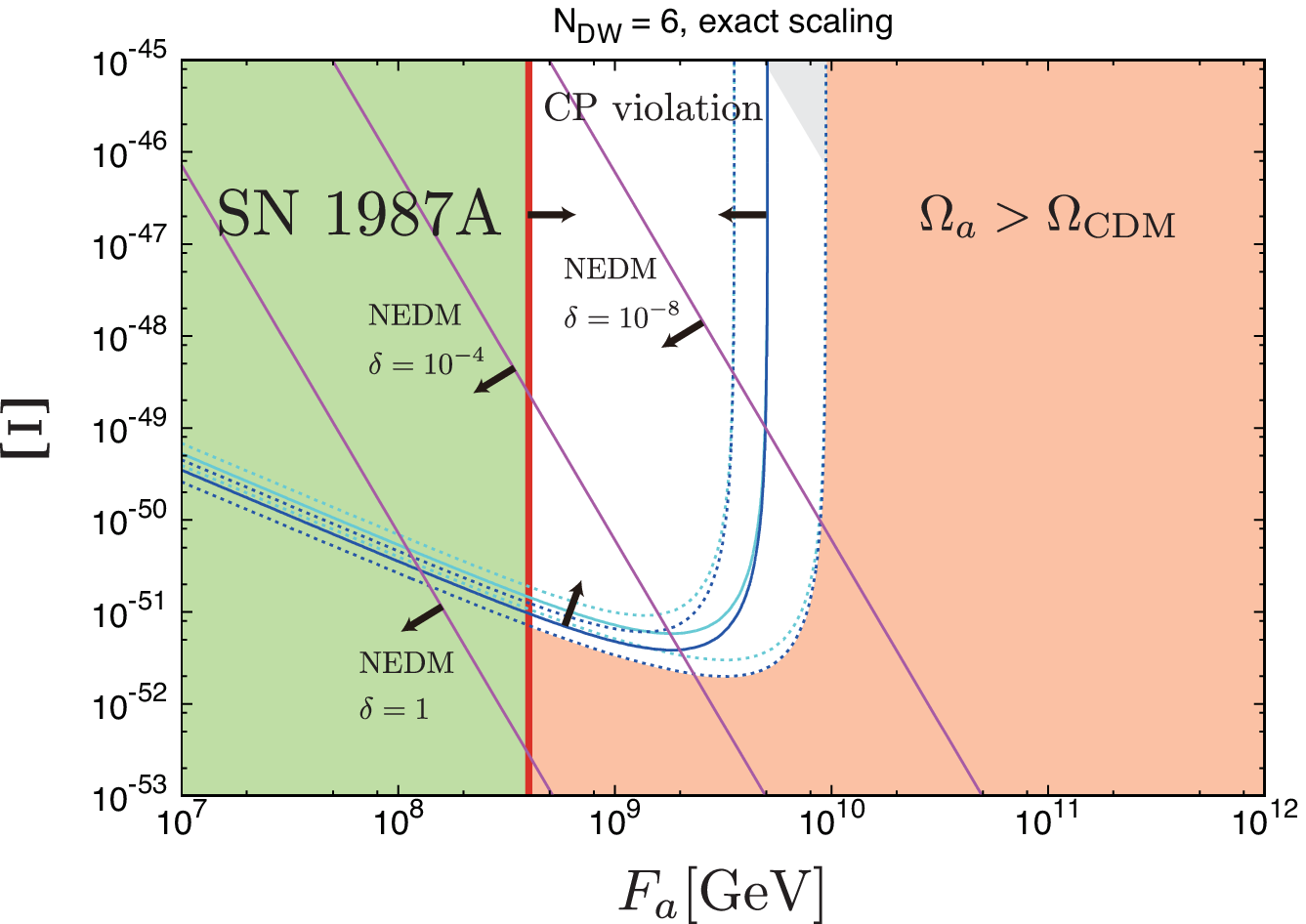}}
\vspace{5mm}
\\
\subfigure[]{
\includegraphics[scale=0.90]{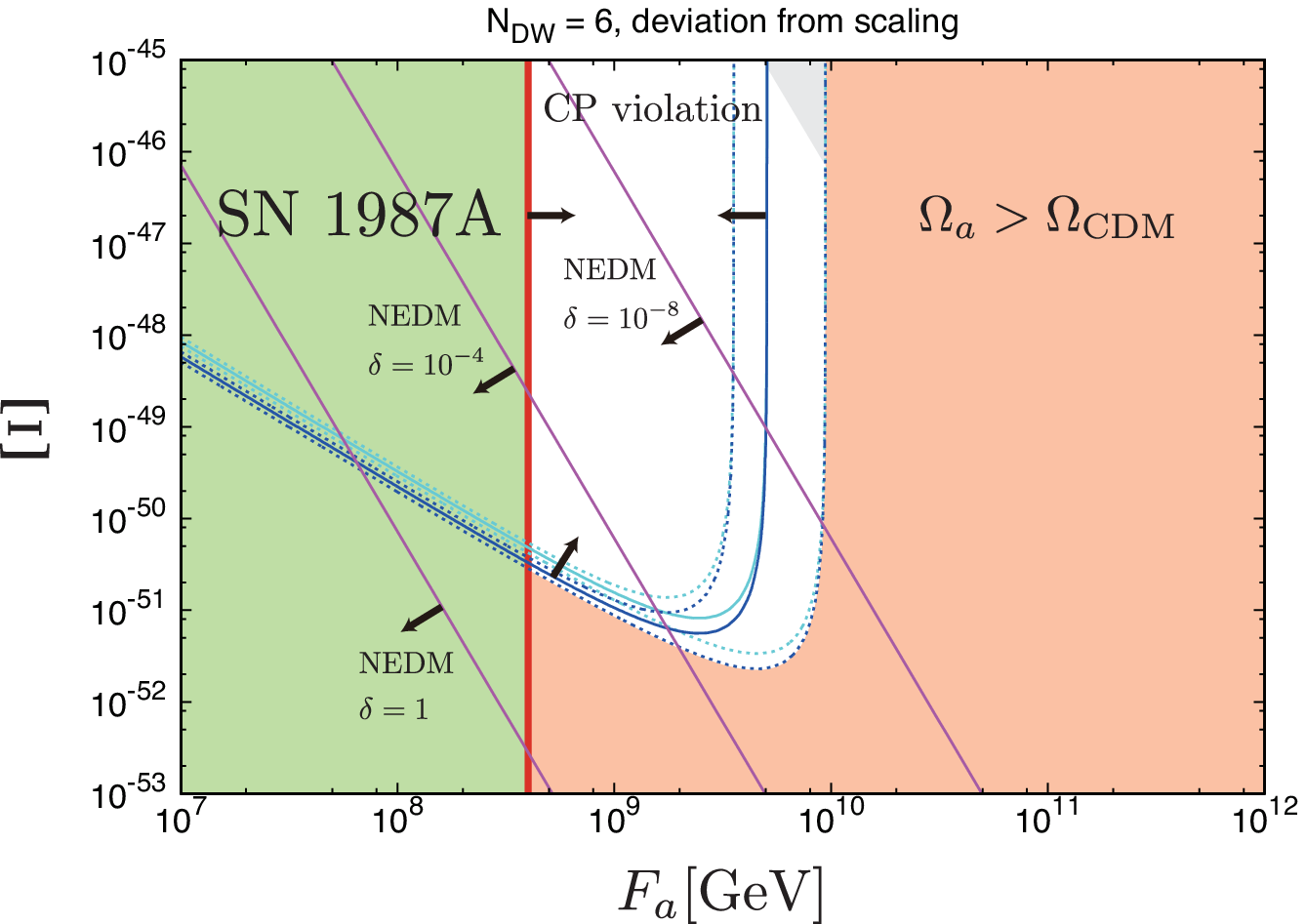}}
\end{array}$
\caption{
The same figure as Fig.~\ref{fig7} but observational constraints are plotted for the model with $N_{\rm DW}=6$ based on
(a) the assumption of exact scaling and (b) that of deviation from scaling.
For parameters required to estimate $\Omega_{a,\mathrm{dec}}h^2$,
we take $\tilde{\epsilon}_a=2.02\pm 0.06$ (the result for $N_{\rm DW}=6$ in Table~\ref{tab7}), $\mathcal{A}=2.23\pm 0.19$ (the result for $N_{\rm DW}=6$ and $N=16384$ in Table~\ref{tab4}),
$\mathcal{A}_{\rm form}=1.73\pm 0.06$, and $p=0.932$ (the result for $N_{\rm DW}=6$ in Table~\ref{tab5}).
The value for $C_d$ is taken from Table~\ref{tab6}, such that $C_d=2.08\pm 0.17$ ($3.14\pm0.26$) for $10\%$ ($1\%$) criterion with the assumption of exact scaling [panel (a)]
and $C_d=2.44\pm 0.20$ ($3.67\pm0.22$) for $10\%$ ($1\%$) criterion with the assumption of deviation from scaling [panel (b)].}
\label{fig8}
\end{figure}

Figure~\ref{fig7} summarizes the constraints for $\Xi$ and $F_a$ given by Eqs.~\eqref{constraint_abundance},~\eqref{constraint_NEDM}, and~\eqref{constraint_astro}
in the model with $N_{\rm DW}=3$.
We also plot the same constraints in the model with $N_{\rm DW}=6$ in Fig.~\ref{fig8}.
We see that the constraint coming from the dark matter abundance [Eq.~\eqref{constraint_abundance}] gives an upper bound $F_a\lesssim\mathcal{O}(10^9\textendash 10^{10})\mathrm{GeV}$
on the axion decay constant and a lower bound $\Xi\gtrsim\mathcal{O}(10^{-52}\textendash 10^{-50})$ on the bias parameter.
These bounds can vary by a factor of $\mathcal{O}(1)$ because of the uncertainties of the parameters $\epsilon$, $\xi$, $\tilde{\epsilon}_a$, $\mathcal{A}$ (or $\mathcal{A}_{\rm form}$),
and $C_d$ estimated from the results of the numerical simulations.
This constraint also depends on the criterion to determine the decay time of domain walls, and the constraint obtained from $10\%$ criterion becomes
weaker than that obtained from $1\%$ criterion.
Furthermore, the assumption of deviation from scaling leads to a more severe constraint on $\Xi$ [Figs.~\ref{fig7} (b) and~\ref{fig8} (b)]
in comparison to the case with the assumption of exact scaling [Figs.~\ref{fig7} (a) and~\ref{fig8} (a)]. 


The white region shown in Figs.~\ref{fig7} and~\ref{fig8} is not excluded by the constraints of SN1987A and the dark matter abundance,
but it can be excluded by the constraint of NEDM [Eq.~\eqref{constraint_NEDM}] since it gives an upper bound on $\Xi$.
Note that this NEDM line also depends on the magnitude of $\delta$, which is the phase appearing in the bias term [Eq.~\eqref{V_phi_bias}].
As shown in Figs.~\ref{fig7} and~\ref{fig8}, the whole parameter region is excluded if $\delta=1$.
On the other hand, a loophole appears if the magnitude of $\delta$ is sufficiently small.
Let us define the critical value $\delta_{\rm crit}$ below which the allowed region arises in the parameter space of $F_a$ and $\Xi$.
The value of $\delta_{\rm crit}$ is affected by the various uncertainties contained in $\Omega_{a,\mathrm{dec}}h^2$.
For instance, in the model with $N_{\rm DW}=6$, we have
\begin{align}
\delta_{\rm crit} &= \left\{
\begin{array}{l l}
(2.1\textendash 3.7)\times 10^{-2} & \mathrm{for\ 10\%\ criterion} \\
(1.4\textendash 2.4)\times 10^{-2} & \mathrm{for\ 1\%\ criterion} 
\end{array}
\right.
\quad (N_{\rm DW}=6, \mathrm{exact\ scaling}) \label{delta_crit_exact_scaling}
\end{align}
for the assumption of exact scaling, and 
\begin{align}
\delta_{\rm crit} &= \left\{
\begin{array}{l l}
(6.9\textendash 9.0)\times 10^{-3} & \mathrm{for\ 10\%\ criterion} \\
(4.8\textendash 6.1)\times 10^{-3} & \mathrm{for\ 1\%\ criterion} 
\end{array}
\right.
\quad (N_{\rm DW}=6, \mathrm{deviation\ from\ scaling}) \label{delta_crit_dev_scaling}
\end{align}
for the assumption of deviation from scaling.

We note that the axion mass is dominated by the bias term rather than the QCD instanton effect
if $\Xi$ is as large as~\cite{Hiramatsu:2012sc}
\begin{equation}
\Xi > 2 \times 10^{-45}\times N_{\rm DW}^{-2}\left(\frac{10^{10}\mathrm{GeV}}{F_a}\right)^4.
\label{condition_Xi_dominate}
\end{equation}
In this regime, the cosmological scenario is drastically altered~\cite{Barr:1992qq,Hiramatsu:2012sc}.
Here we do not consider such an unusual scenario.

It is notable that in the models with $N_{\rm DW}>1$ axions can be responsible for dark matter with the decay constant of order $F_a\approx\mathcal{O}(10^8\textendash 10^{10})\mathrm{GeV}$,
if we allow a mild tuning $\delta<\delta_{\rm crit}$.
This result has an interesting consequence that the mass of axion dark matter becomes $\mathcal{O}(10^{-4}\textendash 10^{-2})\mathrm{eV}$,
and that this mass range has a relevance to experimental studies.
The detection of the axions can be performed by the use of its coupling with electromagnetic fields~\cite{Sikivie:1983ip,Sikivie:1985yu},
and there are various ongoing and planned experiments.
The detection methods can be categorized into two classes, the axion haloscope and the axion helioscope.
The axion haloscope uses the resonant cavity system to detect relic axions distributed in the halos,
while the axion helioscope aims to detect axions radiated by the sun.

As for the haloscope-type experiment, the Axion Dark Matter Experiment (ADMX) is working at the University of Washington,
and it already took data in the mass range $1.9\mu\mathrm{eV}\lesssim m_a\lesssim 3.53\mu\mathrm{eV}$~\cite{Asztalos:2003px,Asztalos:2009yp}.
Currently, the improvement is underway in two directions: One is to reduce the system temperature and cover the mass range up to
$3.7\textendash 8.7\mu\mathrm{eV}$ (ADMX Phase II)~\cite{Asztalos:2011ei}.
The other is to perform experiments by the use of higher harmonic ports (ADMX-HF),
which will cover the range of $4\textendash 40\mathrm{GHz}$ ($16\textendash 160\mu\mathrm{eV}$)~\cite{vanBibber:2013ssa}.

In Europe, the CERN Axion Solar Telescope (CAST) works as the helioscope-type detector.
During the data taking operated in 2003$\textendash$2011, it put the constraint on the QCD axion models in the mass range $0.1\mathrm{eV}\lesssim m_a\lesssim 1.17\mathrm{eV}$~\cite{Arik:2013nya}.
In addition to the search for solar axions at CAST,
the next generation helioscope
called the International Axion Observatory (IAXO) is proposed~\cite{Armengaud:2014gea}.
This experiment aims to improve the signal-to-noise ratio by $4\textendash 5$ orders of magnitude in comparison to CAST,
which enables us to probe the QCD axion models in the mass range $m_a\gtrsim 3\mathrm{meV}$.
The huge magnets furnished in IAXO have potential applications for various axion dark matter searches in addition to the search for solar axions.
In particular, there is a discussion that the cold dark matter axions can be probed in broad mass ranges by the use of ``dish antennas"~\cite{Horns:2012jf,Jaeckel:2013eha}.
It is argued that 8 dishes of IAXO can probe the mass range $0.01\textendash 1\mathrm{meV}$ continuously in the future~\cite{Redondo2014}.

\begin{figure}[htbp]
\begin{center}
\includegraphics[scale=1.0]{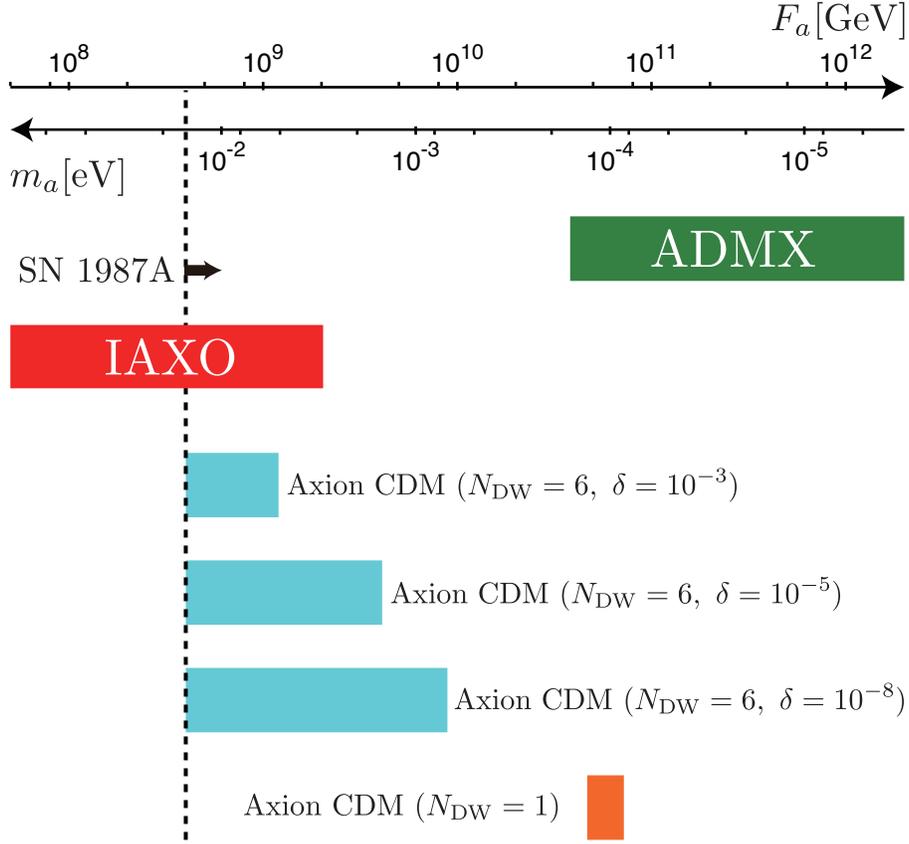}
\end{center}
\caption{The schematics of the parameter ranges where the axion becomes cold dark matter for the models with $N_{\rm DW}=1$ (orange interval)
and $N_{\rm DW}=6$ (light blue interval). The width of the orange interval corresponds to the uncertainty of the mass given by Eq.~\eqref{constraint_ma}.
The blue intervals represent the allowed region for $\delta=10^{-8}$, $10^{-5}$, and $10^{-3}$, and they also contain the uncertainties appearing in the
estimation of $\Omega_{a,\mathrm{tot}}h^2$ (see text for details).
The vertical dotted line corresponds to the bound from the observation of SN1987A [Eq.~\eqref{constraint_astro}].
Red and green intervals represent the regions that will be covered by IAXO and ADMX, respectively.}
\label{fig9}
\end{figure}

Figure~\ref{fig9} shows the predicted mass ranges, in which axions become dominant component of dark matter,
and the ranges covered by the planned detectors.
Here, we also put the mass range predicted by the model with $N_{\rm DW}=1$ [see Eq.~\eqref{constraint_ma}].
For the models with $N_{\rm DW}>1$, the predicted mass range becomes different according to the value of $\delta$.
We also note that the bias parameter $\Xi$ must take a value of $\mathcal{O}(10^{-52}\textendash 10^{-50})$ in this case (see Figs.~\ref{fig7} and~\ref{fig8}).
In these models, the value of $m_a$ ($F_a$) becomes higher (smaller) than that in the models with $N_{\rm DW}=1$,
since the long-lived domain walls copiously produce axions at later times,
which enhances the dark matter abundance compared to the case with short-lived domain walls.
As shown in Fig.~\ref{fig9}, these parameter regions can be probed in the next generation experiments such as IAXO.


\section{\label{sec5}Discussion}
We have considered the scenario where the PQ symmetry is broken after inflation, and
investigated the production of cold dark matter axions from topological defects.
The total dark matter abundance is given by the sum of the misalignment component, the contribution from strings,
and that from string-wall systems.
The contribution from the misalignment mechanism and strings are given by Eq.~\eqref{Omega_a_mis_2}
and Eq.~\eqref{Omega_a_string_1} [or Eq.~\eqref{Omega_a_string_2}], respectively.
The estimation of the contribution from string-wall systems is different according to whether $N_{\rm DW}=1$ or $N_{\rm DW}>1$.

In the models with $N_{\rm DW}=1$, string-wall systems are short-lived, and they decay soon after the formation.
In this case, the contribution from them is estimated by Eq.~\eqref{Omega_a_dec_1} or Eq.~\eqref{Omega_a_dec_2}.
The requirement that the total axion abundance do not exceed the observed cold dark matter abundance
leads to the constraints given by Eqs.~\eqref{constraint_Fa} and~\eqref{constraint_ma}.
These results imply that the values of the axion decay constant and the axion mass become $F_a\simeq (4.6\textendash 7.2)\times 10^{10}\mathrm{GeV}$
and $m_a\simeq (0.8\textendash 1.3)\times 10^{-4}\mathrm{eV}$, respectively, if the axion is dominant component of dark matter.

In the models with $N_{\rm DW}>1$, string-wall systems become stable and long-lived, resulting in a problem in the standard cosmology.
Such a problem can be avoided by introducing the bias term [Eq.~\eqref{V_phi_bias}], which leads to the annihilation of them at a late time.
In this case, the contribution from them is estimated by Eq.~\eqref{Omega_a_dec_long_exact_scaling} or Eq.~\eqref{Omega_a_dec_long_dev_scaling}.
The observational constraints on the model parameters are severe as shown in Figs.~\ref{fig7} and~\ref{fig8},
and the whole parameter region is excluded without a tuning of $\mathcal{O}(10^{-3}\textendash 10^{-2})$ on the phase parameter $\delta$ of the bias term.
On the other hand, if we allow the mild tuning on $\delta$, the axion can be dark matter in the parameter range $F_a\approx\mathcal{O}(10^8\textendash 10^{10})\mathrm{GeV}$
or $m_a\approx\mathcal{O}(10^{-4}\textendash 10^{-2})\mathrm{eV}$, and such a mass range is relevant to the planned experiments.

In order to estimate the abundance of axions, we must solve a complicated dynamics of topological defects.
Therefore, the results suffer from various uncertainties originating in the methodology of numerical simulations.
In Table~\ref{tab8}, we summarize the values and uncertainties of the numerical coefficients estimated in the existing researches.
For the estimation of the contribution from stings, there is a large systematic uncertainty on the estimation of the length parameter $\xi$.
A similar uncertainty exists in the area parameter $\mathcal{A}$ used in the models with $N_{\rm DW}=1$.
In the models with $N_{\rm DW}>1$, a large uncertainty is caused by the procedure to determine
the coefficient $C_d$, whose value slightly varies with $N_{\rm DW}$ and depends on the
criterion for the decay time of domain walls. Furthermore, the expression of the axion abundance becomes different
if we assume the deviation from scaling solution, as shown in Eqs.~\eqref{Omega_a_dec_long_exact_scaling} and~\eqref{Omega_a_dec_long_dev_scaling}.

{\tabcolsep = 2mm
\begin{table}[h]
\begin{center} 
\caption{The values of parameters estimated in numerical simulations and their uncertainties.}
\vspace{3mm}
\begin{tabular}{ l l l }
\hline\hline
Symbol & Definition & Estimated value \\
\hline 
$\xi$ & Eq.~\eqref{rho_string} & $1.0\pm 0.5$ \\
$\epsilon$ & Eq.~\eqref{mean_omega_a_string} & $4.02\pm0.70$ \\
$\mathcal{A}$ ($N_{\rm DW}=1$) & Eq.~\eqref{rho_wall_t1} & $0.50\pm0.25$ \\
$\tilde{\epsilon}_w$ ($N_{\rm DW}=1$) & Eq.~\eqref{tilde_epsilon_w_def} & $3.23 \pm 0.18$ \\
$\mathcal{A}$ ($N_{\rm DW}>1$) & Eq.~\eqref{rho_wall_t} & cf. Table~\ref{tab4} \\
$\mathcal{A}_{\rm form}$ ($N_{\rm DW}>1$) & Eq.~\eqref{rho_wall_t_dev} & cf. Table~\ref{tab5} \\
$C_d$ (exact scaling) & Eq.~\eqref{t_dec_exact_scaling} & cf. Table~\ref{tab6} \\
$C_d$ (deviation from scaling) & Eq.~\eqref{t_dec_dev_scaling} & cf. Table~\ref{tab6} \\
$\tilde{\epsilon}_a$ ($N_{\rm DW}>1$) & Eq.~\eqref{tilde_epsilon_a_def} & cf. Table~\ref{tab7} \\
\hline\hline
\label{tab8}
\end{tabular}
\end{center}
\end{table}
}

In addition to the above issues, there are other causes of uncertainties, which are not addressed in the present work.
First, we note that the axion abundance also depends on the scale $\Lambda_{\rm QCD}$, for which we use $\Lambda_{\rm QCD}=400\mathrm{MeV}$
in this paper. The uncertainty of this QCD scale can affect the estimation of the cold dark matter abundance.\footnote{In Ref.~\cite{Wantz:2009qk},
the QCD scale is evaluated in the IILM as $\Lambda_{\rm QCD}\approx 400\mathrm{MeV}$ with an overall error of $44\mathrm{MeV}$.}
Second, we use the approximation that axions are exactly massless for $t<t_1$, and that the effect of the finiteness of the mass term
becomes relevant only for $t>t_1$. The continuous change of the form of the axion potential around the time $t_1$ might modify the estimation
of the present energy density of axions, but such an effect is not considered in the present analysis.
Finally, in the numerical simulation, we vary the ratio between the axion mass $m_a$ (or $\Lambda_{\rm QCD}$) and 
the PQ scale $\eta$ only in the range of $\mathcal{O}(0.1)$.
In actuality, there is a large hierarchy between these two mass scales, and it is necessary to
confirm that the present results of the numerical simulations are unchanged
even if we set the ratio $m_a/\eta$ (or $\Lambda_{\rm QCD}/\eta$) as a smaller value.
Understanding the consequences of these subtleties is not straightforward, but it is necessary to improve the accuracy of theoretical calculations
with a view to investigating axion physics in the next decades of experimental studies.

\begin{acknowledgments}
The authors gratefully thank Takashi Hiramatsu for discussions on the numerical simulations.
This work is supported by Grant-in-Aid for Scientific research from the Ministry of Education, 
Science, Sports, and Culture (MEXT), Japan, No. 25400248 (M.~K.), 
World Premier International Research Center Initiative (WPI Initiative), MEXT, Japan.
Numerical computation in this work was carried out at the Yukawa Institute Computer Facility.
K.~S.~is supported by the Japan Society for the Promotion of Science through research fellowships.
T.~S.~is supported by the Academy of Finland grant 1263714.
We thank the CSC - IT Center for Science (Finland) for computational resources.
\end{acknowledgments}


\appendix

\section{\label{secA}Energy density of axions from long-lived domain walls}
\renewcommand{\theequation}{A.\arabic{equation}}
\setcounter{equation}{0}
In this appendix, we calculate the present energy density of axions produced from long-lived domain walls, which is used to 
obtain the constraints for the models with $N_{\rm DW}>1$ in Sec.~\ref{sec4-2}.
The following analysis is similar to that performed in Sec.~4.1 of Ref.~\cite{Hiramatsu:2012sc}.
In that work, the radiation of gravitational waves was also considered
as the energy loss mechanism of domain walls, but it turned out that the gravitational radiation is insignificant
in the parameter region of interest [i.e. $\Xi\gg\mathcal{O}(10^{-58})$].
Therefore, we omit the effect of the gravitational radiation in the following analysis.
We also take account of the possibility of the deviation from the scaling solution parameterized in Eq.~\eqref{rho_wall_t_dev}.

After the formation of domain walls, the evolution of their energy density $\rho_{\rm wall}$ and that of axions radiated from them
$\rho_{a,\mathrm{dec}}$ is described by
\begin{align}
\frac{d\rho_{\rm wall}}{dt} &= -H\rho_{\rm wall} - \left.\frac{d\rho_{\rm wall}}{dt}\right|_{\rm emission}, \label{drho_wall_dt}\\
\frac{d\rho_{a,\mathrm{dec}}}{dt} &= -3H\rho_{a,\mathrm{dec}} + \left.\frac{d\rho_{\rm wall}}{dt}\right|_{\rm emission}, \label{drho_a_dec_dt}
\end{align}
where $(d\rho_{\rm wall}/dt)|_{\rm emission}$ is the energy loss rate of the walls due to the radiation of axions, and $\rho_{\rm wall}$ is given by Eq.~\eqref{rho_wall_t_dev}:
\begin{equation}
\rho_{\rm wall}(t) = \frac{\mathcal{A}(t)\sigma_{\rm wall}}{t} = \frac{\mathcal{A}_{\rm form}\sigma_{\rm wall}}{t_{\rm form}}\left(\frac{t_{\rm form}}{t}\right)^p.
\label{rho_wall_t_dev_ap}
\end{equation}
From Eqs.~\eqref{drho_wall_dt} and~\eqref{rho_wall_t_dev_ap}, we obtain
\begin{equation}
\left.\frac{d\rho_{\rm wall}}{dt}\right|_{\rm emission} = (2p-1)\frac{\mathcal{A}(t)\sigma_{\rm wall}}{2t^2}.
\end{equation}
Then, Eq.~\eqref{drho_a_dec_dt} leads to
\begin{align}
R(t)^3\rho_{a,\mathrm{dec}}(t) &= \int^t_{t_r}dt'R(t')^3(2p-1)\frac{\mathcal{A}(t')\sigma_{\rm wall}}{2t'^2} \nonumber\\
&\simeq R(t)^3\frac{2p-1}{3-2p}\frac{\mathcal{A}(t)\sigma_{\rm wall}}{t},
\label{R_3_rho_a_dec}
\end{align}
where $t_r$ is the time at which the walls start to radiate axions.
In the second line of Eq.~\eqref{R_3_rho_a_dec}, we ignored the contribution of $t=t_r$
on the assumption that $t\gg t_r$.

The number of axions in the comoving box at the decay time of domain walls $t_{\rm dec}$ is given by
$R(t_{\rm dec})^3\rho_{a,\mathrm{dec}}(t_{\rm dec})/\bar{\omega}_a$, where $\bar{\omega}_a$
is the mean energy of axions radiated from domain wall networks obtained from Eq.~\eqref{tilde_epsilon_a_def}:
\begin{equation}
\bar{\omega}_a = \tilde{\epsilon}_a m_a(0). \label{mean_omega_a_wall_long}
\end{equation}
Note that we used the zero temperature axion mass $m_a(0)$ [Eq.~\eqref{zero_temp_mass}] in Eq.~\eqref{mean_omega_a_wall_long},
since we consider the case with $t_{\rm dec}\gg t_1$.
Assuming that there is no additional production of axions for $t>t_{\rm dec}$,
we estimate the present energy density of axions as
\begin{align}
\rho_{a,\mathrm{dec}}(t_0) &= m_a(0)\left(\frac{R(t_{\rm dec})}{R(t_0)}\right)^3\frac{\rho_{a,\mathrm{dec}}(t_{\rm dec})}{\bar{\omega}_a} \nonumber\\
&= \frac{1}{\tilde{\epsilon}_a}\left(\frac{R(t_{\rm dec})}{R(t_0)}\right)^3\frac{2p-1}{3-2p}\frac{\mathcal{A}(t_{\rm dec})\sigma_{\rm wall}}{t_{\rm dec}}. \label{rho_a_dec_t0_long}
\end{align}
Furthermore, the decay time of domain walls $t_{\rm dec}$ is explicitly written in Eq.~\eqref{t_dec_dev_scaling}:
\begin{equation}
t_{\rm dec} = C_d\left[\frac{\mathcal{A}_{\rm form}\sigma_{\rm wall}}{t_{\rm form}\Xi\eta^4(1-\cos(2\pi/N_{\rm DW}))}\right]^{1/p}t_{\rm form}. \label{t_dec_dev_scaling_ap}
\end{equation}
Hereafter, we use the approximation $t_{\rm form}\simeq t_1$, where $t_1$ is defined by Eq.~\eqref{condition_t1}.
$\sigma_{\rm wall}$ appearing in Eq.~\eqref{t_dec_dev_scaling_ap} is given by Eq.~\eqref{sigma_wall} with $m_a=m_a(0)$.
Now we have
\begin{align}
\frac{R(t_{\rm dec})}{R(t_0)} &= \frac{R(t_{\rm eq})}{R(t_0)} \frac{R(t_{\rm dec})}{R(t_{\rm eq})} \nonumber\\
&\simeq 4.64\times 10^{-14}\times[7.22\times 10^3]^{1/2p}\times C_d^{1/2}
\left[\frac{\mathcal{A}_{\rm form}}{N_{\rm DW}^4(1-\cos(2\pi/N_{\rm DW}))}\right]^{1/2p} \nonumber\\
&\quad \times\left(\frac{g_{*,1}}{80}\right)^{(1/p-1)n/4(4+n)}\left(\frac{\Xi}{10^{-52}}\right)^{-1/2p}\left(\frac{F_a}{10^{10}\mathrm{GeV}}\right)^{(4p-16-3n)/2p(4+n)}\left(\frac{\Lambda_{\rm QCD}}{400\mathrm{MeV}}\right)^{-1+2/p},
\label{R_dec_R_0}
\end{align}
where $t_{\rm eq}$ is the time of matter-radiation equality.
In the second line of Eq~\eqref{R_dec_R_0}, we used the relations $R(t_{\rm eq})/R(t_0)=4.15\times 10^{-5}(\Omega_{\rm CDM}h^2)^{-1}$,
$R(t_{\rm dec})/R(t_{\rm eq}) = (H(t_{\rm eq})^2/2H(t_{\rm dec})^2)^{1/4}$,
and $H(t_{\rm eq})=1.13\times 10^{-35}(\Omega_{\rm CDM}h^2)^2\mathrm{GeV}$ [see e.g. Ref.~\cite{Weinberg:2008zzc}].
Also, we note that
\begin{equation}
\frac{\mathcal{A}(t_{\rm dec})\sigma_{\rm wall}}{t_{\rm dec}} = \frac{\mathcal{A}_{\rm form}\sigma_{\rm wall}}{t_{\rm form}}\left(\frac{t_{\rm form}}{t_{\rm dec}}\right)^p
= C_d^{-p}\Xi\eta^4\left[1-\cos\left(\frac{2\pi}{N_{\rm DW}}\right)\right]. \label{A_dec_sigma_wall_t_dec}
\end{equation}
Using Eqs.~\eqref{rho_a_dec_t0_long},~\eqref{R_dec_R_0}, and~\eqref{A_dec_sigma_wall_t_dec}, we finally obtain
\begin{align}
\Omega_{a,\mathrm{dec}}h^2 &= \frac{\rho_{a,\mathrm{dec}}(t_0)h^2}{\rho_{c,0}} \nonumber\\
&= 1.23\times 10^{-6}\times[7.22\times 10^3]^{3/2p}\times\frac{1}{\tilde{\epsilon}_a}\frac{2p-1}{3-2p}C_d^{3/2-p}
\mathcal{A}_{\rm form}^{3/2p}\left[N_{\rm DW}^4\left(1-\cos\left(\frac{2\pi}{N_{\rm DW}}\right)\right)\right]^{1-3/2p} \nonumber\\
&\quad\times \left(\frac{g_{*,1}}{80}\right)^{3(1/p-1)n/4(4+n)}\left(\frac{\Xi}{10^{-52}}\right)^{1-3/2p}\left(\frac{F_a}{10^{10}\mathrm{GeV}}\right)^{4+3(4p-16-3n)/2p(4+n)}
\left(\frac{\Lambda_{\rm QCD}}{400\mathrm{MeV}}\right)^{-3+6/p}. \label{Omega_a_dec_long_dev_scaling_A}
\end{align}

If we assume the exact scaling solution ($p=1$), Eq.~\eqref{Omega_a_dec_long_dev_scaling_A} reduces to\footnote{The estimation of $\Omega_{a,\mathrm{dec}}h^2$ shown in
Eq.~\eqref{Omega_a_dec_long_exact_scaling_A} differs from the result of Ref.~\cite{Hiramatsu:2012sc} in some respects.
First, the dependence on $N_{\rm DW}$ is modified since we use a different modeling for $\Delta V$ [Eq.~\eqref{Delta_V}].
Second, the exponent of $\Lambda_{\rm QCD}$ is different
since here we use the expression for $m_a(0)$ given by Eq.~\eqref{zero_temp_mass}
rather than the naive estimation $m_a= 6\times 10^{-4}\mathrm{eV}(10^{10}\mathrm{GeV}/F_a)$ used in Ref.~\cite{Hiramatsu:2012sc}.}
\begin{equation}
\Omega_{a,\mathrm{dec}}h^2 = 0.756\times\frac{C_d^{1/2}}{\tilde{\epsilon}_a}\left[\frac{\mathcal{A}^3}{N_{\rm DW}^4(1-\cos(2\pi/N_{\rm DW}))}\right]^{1/2}
\left(\frac{\Xi}{10^{-52}}\right)^{-1/2}\left(\frac{F_a}{10^{10}\mathrm{GeV}}\right)^{-1/2}\left(\frac{\Lambda_{\rm QCD}}{400\mathrm{MeV}}\right)^3.
\label{Omega_a_dec_long_exact_scaling_A}
\end{equation}
Here, we used $\mathcal{A}_{\rm form}=\mathcal{A}=\mathrm{constant}$, which holds for the case with $p=1$.

\section{\label{secB} Error estimation}
\renewcommand{\theequation}{B.\arabic{equation}}
\setcounter{equation}{0}
In this appendix, we shortly comment on the estimation of the error of the mean energy of radiated axions.
Suppose that for each realization of the simulations we obtain the power spectrum
$P^{(r)}(k_i,\tau)$ and its covariance matrix\footnote{See Eq.~(C.16) of Ref.~\cite{Hiramatsu:2012sc} for a more concrete definition of the covariance matrix.}
$C_{ij}^{(r)}(\tau)\equiv \mathrm{Cov}[P^{(r)}(k_i,\tau),P^{(r)}(k_j,\tau)]$, where the index $r=1,2,\dots, N_r$ indicates a realization and
$N_r$ is the total number of realizations.
Following Eqs.~\eqref{diff_spectrum} and~\eqref{mean_omega}, we compute the mean energy as
\begin{align}
\bar{\omega}_a^{(r)} &= \frac{X}{Y}, \\
X &\equiv \sum_{i=1}^{n_{\rm bin}}\big\{P^{(r)}(k_i,\tau_B) - \mathcal{R}(k_i,\tau_A,\tau_B) P^{(r)}(k_i,\tau_A)\big\}, \\
Y &\equiv \sum_{i=1}^{n_{\rm bin}}\frac{1}{\omega_a(k_i,\tau_B)}\big\{P^{(r)}(k_i,\tau_B) - \mathcal{R}(k_i,\tau_A,\tau_B) P^{(r)}(k_i,\tau_A)\big\}.
\end{align}
Then, the variance of the mean energy is given by
\begin{align}
\mathrm{Var}\left[\bar{\omega}_a^{(r)}\right] 
&= \mathrm{Cov}\left[\bar{\omega}_a^{(r)},\bar{\omega}_a^{(r)}\right] \nonumber\\
&= \frac{1}{Y^2}\left(\mathrm{Cov}[X,X]-2\frac{X}{Y}\mathrm{Cov}[X,Y]+\frac{X^2}{Y^2}\mathrm{Cov}[Y,Y]\right) \nonumber\\
&= \frac{1}{Y^2}\sum_{i=1}^{n_{\rm bin}}\sum_{j=1}^{n_{\rm bin}}\left(1-\frac{\bar{\omega}_a^{(r)}}{\omega_a(k_i,\tau_B)}\right)\left(1-\frac{\bar{\omega}_a^{(r)}}{\omega_a(k_j,\tau_B)}\right)\left\{\mathcal{R}(k_i,\tau_A,\tau_B)\mathcal{R}(k_j,\tau_A,\tau_B)C_{ij}^{(r)}(\tau_A)+C_{ij}^{(r)}(\tau_B)\right\}.
\label{var_k}
\end{align}
In the last line of the above equation, we simply assumed that there is no correlation between $\tau_A$ and $\tau_B$.

In the previous studies~\cite{Hiramatsu:2010yu,Hiramatsu:2012gg,Hiramatsu:2012sc}, we obtained the final results for the mean momentum
and its error by simply averaging over the results of $N_r$ realizations.
However, such a simple average might lead to inappropriately large uncertainties in the case where the variance $\mathrm{Var}[\bar{k}^{(r)}]$ (or $\mathrm{Var}[\bar{\omega}_a^{(r)}]$)
changes significantly for each realization.
Instead of using such a simple average, in this work we use the following weighted averages
\begin{align}
\bar{\omega}_a = \frac{\displaystyle{\sum_{r=1}^{N_r}}\frac{\bar{\omega}_a^{(r)}}{\mathrm{Var}[\bar{\omega}_a^{(r)}]}}{\displaystyle{\sum_{r=1}^{N_r}}\frac{1}{\mathrm{Var}[\bar{\omega}_a^{(r)}]}} \qquad \mathrm{and} \qquad
\Delta\bar{\omega}_a = \frac{N_r}{N_r-1}\frac{\displaystyle{\sum_{r=1}^{N_r}}\frac{(\bar{\omega}_a^{(r)}-\bar{\omega}_a)^2}{\mathrm{Var}[\bar{\omega}_a^{(r)}]}}{\displaystyle{\sum_{r=1}^{N_r}}\frac{1}{\mathrm{Var}[\bar{\omega}_a^{(r)}]}}.
\label{new_average}
\end{align}
The outcomes of these new averaging methods are discussed in Sec.~\ref{sec3-3}.

\end{document}